\documentclass[pre,onecolumn]{revtex4-1}
\usepackage{amsmath,amssymb}
\usepackage{color}
\usepackage{graphicx}
\include{epsf}
\usepackage{clrscode}
\usepackage{dsfont}

\def\(({\left(}
\def\)){\right)}
\def\[[{\left[}
\def\]]{\right]}
\def\e{{\rm e}}

\newcommand{\bF}{{\textbf {F}}}

\newcommand{\bx}{{\textbf {x}}}

\newcommand{\bs}{{\textbf {s}}}
\newcommand{\bv}{{\textbf {v}}}
\newcommand{\by}{{\textbf {y}}}

\newcommand{\one}{{\mathds{1}}}

\newcommand{\be}{\begin{equation}}
\newcommand{\ee}{\end{equation}}
\newcommand{\bea}{\begin{eqnarray}}
\newcommand{\eea}{\end{eqnarray}}

\newcommand{\ox}{\overline{x}}

\def\expect{\mathbb E}

\begin{document}

\title{Probabilistic Reconstruction in Compressed Sensing:\\ Algorithms,
  Phase Diagrams, and Threshold Achieving Matrices}

\author{Florent Krzakala$^{1,*}$, Marc  M\'ezard$^2$, Francois Sausset$^2$,  Yifan Sun$^{1,3}$  and Lenka Zdeborov\'a$^4$}

\affiliation{
 $^1$ CNRS and ESPCI ParisTech, 10 rue Vauquelin, UMR 7083 Gulliver, Paris 75005, France. \\
$^2$ Univ. Paris-Sud \& CNRS, LPTMS, UMR8626,  B\^{a}t.~100, 91405
Orsay, France. \\
$^3$ LMIB and School of Mathematics and Systems Science, Beihang University, 100191 Beijing, China. \\
$^4$ Institut de Physique Th\'eorique, IPhT, CEA Saclay, and URA 2306,
CNRS, 91191 Gif-sur-Yvette, France.\\
$^*$ To whom correspondence shall be sent: fk@espci.fr\\
}

\begin{abstract}
  Compressed sensing is a signal processing method that acquires data
  directly in a compressed form. This allows one to make less
  measurements than what was considered necessary to record a signal,
  enabling faster or more precise measurement protocols in a wide
  range of applications.  Using an interdisciplinary approach, we have
  recently proposed in \cite{KrzakalaPRX2012} a strategy that allows
  compressed sensing to be performed at acquisition rates approaching to
  the theoretical optimal limits. In this paper, we give a more
  thorough presentation of our approach, and introduce many new
  results. We present the probabilistic approach to reconstruction and
  discuss its optimality and robustness. We detail the derivation of
  the message passing algorithm for reconstruction and expectation
  maximization learning of signal-model parameters. We further develop the
  asymptotic analysis of the corresponding phase diagrams with and
  without measurement noise, for different distribution of signals,
  and discuss the best possible reconstruction performances
  regardless of the algorithm. We also present new efficient seeding
  matrices, test them on synthetic data and analyze their performance
  asymptotically.
\end{abstract}

\date{\today}
\maketitle

\tableofcontents

\newpage
\section{Introduction}

\subsection{Background on compressed sensing}

When acquiring a signal, one needs to perform as many measurements as
the number of unknowns. For a continuous signal, for instance, this
translates into the Nyquist's law: in order to sample perfectly, the
sampling rate must be at least twice the maximum frequency present in
the signal. This conventional approach underlies virtually all signal
acquisition protocols used in physics experiments, in audio and visual
electronics, in medical imaging devices and so on.  The compressed sensing
(CS) approach is triggering a major evolution in signal acquisition
that goes against this common wisdom: According to CS, one can recover
signals and images perfectly using far fewer measurements, and this
results in a gain of time, cost, and precision. To make this possible,
CS relies on the fact that many signals of interest contain redundancy
and thus are sparse in some basis (i.e they contain many coefficients
close to or equal to zero when represented in some domain). This is
the same insight used in data compression: the pictures we take with
our cameras can be strongly compressed in the wavelet basis (almost)
without the loss of quality, and this idea is for instance behind the
JPEG 2000 algorithm. It would thus be convenient to record signals
directly in a compressed format (thus the origin of the name
``compressed sensing'') to save both in memory space and in number of
measurements. The CS approach aims to design measurement protocols that
acquire only the necessary information about the signal, in some
compressed form. In a second step, one uses computational power to
reconstruct the original signal exactly
\cite{Candes:2008,Donoho:06}. The inverse problem posed by this second
step is in general highly non-trivial.

Mathematically, the CS problem can be posed as follows: given an
$N$-component signal $\bs$, one makes $M$ measurements that are grouped
into an $M$-component vector $\by$, obtained from $\bs$ by a linear
transformation using a $M\times N$ measurement matrix
$\bF$,   given by  $y_{\mu} = \sum_{i=1}^N F_{\mu i}s_i$ with $\mu=1,2,\ldots, M$.
 The observer has freedom in the choice of the measurement protocol,
 and he knows the results of the measure (the $M$ values in vector $\by$) and
the $M\times N$ matrix $\bF$ (in general various kinds of noise are
present, as we shall discuss below). The aim is then to reconstruct the signal
$\bs$ from the knowledge of $\bF$ and $\by$. This amounts to inverting the
linear system $\by = \bF\bs$. However, we want to have $M$ as small as
possible and when $M<N$ there are fewer equations than unknowns. The
system is under-determined and the inverse problem is ill-defined. CS,
however, deals with sparse signals $\bs$, in the sense that only $K<N$
of the components are non-zero. In the noiseless case, an exact
reconstruction is possible for such signals as soon as $M>K$, and this
condition is also a necessary one for instance in the case where the
non-zero component of the signal are independent identically
distributed  (iid) real variables, drawn from a distribution
with a continuous part. This
ability to recover signals using only a limited number of measurements
is crucial in many fields ranging from experimental physics and image
processing to astronomy or systems biology, making CS a very
attractive concept.


The most widely used technique in CS is based on a development that took place six years ago thanks
to the works of Cand\`{e}s, Tao, Donoho and
collaborators \cite{CandesTao:05,Candes:2008,Donoho:06,CandesRombergTao06}:
they proposed to find the vector satisfying the constraints
${\by} = {\bF\bx}$ which has the smallest $\ell_1$ norm, defined as
$||x||_{\ell_1}=\sum_{i=1}^N |x_i|$. This optimization problem is convex and
can be solved using efficient linear programming techniques. They have
also suggested the use of a random measurement matrix $\bF$ with iid entries. This is a crucial point, as it makes the
$M$ measurements random and incoherent. Incoherence expresses the idea
that objects having a sparse representation must be spread out in the
domain in which they are acquired, just as a Dirac function or a spike
in the time domain is spread out in the frequency domain after a
Fourier transform. These ideas have led to fast and efficient
algorithms, and the $\ell_1$-reconstruction is now widely used, and has
been at the origin of the burst of interest in CS over the last few
years. It is possible to compute exactly the performance of the $\ell_1$
reconstruction in the limit $N\to \infty$, and the analytic study
shows the appearance of a sharp phase transition
\cite{Donoho:2005wq}. For any
signal with density $\rho=K/N$, the $\ell_1$ reconstruction gives indeed
the exact result $x = s$ with probability one only if
$\alpha=M/N>\alpha_{\ell_1}(\rho)$ where $\alpha_{\ell_1}(\rho)$ is, however,
larger than $\rho$. The $\ell_1$ reconstruction is thus sub-optimal: it requires more
measurements than theoretically necessary.

\subsection{Our main results}
\label{block_matrices}
In this paper we analyze a probabilistic reconstruction of the signal
in compressed sensing, which we have introduced  in
\cite{KrzakalaPRX2012}. We provide here a more detailed presentation,
and we include several new results. We use a simplification of the belief
propagation (BP) algorithm, also known as approximate message passing
(AMP) \cite{DonohoMaleki09} or generalized approximate message passing
(G-AMP) \cite{Rangan10b} in the context of CS. The
probabilistic approach is combined with an expectation-maximization
type of learning of parameters as in \cite{KrzakalaPRX2012} (which has
been independently proposed in the context of G-AMP in
\cite{VilaSchniter11}). We use the replica and cavity methods to
analyze on one hand the asymptotic performance of the BP algorithm and
on the other hand the information theoretical limits for signal
reconstruction, and the associated phase transitions. For sensing
matrices with iid entries there is a region of parameters (signal
sparsity, undersampling rate and measurement noise) in which there is
a gap between the BP reconstruction and the optimal reconstruction. In
this hard region, BP iterations are blocked in a suboptimal fixed
point. We also study in detail the phase diagram in the presence of
measurement noise and observe that the region where BP is suboptimal
persists, but becomes smaller and eventually disappears as the noise
variance grows.  Analyzing the origin of this algorithmic barrier and
thinking about an analogy with crystal nucleation in
\cite{KrzakalaPRX2012} we designed and tested BP reconstruction with
seeded measurement matrices for which this gaps shrinks or entirely
disappears. The implementation of our matrices and of the algorithm is
available at http://aspics.krzakala.org/.

We  now describe the organization of this paper and list its main contributions:

\begin{itemize}

\item{\bf Optimality of the probabilistic reconstruction} 
  We review in Sec~\ref{optimal_B} the well known fact that probabilistic inference
  is optimal when the signal model matches the actual signal
  distribution. For good performance one usually requires a signal
  model that is ``close enough'' to the actual signal to be
  inferred. The unavailability of such a good signal model is often at
  the basis of the critics of this probabilistic - Bayesian -
  inference. The situation is much more favorable in the case of noiseless compressed
  sensing. We noticed, and
  proved, in \cite{KrzakalaPRX2012} that in the case of noiseless
  CS probabilistic inference is optimal even if
  the signal model mismatches seriously the actual signal, details are in
  Sec.~\ref{Optimality}. This property makes our approach very
  robust. In our numerical experiments we successfully use the
  Gauss-Bernoulli model even for signals that are far from having iid
  Gauss-Bernoulli components. Despite this result, in practice it turns out to be
  useful to incorporate expectation-maximization learning of
  parameters of the signal model, as described in Sec.~\ref{EM_learning}.

\item{\bf The message passing reconstruction algorithm} We derive in
  detail the reconstruction algorithm and discuss how it is related to
  the existing ones. In section \ref{BP_derivation} we give it in a
  form where the messages are being sent between signal components and
  measurement components and back - this being equivalent to the
  relaxed-BP algorithm \cite{GuoWang06,Rangan10}. In section
  \ref{sec:TAP} we then derive a simplified form where messages
  ``live'' only on the signal components and on the measurement
  component. This form is related to the seminal
  Thouless-Anderson-Palmer (TAP) \cite{ThoulessAnderson77} equations
  in spin glass theory, and is equivalent to the AMP algorithm in the
  context of CS \cite{DonohoMaleki10,Rangan10b}. For
  measurement matrices with iid entries further simplifications of the
  algorithm exist, and are useful for a more efficient 
  implementation, and this is shown in
  Sec.~\ref{TAP_simpl}. We also derive the BP equations for
  expectation maximization learning of parameters in Sec.~\ref{BP_learning}. 

\item{\bf Asymptotic analysis of the algorithm and of the
    probabilistic approach} We use the cavity and replica methods to
  perform two types of asymptotic analysis. On one hand using the
  density evolution we describe
  the behavior of the belief propagation algorithm in the limit of
  large systems (Sec.~\ref{sec:DE}), on the other hand using the
  replica method we compute the theoretical limits
  for reconstruction (Sec.~\ref{replicas}), which are non-trivial in particular in the
  presence of noise and by definition do not depend on the
  algorithm. We derive the asymptotic evolution for measurement matrices having iid (or iid per
  block) entries of zero mean and variance $1/N$. The equations are
  independent of the other details of the distribution of matrix
  elements, and these predictions thus hold for many type of matrices
  (for instance, Gaussian or discrete binary ones). This makes
  our results very robust. In Sec.~\ref{sec:Bayes} we then discuss the
  simplifications that appear in the Bayes optimal case of matching
  signal model and signal distribution. In Sec.~\ref{DE_learning} we
  derive the asymptotic evolution of the parameters in expectation
  maximization learning. Finally in Sec.~\ref{replica_1D} we summarize
  all these previous equations in the case of block measurement
  matrices. 

\item{\bf Phase transitions, phase diagrams} Using both the BP
  reconstruction algorithm and the asymptotic analysis we study the
  phase diagram and associated phase transitions for reconstruction of
  the signal. We study several settings: The optimal Bayesian
  inference when the signal model matches the signal distribution in
  Sec.~\ref{Res:Bayes}, the case when the signal model does not match
  the signal distribution and the phase diagram after expectation
  maximization learning in Sec.~\ref{Res:learning}, the phase diagram
  in the presence of measurement noise in Sec.~\ref{Sec:Noisy}, and
  the reconstruction with seeding block matrices in
  Sec.~\ref{Sec:Seeding}. Note that in doing the optimal Bayesian
  inference case, we thus study the best possible reconstruction
  performance, regardless of the algorithm.

\item{\bf Optimality achieving measurement matrices} In
  \cite{KrzakalaPRX2012} we introduced a new type of ``seeding''
  measurement matrices with which theoretically optimal reconstruction
  can be obtained using the BP algorithm. Such a ``threshold
  saturation'' was later on proved for this type of matrices in
  \cite{DonohoJavanmard11} (called ``spatial coupling''). In Sec.~\ref{whywhen} we
  discuss again our motivation for the design of seeding matrices and
  show that there is relatively a lot of freedom in implementing the
  concept of seeding. We give new examples of efficient seeding
  matrices, which are actually simpler and more efficient than the one
  we have introduced earlier. In Sec.~\ref{1D_no_noise} we also show that these matrices are
  effective even when the model signal in the prior is different from
  the actual ones. In Sec.~\ref{1D_noise} we illustrate that one can approach the optimal
  reconstruction limit, even in the case of noisy measurements.

\item{\bf Noise-sensitivity} We discuss in detail the phase diagram
  and the performance of the algorithm in the presence of measurement
  noise in Sec.~\ref{Sec:Noisy}.  We show that there are two regions in the phase
  diagram. Either the BP approach is optimal, i.e. it provides the
  same mean-squared-error as would be obtained by an intractable
  exhaustive search algorithm. Or BP is suboptimal due to an existence
  of a metastable state - in this case optimality can be restored
  using the seeding matrices as we show in Sec.~\ref{1D_noise}. Overall this shows that the present
  approach has the best achievable noise stability.

\item{\bf Rigorous versus exact} It is important to notice that
  the density evolution that we use for asymptotic analysis of BP was
  proven to be exact for the homogeneous matrices
  \cite{BayatiMontanari10}. According to a private communication with
  the authors a proof for the block matrices also exists
  \cite{BayatiLelarge12}.  Therefore, our predictions on the behavior
  of the algorithm are exact. As far as our predictions for
  the optimal inference are concerned, although our presentation here
  is not rigorous, the predictions are
  exact in the context of the series of works
  \cite{WuVerdu10,WuVerdu11,WuVerdu11b}.

\end{itemize}

Let us define here the block measurement matrices that we use in this
paper to implement the seeding concept. Note however, that the seeding
measurement matrices do not have to be block matrices. Other
implementations are possible. We leave for future work an
investigation into 
the optimal design for practical situations.

The block measurement matrices $F_{\mu i}$ are constructed as
follows: The $N$ variables are divided into $L_c$ groups of $N_p$,
$p=1,\dots,L_c$, variables in each group. We denote $n_p=N_p/N$. And
the $M$ measurements are divided into $L_r$ groups of $M_q$,
$q=1,\dots,L_r$, measurements in each group, define
$\alpha_{qp}=M_q/N_p$. Then the matrix $F$ is composed of $L_r\times
L_c$ blocks and the matrix elements $F_{\mu i}$ are generated
independently, in such a way that if $\mu$ is in group $q$ and $i$ in
group $p$ then $F_{\mu i}$ is a random number with zero mean and
variance $J_{q,p}/N$. Thus we obtain a $L_r\times L_c$ coupling matrix
$J_{q,p}$. For the asymptotic analysis we assume that $N_p\to \infty$,
for all $p=1,\dots,L_c$ and $M_q\to \infty$ for all $q=1,\dots,L_r$.
We define $I(\mu)$ ($I(i)$) to be the index of the block to which
$\mu$ ($i$) belongs, $B_q$ is the set of indices in block $q$.  The
case of homogeneous matrix can easily be recovered by setting
$L_c=L_r=1$. Note that not all block matrices are good seeding
matrices, the parameters have to be set in such a way that seeding is
implemented (i.e. existence of the seed and interaction such that the
seed grows). The choice of parameters is discussed in
Sec.~\ref{Sec:Seeding}.

Let us note that for both the homogeneous and the block matrices the
results do not depend on the details of the distribution of its
entries, as far as its mean and variance are fixed.  In our
simulations we mostly use Gaussian distributed random entries, or $\pm
1/N$. The later has the advantage of taking less memory space, since
we can store them with bits and deal with the $\sqrt{N}$ separately
(memory space to store the matrix is the main limitation of our
simulations).

Note also that throughout the paper we use matrix entries of zero
mean. Physical constraints might require the mean to be non-zero, but our
algorithm would have to be modified for such cases. The problem,
however, can be transformed rather easily to one of zero mean.  Consider
indeed the system $\by = \bF\bs$. Summing all $M$ values of the vector
$\by$ (and denoting $\overline{y} =(1/M) \sum_{\mu} y_{\mu}$ and
$\overline{F_i}=(1/M) \sum_{\mu} F_{\mu i}$) one finds
$
 M\overline{y} = \sum_{\mu} \sum_i F_{\mu i} x_i = \sum_{i}
 (\sum_{\mu} F_{\mu i} ) s_i = M \sum_{i} \overline{F_{i}} s_i
$.
Denote $\overline{\by}$ the vector whose all $M$ components are equal to
$\overline{y}$ and $\overline{\bF}$ the $M \times N$ matrix where the
$i$-th column is given by the values
$\overline{F_{i}}$. Then the system
$
\by - \overline{\by} = (\bF-\overline{\bF}) \bs
$
has a matrix with zero mean entries.

\subsection{Related works}

Here we discuss some interesting connections to other works on
compressed sensing. It is important to realize that our main result,
namely the joint design of an algorithm and a class of measurement
matrices that lead to optimal CS reconstruction, and
their analysis, is based on three main ingredients that were previously explored
in the literature. These ingredients are the probabilistic approach,
the use of message passing algorithm for sampling from the probability
distribution, and the design of seeding matrices. It is only the joint
use of these three ingredients that achieves optimal reconstruction,
and the understanding of the reasons owes a lot to accumulated
knowledge from statistical physics of disordered systems (for
instance, using seeding matrices with $\ell_1$ reconstruction is
 useless, because $\ell_1$ reconstruction is not limited by a
glass transition, its limitation is intrinsic to the use of the
$\ell_1$ norm).

\begin{itemize}
\item The state-of-the-art method for signal reconstruction in
  CS is based on the minimization of the $\ell_1$ norm
  of the signal under the linear constraint, for an overview of this
  technique see \cite{Donoho:2005wq,Candes:2008}. A number of works
  also adopted a probabilistic or Bayesian approach
  \cite{JiXue2008,Seeger2008,BaronSarvotham10}. Generically, one
  disadvantage of the probabilistic approach is that no exact
  algorithm is known for evaluation of the corresponding
  expectations. Whereas $\ell_1$ minimization is done exactly using
  linear programming.  In our approach, this problem is resolved with the use of
  belief propagation that turns out to be an extremely efficient
  heuristic. Another issue of the Bayesian approach is the choice of
  the signal model. Whereas the performance of the $\ell_1$
  reconstruction is independent of the signal distribution, this is
  not the case for the Bayesian approach in general. We show that
  actually for the noiseless CS the optimal exact
  reconstruction is possible even if the signal model does not match
  the signal distribution.

\item In the noiseless case of CS it is very intuitive
  that exact reconstruction of the signal is in principle possible if
  and only if the number of measurements is larger than the number of
  non-zero component of the signal, $\alpha>\rho_0$. In a more generic
  case, for instance in the presence of the measurement noise it is
  not straightforward to compute the best achievable mean-squared
  error in reconstruction. These theoretical optimality limits were
  analyzes rigorously in very general cases by
  \cite{WuVerdu11,WuVerdu11b}. These results agree with the
  non-rigorous replica method as developed for CS e.g.
  in
  \cite{RanganFletcherGoyal09,KabashimaWadayama09,GanguliSompolinsky10}. Here
  we analyze the theoretically optimal reconstruction using as well
  the replica method (and explicit its connection with the density
  evolution).

\item The belief propagation (BP) is an inference algorithm that is
  exact on tree graphical model and that is a powerful heuristic also
  on loopy graphical models. It was discovered independently by
  several communities, in coding \cite{Gallager62}, in inference
  \cite{Pearl82}, or in statistical physics
  \cite{ThoulessAnderson77}. See
  \cite{YedidiaFreeman03,KschischangFrey01} for good overviews.
  Belief propagation was used for CS with sparse
  measurement matrices by several authors, see
  e.g. \cite{ZhangPfister08,BaronSarvotham10,KabashimaWadayama11}.  In
  the usual setting, however, CS corresponds to dense
  measurement matrices, hence a fully connected graphical model with
  continuous variables, the canonical form of BP iterations is
  intractable for such a case. However, neglecting only factors that
  go to zero in the large system size limit, the iterative equations
  can be written only for the means and variances of the corresponding
  probability distributions.  Such a belief propagation algorithm was
  used in CS under the name relaxed BP (rBP)
  \cite{GuoWang06,Rangan10}. Again, by neglecting only $o(1)$ terms
  rBP can be further simplified as shown for $\ell_1$ reconstruction
  in \cite{DonohoMaleki09,DonohoMaleki10} this version of the message
  passing was called approximate message passing (AMP), it is
  equivalent to the Thouless-Anderson-Palmer (TAP) \cite{ThoulessAnderson77} equations in spin
  glass theory. The AMP was further generalized (G-AMP) to the case of
  a general signal model in \cite{DonohoMaleki10,Rangan10b}. The
  algorithm that we use here is equivalent to G-AMP. We, however, provide an independent derivation. 

\item The performance of the belief propagation algorithm can be
analyzed analytically in the large system limit. This can be done
either using the replica method, as in \cite{GuoBaron09}, or using density
evolution. An asymptotic density-evolution-like
analysis of the AMP algorithm, called state evolution,
was developed in \cite{DonohoMaleki09}, and more generally in~\cite{BayatiMontanari10}. State evolution is the analog of
density evolution for dense graphs. General analysis of algorithmic
phase transitions for G-AMP was presented in \cite{DonohoJohnstone11}.
In this paper we perform the same density evolution analysis for other variants of the problem (with
learning, where the signal model does not match the signal
distribution, with noise, etc.),
without the rigorous proofs. Our main point is to analyze and
understand the phase transitions that pose algorithmic barriers to the message
passing reconstruction.

\item In cases when the signal distribution is not known, we can use
  expectation maximization (EM) to learn the parameters of the signal
  model \cite{Dempster}. EM learning with the expectation step being
  done with BP was done in e.g.~\cite{DecelleKrzakala11b}.  In the
  context of CS, the EM was applied together with
  message passing reconstruction in~\cite{KrzakalaPRX2012}. An
  independent implementation along the same ideas also appeared in
  \cite{VilaSchniter11} under the name EM-GAMP algorithm (where EM
  stands for expectation-maximization). All the predictions made in
  the present paper thus also apply to the EM-GAMP algorithm.

\item Based on our understanding of the properties of the algorithmic
  barrier encountered by the message passing reconstruction algorithm,
  we have design special seeded measurement matrices for which
  reconstruction is possible even for close-to-optimal measurement
  rates. These matrices are based on the idea of spatial coupling that
  was developed first in error correcting codes
  \cite{FelstromZigangirov99,LentmaierFettweis10}, see
  \cite{KudekarRichardson10} for more transparent understanding and
  results. Several other applications of the same idea exist, in
  different contexts. For an overview see \cite{KudekarRichardson12}.

\item The use of spatial coupling was first suggested for compressed
  sensing in \cite{KudekarPfister10}, where the authors
  observed an improvement over the
  reconstruction with homogeneous measurement matrices (see Fig.~5 in
  \cite{KudekarPfister10}). They, however, did not
  combine all the key ingredients to achieve reconstruction up to
  close to the theoretical limit $\alpha=\rho_0$, as we did in
  \cite{KrzakalaPRX2012}.  Their implementation of belief
  propagation was also not using the simplification under which only
  mean and variance of the messages are needed, hence the algorithm
  was not competitive speed-wise.

  We introduced seeded measurement matrices for CS in
  \cite{KrzakalaPRX2012}, and showed there, both numerically and using the density
  evolution, that with such matrices it is possible to achieve the
  information theoretically optimal measurement rates. The design was
  motivated by the idea of crystal nucleation and growth in statistical
  physics. Subsequent work \cite{DonohoJavanmard11} justified this
  threshold saturation rigorously in the special case when the signal
  model corresponds to the signal distribution, but also more
  generally using the concept of R\'enyi information dimension instead
  of sparsity, as in \cite{WuVerdu10,WuVerdu11b}. Numerical experiments
  with seeded non-random (Gabor-type) matrices were also performed in
  \cite{JavanmardMontanari12}.
\end{itemize}

\newpage
\section{Probabilistic reconstruction in compressed
  sensing} \label{Sec:Probabilistic}

The definition of the compressed sensing problem as studied in this
paper is as follows
\be
     y_\mu = \sum_{i=1}^N  F_{\mu i } s_i  + \xi_\mu  \quad
     \mu=1,\dots,M\, ,  \label{def}
\ee
where $s_i$ are the signal elements, out of which only $K=\rho_0 N$ are
non-zero, $0<\rho_0<1$. We denote by $\phi_0$ the asymptotic empirical distribution of the
non-zero elements.
$F_{\mu i}$ are the elements of a known
measurement matrix, $y_\mu$ are the known result of measurements,  and
$\xi_\mu$ is Gaussian white noise on the measurement with variance
$\Delta_\mu$.
We denote by $\alpha=M/N$ the number of measurements per variable. The goal of
CS is to find an approach (i.e. measurement matrix and
a reconstruction algorithm) that allows reconstruction with as low
values of $\alpha$ as possible.

In the asymptotic theoretical analysis we will be interested in the
case of large signals $N\to \infty$, we will keep signal density
$\rho_0$ and measurement rate $\alpha$ of order one. We also want to
keep the components of the signal and of the measurements of order
one, hence we consider the elements of the measurement matrix to have
mean and variance of order $O(1/N)$.

We shall adopt a probabilistic inference approach to reconstruct the
signal. The aim is to sample a vector $\bx$ from the following probability measure
\be
\hat{P}({\bx}) = \frac{1}{Z}\prod_{i=1}^N \left[ (1-\rho) \delta( x_i)
  + \rho \phi(x_i) \right] \prod_{\mu=1}^M \frac{1}{\sqrt{2\pi
    \Delta_\mu}} e^{-\frac{1}{2\Delta_\mu}(y_\mu - \sum_{i=1}^N F_{\mu
    i}x_i)^2} \, ,\label{p_hat}
\ee
where $Z$, the partition function,  is a normalization constant.
Here we model the signal as stochastic with iid entries, the fraction
of non-zero entries being $\rho>0$ and their distribution being $\phi$, we
restrict ourselves to functions $\phi(x)<\infty$ with finite variance.

We stress that in general the signal properties are not known and hence
(unless stated otherwise) we do not assume that the signal model
matches the empirical signal distribution, $\rho = \rho_0$, $\Delta=\Delta_0$, nor $\phi =
\phi_0$.  Most previous approaches to reconstruction in CS can be
stated in the form (\ref{p_hat}), e.g. the $\ell_1$ minimization is
equivalent to $\rho=1$ and Laplace function $\phi$. One crucial point
in our approach is using $\rho<1$ which includes
the fact that one searches a sparse signal in the model of the signal.

Eq.~(\ref{p_hat}) can be seen as the Boltzmann measure
on the disordered system with Hamiltonian
\be
H(\bx)= - \sum_{i=1}^N \log  \left[ (1-\rho) \delta( x_i) + \rho
  \phi(x_i) \right]  +  \sum_{\mu=1}^M\frac{(y_\mu - \sum_{i=1}^N
  F_{\mu i}x_i)^2}{2\Delta_{\mu}} \, ,\label{Hamiltonian}
\ee
where the ``disorder'' comes from the randomness of the
measurement matrix $F_{\mu i}$ and the results $y_{\mu}$. Stated this
way, the problem is similar to a spin glass with $N$ particles interacting with a long-range
disordered potential. The signal $\bx = \bs$ is a
very special configuration of these particles, that we can call
``planted'', which was used to generate the problem (i.e. the value of the vector
$\by$). In this sense all inference problems are equivalent to planted
spin glass models.

\subsection{Optimality in the noiseless case}
\label{Optimality}

In the noiseless case, $\Delta_\mu\to 0$, sampling from the measure $\hat
P(\bx)$ leads to exact reconstruction as long as
$\alpha>\rho_0$ and the support of the function $\phi$ contains all the
non-zero elements of the signal (i.e. an arbitrary finite function
of finite variance supported on
real numbers for general real entry signals). In particular the
density and the distribution of the true signal does not need to be known,
i.e. $\rho \neq \rho_0$ and $\phi \neq \phi_0$. This is a strong
optimality property that was proven in the large size limit $N\to
\infty$ in \cite{KrzakalaPRX2012} and that can be seen as follows.

Define an auxiliary partition function $Y(D)$ as the normalization of the measure
$\hat P(\bx)$ restricted to configurations at a mean-squared distance $D$ from the signal $\bs$, i.e.
\be
   Y_{\Delta}(D) \equiv  \int_{B_D({\bs})} \prod_{i=1}^N {\rm d}x_i \,
    \prod_{i=1}^N
   \left[ (1-\rho) \delta( x_i) + \rho \phi(x_i) \right]
   \prod_{\mu=1}^M \frac{1}{\sqrt{2\pi \Delta}}
   e^{-\frac{1}{2\Delta}[\sum_{i=1}^N F_{\mu i}(x_i-s_i)]^2}\, , \label{YD}
\ee
where $B_D({\bs})$ is the sphere centered on $\bs$, defined by $\left(\frac{1}{N}\sum_{i=1}^N (x_i-s_i)^2= D\right)$.
When $D\to 0$ and $\Delta\to 0$, the $N$ dimensional integral in
(\ref{YD}) involves a product of $(1-\rho_0+\alpha)N$ Dirac delta
functions. Hence as long as $\alpha>\rho_0$ the function $Y_{\Delta}(D)$ diverges
as $D\to 0$, $\Delta\to 0$. This holds for every matrix $F$ and every
function $\phi$ as long as it is supported on all the elements of $\bs$.

In a second part of the optimality argument one needs to
show that $\lim_{\Delta\to 0}Y_{\Delta}(D)/Y_{\Delta}(0)=0$ whenever $D>0$. First note that only
configurations that solve all the $M$ linear equations give a non-zero
contribution to (\ref{YD}). Second, it is known that the signal
$\bs$ is the solution of the linear system with the largest number of
zero elements \cite{CandesRombergTao06}
hence all the other solutions of the linear system have a negligible
contribution to the integral (necessarily, a smaller number of Dirac delta
functions remains after the integration).


Given this result, it then follows that for any $\rho_0$-dense
original signal $\bs$, and any $\alpha>\rho_0$, the probability $\hat
P(\bs)$ of the original signal goes to one when $\Delta\to 0$. This
result holds as long as the configuration minimizing the $\ell_0$ norm
equals the original signal $\bs$. Remarkably this optimality holds
independently of the distribution $\phi_0$ of the original signal,
which does not even need to be iid.  Hence in the noiseless case,
sampling $\bx$ proportionally to the measure $\hat
P(\bx)$ gives the exact reconstruction in the whole region $\alpha >
\rho_0$.


\subsection{The Bayesian optimality and the Nishimori conditions}
\label{optimal_B}
The probabilistic approach can also be recovered from a Bayesian
point of view. Indeed, given $\bF$ and $\by$, from Bayes theorem,
we have
\be
P(\bx|\bF, \by)  = \frac{P(\bx |\bF)P(\by|\bF,\bx)  }{P(\by|\bF)} \, .
\ee
The value of measurements $\by$ given the knowledge of the matrix
$\bF$ and the signal $\bx$ is, by
definition of the problem, given by
$
P(\by|\bF,\bx) =
\prod_{\mu=1}^M \delta (y_\mu - \sum_{i=1}^N F_{\mu i}x_i )
$
in the noiseless case, and by
\be
P(\by|\bF,\bx) =
\prod_{\mu=1}^M
\frac{1}{\sqrt{2\pi \Delta_\mu}}e^{-\frac{1}{2\Delta_\mu}(y_\mu -
  \sum_{i=1}^N F_{\mu i}x_i)^2}\, , \label{P_yx}
\ee
with random Gaussian measurement noise of variance $\Delta_\mu$, for
measurement $\mu$.
To express the probability $P(\bx | \bF )$ we consider that the
signal does not depend on the measurement matrix (which is true in
all practical situations we are aware of). Further, in this paper, we do not aim to
exploit possible correlations in signal entries (which could only
improve the result of inference) and
hence we model the signal as an iid:
\be
P(\bx | \bF ) =
\prod_{i=1}^N \left[ (1-\rho) \delta( x_i) + \rho \phi(x_i) \right]\,
. \label{P_x}
\ee
Thus the posterior probability of $\bx$ after the measurement of $\by$ is
given by
\be
P(\bx | \bF, \by)  = \frac{1}{Z(\by,\bF)}\prod_{i=1}^N
\left[ (1-\rho) \delta( x_i) + \rho \phi(x_i) \right] \prod_{\mu=1}^M
\frac{1}{\sqrt{2\pi \Delta_\mu}} e^{-\frac{1}{2\Delta_\mu}(y_\mu -
  \sum_{i=1}^N F_{\mu i}x_i)^2} \, ,\label{p_bayes}
\ee
where $Z(\by,\bF)=P(\by|\bF)$ is again the normalization constant. This is nothing else
than the $\hat P(\bx)$ in Eq.~(\ref{p_hat}).

We remind the reader that in the noiseless case,
$\Delta_0=\Delta_\mu=0$, we have the optimality
result for an arbitrary signal, as described in the previous
section. However, for the case with noise, if the true density
of the signal, $\rho_0$, the measurement noise, $\Delta_0$, and the asymptotic empirical distribution of
the signal,  $\phi_0$, are not known then sampling from (\ref{p_bayes}) is in general not
optimal.

However, if we assume knowledge of the true density of the signal,
$\rho=\rho_0$, the measurement noise, $\Delta=\Delta_0$, and the
asymptotic empirical distribution of the signal,  $\phi=\phi_0$, then we just described the Bayes-optimal way
to infer the signal $\bs$ from the knowledge of the matrix $\bF$ and
the measurements $\by$. In
particular, an estimator $\bx^{\star}$ that minimizes the mean-squared
error with respect to the original signal $\bs$, defined as $E=
\sum_{i=1}^N(x_i-s_i)^2/N$, is then obtained from averages of $x_i$ with
respect to the probability measure $P(\bx | \bF, \by)$, i.e.,
\be
x^{\star}_i=\int \text{d}  x_i \,  x_i\,  \nu_i(x_i) \, ,\label{average_marginal}
\ee
where $\nu_i(x_i)$ is the marginal probability distribution of the variable $i$
\be
\nu_i(x_i) \equiv \int_{\{x_j\}_{j\neq i}} P(\bx| \bF, \by) \, .
\ee
In the remainder of this article we will be using this estimator.
To give another example, the optimal estimator that minimizes the mean ``absolute value'' error ${\rm AVE} =
\sum_{i=1}^N|x_i-s_i|/N$ is given by the median of the marginal
probability $\nu_i(x_i)$.

There are important identities that hold for the Bayes-optimal
inference and that simplify many of the calculations that follow. In
the physics of disordered systems these identities are known as the
Nishimori conditions  \cite{Iba99,NishimoriBook}. Basically, the Nishimori conditions
follow from the fact that the planted configuration (i.e. the original
signal) is an equilibrium configuration with respect to the Boltzmann
measure (\ref{p_bayes}). Hence many properties of the planted
configuration can be computed without its knowledge by averaging over
the distribution (\ref{p_bayes}).

To derive the Nishimori conditions, consider the measurement matrix
$\bF$ fixed and for simplification let us drop the dependence on $\bF$
from the notation. Consider a function $A(\bx)$ depending on a ``trial''
configuration $\bx$. We define the ``thermodynamic average'' of $A$ as
\be
       \langle A(\bx) \rangle \equiv \int {\rm d} \bx\,  A(\bx) P(\bx
       |\by)\, ,
\ee
where $ P(\bx |\by)$ is given by Eq.~(\ref{p_bayes}). Similarly, for a
function $A(\bx_1,\bx_2)$ that depends on two trial configurations
$\bx_1$ and $\bx_2$, we define 
\be
       \langle \langle A(\bx_1,\bx_2) \rangle \rangle \equiv \int {\rm d} \bx_1 \int {\rm d} \bx_2 \,  A(\bx_1,\bx_2) P(\bx_1
       |\by) P(\bx_2
       |\by)\, ,
\ee
For a function $B$
that depends on the measurement $\by$ and on the signal $\bs$ we define the ``disorder average'' as
\be
             [B(\bs,\by)]  \equiv \int {\rm d}\by \int {\rm d}\bs
             \,  P(\bs) \, P(\by|\bs) B(\bs,\by) \, ,
\ee
where the signal distribution $P(\bs)$ is given by Eq.~(\ref{P_x}),
and $P(\by|\bs)$ is the probability of a measurement $\by$ given the
signal $\bs$, as in Eq.~(\ref{P_yx}). Note that if $B$ does not
explicitly depend on $\bs$ then we have $[B(\by)]  \equiv \int
{\rm d}\by Z(\by) B(\by)$ because $Z(\by) = \int {\rm d}\bs \,  P(\bs) \, P(\by|\bs) $.
Using these definitions we obtain
\bea
 [ \langle A(\bx,\bs) \rangle ] &=& \int {\rm d}\by \int {\rm d}\bs
 \, P(\bs) P(\by|\bs)   \int {\rm d} \bx \, A(\bx,\bs)\,
   P(\bx|\by) =
\int {\rm d}\by\,  Z(\by) \int {\rm d}\bs \int {\rm
  d}\bx \,  A(\bx,\bs) \frac{ P(\bs) P(\by|\bs)
}{Z(\by)} P(\bx|\by) \nonumber \\
&=&\int {\rm d}\by\,  Z(\by)\int {\rm d}\bx_1 \int {\rm
  d}\bx_2 \, A(\bx_1,\bx_2)  P(\bx_2|\by)  P(\bx_1|\by)
= [ \langle \langle A(\bx_1,\bx_2) \rangle \rangle ] \, , \label{Nishi_gen}
\eea
where in the 3rd equality we renamed variables as $s=x_2$ and
$x=x_1$. Eq.~(\ref{Nishi_gen}) is the general form of the Nishimori
condition.

We remind the reader that for many thermodynamic quantities the
self-averaging property holds, i.e. for large system sizes the quantity
$\langle A(\bx,\bs) \rangle$ converges to the average over disorder of the
same quantity. Eq.~(\ref{Nishi_gen}) provides a rather general form of
the Nishimori condition that holds for inference problems where the
model for signal generation is known.

To give specific examples, let us define $m=\sum_{i=1}^N s_i  x_i /N \equiv \bs \cdot \bx$ and $q=  \bx_1 \cdot \bx_2$. Then we have
in the thermodynamic limit $[\langle m\rangle]=[\langle q\rangle]$. Due to self-averaging we also
have $m=q$ if $\bx,\bx_1$, and $\bx_2$ were samples from the distribution
$P(\bx|\by)$. Defining $Q= \bx \cdot \bx$, and using the Nishimori
condition, we get $Q= \rho\,  {\rm var}\phi$.

\subsection{Expectation maximization learning}
\label{EM_learning}

In general, one does not know the true density of the signal, $\rho_0$,
the measurement noise, $\Delta_0$, nor the asymptotic empirical
distribution of the signal, $\phi_0$ (or its parameters).
These parameters can be learned within the Bayesian
approach, in a way similar to the expectation maximization algorithm
\cite{Dempster,Iba99,DecelleKrzakala11}. Let us
denote $\theta$ as the ensemble of these unknown parameters.
Given the matrix $\bF$ and
measurement vector $\by$, the probability that the parameters
take a given set of values $\theta$ is
\be
    P(\theta|\bF,\by) = \frac{P(\theta|\bF)}{P(\by|F)} \int {\rm d}\bx
    P(\by,\bx|F,\theta) \propto P(\theta|\bF) Z(\theta)  \, ,
\ee
where $Z(\theta)$ is the normalization from (\ref{p_bayes}) with a given set
of parameters $\theta$
\be
Z(\rho,\overline x,\sigma,\Delta)= \int \prod_{i=1}^N {\rm d}x_i
\prod_{i=1}^N \left[ (1-\rho) \delta(x_i) +
  \frac{\rho}{\sqrt{2\pi}\sigma} e^{-\frac{(x_i-\overline
      x)^2}{2\sigma^2}} \right] \prod_{\mu=1}^M \frac{1}{\sqrt{2\pi
    \Delta}} e^{-\frac{1}{2\Delta}(y_\mu - \sum_{i=1}^N F_{\mu
    i}x_i)^2}\, .
\ee
Considering that without knowing the
measurements $\by$ we have no prior knowledge of $\theta$, looking for
the most probable value of parameters is equivalent to maximizing the
partition function with respect to the parameters. Even if we do have
a prior knowledge of $\theta$, in the situations considered in this
article the partition function scales exponentially in $N$ and hence
for large $N$ and function $ P(\theta|\bF)$ independent of $N$,
and maximizing $Z(\theta)$ is still the right thing to do.

In what follows, in order to learn parameters $\theta$ we will hence
derive stationary equations for the partition function $Z(\theta)$ (or
its logarithm). Remarkably, in many settings this leads to simple iterative
equations for learning of parameters.


\section{The belief propagation reconstruction algorithm for
  compressed sensing}
\label{BP_der}

Exact computation of the averages (see
Eq.~(\ref{average_marginal})) requires exponential time and is thus
intractable \cite{Natarajan95}. To approximate the expectations we
will use a variant of the belief propagation (BP) algorithm
\cite{KschischangFrey01,YedidiaFreeman03,MezardMontanari09}. Indeed,
message passing has been shown very efficient in terms of both
precision and speed for the CS problem. Our form of
the message passing algorithm is closely related to the approximate
message passing of \cite{DonohoMaleki09} and is a special case of the generalized
AMP of \cite{DonohoMaleki10,Rangan10b}. We provide here an independent derivation of the algorithm.

\subsection{Belief Propagation recursion}
\label{BP_derivation}
The canonical BP equations for the probability measure $P(\bx|\bF,\by)$, Eq.~(\ref{p_hat}),
are expressed in terms of $ 2 M N$ ``messages'', $ m_{j\to \mu}(x_j)$
and $m_{j\to \mu}(x_j)$, which are probability distribution
functions. They read:
\bea
m_{\mu \to i}(x_i) &=& \frac{1}{Z^{\mu \to i}} \int \prod_{j\neq i} {\rm d}x_j e^{-\frac{1}{2\Delta_\mu}(\sum_{j\neq i}F_{\mu j} x_j + F_{\mu i}x_i - y_\mu)^2} \prod_{j\neq i} m_{j\to \mu}(x_j) \, , \label{BP_1}\\
m_{i\to \mu}(x_i) &=& \frac{1}{Z^{i \to \mu}} \left[ (1-\rho)
  \delta(x_i) + \rho \phi(x_i) \right] \prod_{\gamma \neq \mu}
m_{\gamma \to i}(x_i) \, ,  \label{BP_2} \eea
where $Z^{\mu \to i}$ and $Z^{i \to \mu}$ are normalization factors
ensuring that $\int {\rm d}x_i \, m_{\mu \to i}(x_i) = \int {\rm d}x_i
\, m_{i \to \mu}(x_i) =1$. These coupled integral equations for the
messages are too complicated to be of any practical use.
However, in the large $N$ limit, when the matrix elements
$F_{\mu i}$ scale like $1/\sqrt{N}$, one can simplify these canonical
equations.

Using the Hubbard-Stratonovich
transformation
\be e^{-\frac{\omega^2}{2\Delta}} =\frac{1}{\sqrt{2\pi \Delta}} \int
{\rm d}\lambda \; e^{-\frac{\lambda^2}{2\Delta}+\frac{i\lambda
    \omega}{\Delta}}\, ,
\ee
 for $\omega= (\sum_{j\neq i} F_{\mu j}x_j)$
we can simplify Eq.~(\ref{BP_1}) as \be m_{\mu \to i}(x_i) =
\frac{1}{Z^{\mu \to i}\sqrt{2\pi \Delta}}
e^{-\frac{1}{2\Delta_\mu}(F_{\mu i}x_i - y_\mu)^2} \int {\rm d}\lambda
e^{-\frac{\lambda^2}{2\Delta_\mu}} \prod_{j\neq i} \left[ \int {\rm
    d}x_j m_{j\to \mu}(x_j) e^{\frac{F_{\mu j} x_j}{\Delta_\mu}(y_\mu
    - F_{\mu i}x_i + i\lambda)} \right] \, .  \ee Now we expand the
last exponential around zero because the term $F_{\mu j}$ is small in
$N$, we keep all terms that are of $O(1/N)$. Introducing means and
variances as new "messages" \bea
a_{i\to \mu } &\equiv&   \int {\rm d}x_i \, x_i \,  m_{i\to \mu}(x_i) \, ,  \label{a_imu}\\
v_{i\to \mu } &\equiv& \int {\rm d}x_i \, x^2_i \, m_{i\to \mu}(x_i) -
a^2_{i\to \mu } \, , \label{c_imu} \eea we obtain \be m_{\mu \to
  i}(x_i) = \frac{1}{Z^{\mu \to i}\sqrt{2\pi \Delta_\mu}}
e^{-\frac{1}{2\Delta_\mu}(F_{\mu i}x_i - y_\mu)^2} \int {\rm d}\lambda
e^{-\frac{\lambda^2}{2\Delta_\mu}} \prod_{j\neq i} \left[
  e^{\frac{F_{\mu j}a_{j\to \mu}}{\Delta_\mu} (y_\mu - F_{\mu i}x_i +
    i\lambda) + \frac{F^2_{\mu j}v_{j\to \mu}}{2\Delta^2_\mu}(y_\mu -
    F_{\mu i}x_i + i\lambda)^2} \right] \, .  \ee Performing the
Gaussian integral over $\lambda$, we obtain \be m_{\mu \to i}(x_i) =
\frac{1}{\tilde Z^{\mu \to i}} e^{-\frac{x^2_i}{2}A_{\mu\to i} +
  B_{\mu \to i} x_i}\, , \quad \quad \tilde Z^{\mu \to i} =
\sqrt{\frac{2\pi}{A_{\mu \to i}}} e^{\frac{B^2_{\mu \to i}}{2A_{\mu
      \to i}}}\, ,\label{m_mui} \ee
 where the normalization $\tilde Z^{\mu
  \to i}$ contains all the $x_i$-independent factors, and we have
introduced the scalar messages:
\bea
A_{\mu\to i} &=& \frac{F^2_{\mu i}}{\Delta_\mu + \sum_{j\neq i} F^2_{\mu j} v_{j\to \mu}}  \, , \label{A_mu} \\
B_{\mu \to i} &=& \frac{F_{\mu i}(y_\mu - \sum_{j\neq i} F_{\mu
    j}a_{j\to \mu})}{\Delta_\mu + \sum_{j\neq i} F^2_{\mu j} v_{j\to
    \mu}} \, , \label{B_mu} \eea
 The noiseless
case corresponds to $\Delta_\mu=0$.

To close the equations on messages $a_{i\to \mu}$ and $v_{i \to \mu}$ we notice that
\be
     m_{i\to \mu}(x_i) = \frac{1}{\tilde Z^{i\to \mu}} \left[ (1-\rho) \delta(x_i) + \rho \phi(x_i) \right] e^{-\frac{x^2_i}{2}\sum_{\gamma\neq \mu}A_{\gamma \to i} + x_i\sum_{\gamma\neq \mu}B_{\gamma \to i}} \, . \label{m_imu}
\ee
Messages $a_{i\to \mu}$ and $v_{i \to \mu}$ are respectively the mean and variance of the probability distribution $m_{i\to \mu}(x_i)$.
It is also useful to define the local beliefs $a_i$ and $v_i$ as
\bea
      a_{i } &\equiv&   \int {\rm d}x_i \, x_i \,  m_{i}(x_i) \, ,  \label{a_i}\\
      v_{i } &\equiv&   \int {\rm d}x_i \, x^2_i \,  m_{i}(x_i) - a^2_{i }   \, , \label{c_i}
\eea
where
\be
     m_{i}(x_i) = \frac{1}{\tilde Z^{i}} \left[ (1-\rho) \delta(x_i) + \rho \phi(x_i) \right] e^{-\frac{x^2_i}{2}\sum_{\gamma}A_{\gamma\to i} + x_i\sum_{\gamma}B_{\gamma \to i}} \, . \label{m_i}
\ee

For a general function $\phi(x_i)$ let us define the probability distribution
\be
         {\cal M}_\phi(\Sigma^2,R,x) =  \frac{1}{\hat Z(\Sigma^2,R)} \left[ (1-\rho) \delta(x) + \rho
           \phi(x) \right]  \frac{1}{\sqrt{2\pi} \Sigma} e^{-\frac{(x-R)^2}{2\Sigma^2}} \, ,
\ee
where $\hat Z(\Sigma^2,R)$ is a normalization.
We define the average and
variance of ${\cal M}_\phi$ as
\bea
    f_a(\Sigma^2,R) &\equiv& \int {\rm d}x\,  x \,  {\cal M}(\Sigma^2,R,x)\, , \label{f_a_gen}\\
    f_c(\Sigma^2,R) &\equiv& \int {\rm d}x\,  x^2 \,  {\cal M}(\Sigma^2,R,x) -
    f^2_a(\Sigma^2,R) \, , \label{f_c_gen}
\eea
(where we do not write explicitly the dependence on  $\phi$).
We give an explicit form for these two functions for the
Gauss-Bernoulli signal model, Eqs.~(\ref{f_a}-\ref{f_c}), and for the
mixture of Gaussians signal model in Appendix \ref{appendix:mixture}.
Notice that:
\bea
      f_a (\Sigma^2,R) &=& R  + \Sigma^2 \frac{{\rm d}}{{\rm d} R}
      \log{\hat Z(\Sigma^2,R) }  \, , \\
      f_c (\Sigma^2,R) &=& \Sigma^2  \frac{{\rm d}}{{\rm d} R}   f_a
      (\Sigma^2,R) \, .
\eea


The closed form of the BP update is
\begin{eqnarray}
a_{i\to \mu}&=&f_a\left(\frac{1}{\sum_{\gamma\neq \mu}A_{\gamma \to
    i}},\frac{\sum_{\gamma\neq \mu}B_{\gamma \to i}}{ \sum_{\gamma\neq \mu}A_{\gamma \to
    i}   }\right)\, , \quad \quad a_{i}=f_a\left(\frac{1}{\sum_{\gamma}A_{\gamma \to
    i}},\frac{\sum_{\gamma}B_{\gamma \to i}}{ \sum_{\gamma}A_{\gamma \to
    i}   }\right)\, ,
\label{BP_a_closed}  \\
v_{i\to \mu}&=&f_c\left(\frac{1}{\sum_{\gamma\neq \mu}A_{\gamma \to
    i}},\frac{\sum_{\gamma\neq \mu}B_{\gamma \to i}}{ \sum_{\gamma\neq \mu}A_{\gamma \to
    i}   }\right) \, , \quad \quad v_{i}=f_c\left(\frac{1}{\sum_{\gamma}A_{\gamma \to
    i}},\frac{\sum_{\gamma}B_{\gamma \to i}}{ \sum_{\gamma}A_{\gamma \to
    i}   }\right) \, . \label{BP_v_closed}
\end{eqnarray}

For a general signal model $\phi(x_i)$ the functions $f_a$ and $f_c$ can be
computed using a numerical
integration over $x_i$. In special cases, like the case of Gaussian
$\phi$ which we use in practice, these functions are easily computed
analytically and are given in Eqs.~(\ref{f_a}-\ref{f_c}).
 Eqs.~(\ref{a_imu}-\ref{c_imu}) together with
(\ref{A_mu}-\ref{B_mu}) and (\ref{m_imu}) then lead to closed
iterative message passing equations, which can be solved by iterations. There equations can be used for any
signal $\bs$, and any matrix $\bF$. When a fixed point of the BP equations is reached, the elements of the
original signal are estimated as $x_i^* = a_{i}$, and
the corresponding variance $v_i$ can be used to quantify the correctness of this estimate.  Perfect
reconstruction is found when the messages converge to a fixed point
such that
$a_{ i }=s_i$ and $v_{i }=0$.

A message passing algorithm equivalent to the one that we have just described was used in \cite{Rangan10}, where it
was called ``relaxed belief propagation''. In \cite{Rangan10}, it was
used as an approximate algorithm
for the case of a sparse measurement matrix $\bF$.  In our case, the
matrix is not sparse, and the use of mean and variances instead of the
canonical BP messages is exact in the large $N$ limit, thanks to the
fact that the
matrix is not sparse (a sum like $\sum_i F_{\mu_i} x_i$ contains of
order $N$ non-zero terms), and each element of the matrix $F$ scales as $O(1/\sqrt{N})$.

\subsection{The TAP form of the message passing
  algorithm} \label{sec:TAP} In the message-passing form of BP
described above, $2M\times N$ messages are sent, one between each variable
component $i$ and each measurement, in each iteration. In fact, it is possible to rewrite the BP
equations in terms of $N+M$ messages instead of $2\; M\times N$, always within
the assumption that the $F$ matrix  is not sparse, and that all its
elements scale as $O(1/\sqrt{N})$.  In
statistical physics terms, this corresponds to the
Thouless-Anderson-Palmer equations (TAP) \cite{ThoulessAnderson77}
used in the study of
spin glasses. In
the large $N$ limit, these are asymptotically equivalent (only $o(1)$
terms are neglected) to the BP equations.  Going from BP to TAP is, in the compressed
sensing literature, the step to go from the rBP
\cite{Rangan10} to the AMP \cite{DonohoMaleki09} algorithm. Let us now
show how to take this step.

In the large $N$ limit, it is clear from (\ref{BP_a_closed}-\ref{BP_v_closed})
that the messages $a_{i\to \mu}$ and $v_{i\to \mu}$
are nearly independent of $\mu$. However, one must be careful to keep the
correcting ``Onsager reaction terms''.  Let us define
\begin{eqnarray}
\omega_\mu&=& \sum_i F_{\mu i} a_{i \to\mu}\, , \quad \quad V_\mu= \sum_i F_{\mu i}^2  v_{i \to\mu}\, , \label{OG}\\
\Sigma^2_i &= & \frac{1}{\sum_{\mu}A_{\mu\to i}}\, , \quad \quad R_i =\frac{
\sum_{\mu}B_{\mu\to i}} {\sum_{\mu}A_{\mu\to i}}  \, . \label{UV}
\end{eqnarray}
Then we have
\begin{eqnarray}
\Sigma^2_i&=& \left[ \sum_\mu \frac{F^2_{\mu i}}{\Delta_\mu + V_\mu - F^2_{\mu
    i} v_{i\to \mu}} \right]^{-1}= \left[ \sum_\mu \frac{F^2_{\mu i}}{\Delta_\mu +
  V_\mu} \right]^{-1} \, , \\
R_i&=&  \left[  \sum_\mu
\frac{F_{\mu i}(y_\mu - \omega_\mu+ F_{\mu i}a_{i\to \mu})}{\Delta_\mu + V_\mu - F^2_{\mu
    i} v_{i\to \mu}}\right] \left[ \sum_\mu \frac{F^2_{\mu i}}{\Delta_\mu + V_\mu - F^2_{\mu
    i} v_{i\to \mu}} \right]^{-1} =  a_i + \frac{
\sum_\mu F_{\mu i}
\frac{(y_\mu - \omega_\mu)}{\Delta_\mu +V_\mu}}{  \sum_\mu F_{\mu i}^2
\frac{1}{\Delta_\mu + V_\mu}   }
\, .
\end{eqnarray}
In order to compute $\omega_\mu= \sum_i F_{\mu i} a_{i\to \mu}$, we see
that when expressing $a_{i\to \mu}$ in terms of $a_i$ we need to keep
all corrections that are linear in the
matrix element $F_{\mu i}$
\begin{eqnarray}
a_{i\to \mu}= f_a\left(\frac{1}{\sum_{\nu}A_{\nu\to i}  -A_{\mu\to i}},\frac{ \sum_{\nu}B_{\nu\to i} -B_{\mu\to
      i}}{ \sum_{\nu}A_{\nu\to i} -A_{\mu\to i} }\right)=
a_i - B_{\mu\to i}  \Sigma^2 \frac{\partial f_a}{\partial R}\left(\Sigma^2_i,R_i\right)\, .
\end{eqnarray}
Therefore
\begin{eqnarray}
\omega_\mu
= \sum_i F_{\mu i} a_i -\frac{
  (y_\mu-\omega_\mu)}{\Delta_\mu +V_\mu} \sum_i F_{\mu i}^2
v_i \, .
\end{eqnarray}
The computation of $V_\mu$ is similar, this time all the
corrections are negligible in the limit $N\to \infty$.

Finally, we get the following closed system of iterative TAP equations that involve only matrix
multiplication:
\bea
 V^{t+1}_\mu &=& \sum_i F_{\mu i}^{2} \, v^{t}_i \, , \label{TAP_ga} \\
\omega^{t+1}_\mu  &=& \sum_i F_{\mu i} \, a^t_i -\frac{
  (y_\mu-\omega^t_\mu)}{\Delta_\mu +V^t_\mu} \sum_i F_{\mu i}^2\,
v^t_i \, , \label{TAP_al}  \\
   (\Sigma^{t+1}_i)^2&=&\left[ \sum_\mu \frac{F^2_{\mu i}}{\Delta_\mu + V^{t+1}_\mu} \right]^{-1}\, ,
      \label{TAP_U}\\
      R^{t+1}_i&=& a^t_i + \frac{\sum_\mu F_{\mu i} \frac{(y_\mu - \omega^{t+1}_\mu)}{\Delta_\mu
        + V^{t+1}_\mu}}{ \sum_\mu \frac{ F_{\mu i}^2}{\Delta_\mu + V^{t+1}_\mu}}\, , \label{TAP_V}\\
      a^{t+1}_i &=&   f_a\left((\Sigma^{t+1}_i)^2,R^{t+1}_i\right) ,  \label{TAP_a}\\
      v^{t+1}_i &=& f_c\left((\Sigma^{t+1}_i)^2,R^{t+1}_i\right) \, . \label{TAP_v}
\eea
We see that the signal model $P(x_i) = (1-\rho)\delta(x_i) +\rho
\phi(x_i)$ assumed in the probabilistic approach appears only through
the definitions (\ref{f_a_gen}-\ref{f_c_gen}) of the two functions
 $f_a$ and $f_c$ . In the case where the signal model is chosen as
 Gauss-Bernoulli, these functions are given explicitly by
Eqs.~(\ref{f_a}-\ref{f_c}). Equations (\ref{TAP_ga}-\ref{TAP_v}) are equivalent to the (generalized) approximate
message passing of \cite{DonohoMaleki09,Rangan10b}.

A reasonable initialization of these
equations is
\bea
a^{t=0}_i&=& \rho \int {\rm d}x \, x \, \phi(x)\, , \label{ainit}\\
v^{t=0}_i &=& \rho \int {\rm d}x \, x^2 \, \phi(x)  - \left(a^{t=0}_i\right)^2\, , \label{vinit}\\
\omega^{t=0}_\mu &=& y_\mu\, .\label{omegainit}
\eea

\subsection{Further simplification for measurement matrices with random entries}
\label{TAP_simpl}

For some special classes of random measurement matrices $\bF$, the TAP equations (\ref{TAP_ga}-\ref{TAP_V}) can be
simplified further. Let us start with the case of a homogenous matrix $\bF$ with iid random entries of zero mean and variance $1/N$ (the
distribution can be anything as long as the mean and variance are
fixed). The simplification can be understood as follows.  Consider for instance the quantity
$V_\mu$. Let us define $\overline V$ as the average of $V_\mu$
with respect to different realizations of the measurement matrix $F$.
\be
        \overline V
        = \sum_{i=1}^N \overline{ F^2_{\mu i}} v_i = \frac{1}{N}
        \sum_{i=1}^N v_i\, .
\ee
The variance is
\bea
      {\rm var}\,  V &\equiv& \overline{ (V_\mu -
      \overline V)^2} =
      \sum_{i \neq j}  \overline{ \left(F^2_{\mu i}
        - \frac{1}{N}\right) \left(F^2_{\mu j}
        - \frac{1}{N}\right) }\, v_i v_j   + \sum_{i=1}^N \overline{ \left(F^2_{\mu i}
        - \frac{1}{N}\right)^2}\,  v_i^2 \nonumber \\
    &=& 0 + \frac{2}{N} \left(
        \frac{1}{N} \sum_{i=1}^N v_i^2 \right) = O\left(\frac{1}{N}\right) \, .
\eea
Since the average is of order one and the variance of order $1/N$, in
the limit of large $N$ we can hence neglect the dependence on the
index $\mu$ and consider all $V_\mu$ equal to their average. The
same argument can be repeated for all the terms that contain
$F_{\mu i}^2$. Hence for the homogenous matrix $\bF$ with iid random entries of zero mean and variance $1/N$, one can effectively
``replace'' every $F^2_{\mu i}$ by $1/N$ in
Eqs.~(\ref{TAP_U}-\ref{TAP_V}) and (\ref{TAP_ga}-\ref{TAP_al}).
The iteration equations then take the simpler form (assuming for
simplicity that $\Delta_{\mu}=\Delta$)
%
\begin{eqnarray}
V & = & \frac{1}{N}\sum_i v_i \, , \label{TAP_ga_fully} \\
\omega_\mu & =  & \sum_i F_{\mu i} a_i -\frac{
  (y_\mu-\omega_\mu)}{\Delta +V} \[[\frac{1}{N} \sum_i v_i\]]
\, , \label{TAP_al_fully}\\
\Sigma^2 & = & \frac{\Delta  + V}{\alpha}\, , \label{TAP_U_fully}\\
R_i & = & a_i + \sum_\mu F_{\mu i}
\frac{(y_\mu - \omega_\mu)}{\alpha}  \, . \label{TAP_V_fully}\\
 a_i &=&   f_a\left(\Sigma^2,R_i\right) ,  \label{TAP_a_fully}\\
      v_i &=& f_c\left(\Sigma^2,R_i\right) . \label{TAP_v_fully}
\end{eqnarray}
These equations can again be solved by iteration. They only involve  $2( M+N+1 )$ variables.
For a general matrix $\bF$ one iteration of the above algorithm takes
$O(NM)$ steps (and in practice we
observed that  the number of iterations needed for convergence is
basically independent of $N$).
For matrices that can be computed recursively (i.e. without storing all
their $NM$ elements) a speed up of this algorithm is possible, as the message passing
loop takes only $O(M+N)$ steps.

A second class of matrices for which a similar simplification exists
is the case of the block matrices
defined in Sec.~\ref{block_matrices}.  For simplicity, we consider
the case when the noise only depends on the block, i.e.,
$\Delta_{\mu}=\Delta_{q}$ for all $\mu$ in block~$q$.  For the block
measurement matrix with random entries of variance $J_{q,p}/N$ the
simplified TAP equations read
\begin{eqnarray}
V_{q} & = &
\frac{1}{N}\sum_{p=1}^{L_c}J_{q,p}\sum_{i\in B_p}v_i\, ,\label{TAP3_ga}
\\
\omega_{\mu} & = & \sum_{p=1}^{L_c}\sum_{i \in B_p}F_{\mu
  i}a_i-\frac{y_{\mu}-\omega_{\mu}}{\Delta_{I(\mu)}+V_{I(\mu)}}\frac{1}{N}\sum_{p=1}^{L_c}
J_{I(\mu),p}\sum_{i\in B_p}
v_i\, , \label{TAP3_w}\\
\Sigma^2_p & = & \left[ n_p \sum_{q=1}^{L_r}\frac{\alpha_{qp}
  J_{q,p}}{\Delta_q+V_q} \right]^{-1}\, ,\label{TAP3_U}\\
R_i & = & a_i +
\frac{ \sum_{q=1}^{L_r}\sum_{\mu \in B_q} F_{\mu i}
  \frac{y_{\mu}-\omega_{\mu}}{\Delta_{q}+V_{q}}} {
n_{I(i)} \sum_{q=1}^{L_r}\frac{\alpha_{qI(i)}
  J_{q,I(i)}}{\Delta_q+V_q} }\, , \label{TAP3_V}\\
a_i &=&   f_a\left(\Sigma_{I(i)}^2,R_i\right) ,  \label{TAP3_a}\\
      v_i &=& f_c\left(\Sigma_{I(i)}^2,R_i\right) , \label{TAP3_v}
\end{eqnarray}
where $p=1,2,\ldots L_c $, $q=1,2,\ldots L_r$. $I(\mu)$ (and $I(i)$)
is defined as the index of the block to which $\mu$ (i) belongs, $B_q$
is the set of indices in block $q$. We remind that
$\alpha_{qp}=M_q/N_p$ and $n_p=N_p/N$.

\subsection{Parameter learning with expectation maximization}
\label{BP_learning}

In our practical implementation, we use as signal model a   Gauss-Bernoulli distribution. That is, the function  $\phi(x)$ is Gaussian with mean $\ox$ and
variance $\sigma^2$. The functions $f_a$ and $f_c$ are in this case:
\bea
f_a(\Sigma^2,R) &=& \frac{  \rho\,
e^{-\frac{(R-\overline x)^2}{2(\Sigma^2+\sigma^2)}}
\frac{\Sigma}{(\Sigma^2+\sigma^2)^{\frac{3}{2}}} (\overline x \Sigma^2
+ R \sigma^2) }{     (1-\rho)
e^{-\frac{R^2}{2\Sigma^2}} + \rho \frac{\Sigma}{\sqrt{\Sigma^2+\sigma^2}}
e^{-\frac{(R-\overline x)^2}{2(\Sigma^2+\sigma^2)}} } \, ,  \label{f_a}\\
f_c(\Sigma^2,R) &=&
\frac{  \rho\,  (1-\rho)  e^{-\frac{R^2}{2\Sigma^2}-\frac{(R-\overline x)^2}{2(\Sigma^2+\sigma^2)}}
\frac{\Sigma}{(\Sigma^2+\sigma^2)^{\frac{5}{2}}} \left[ \sigma^2
  \Sigma^2 (\Sigma^2 + \sigma^2)+ (\overline x \Sigma^2
+ R \sigma^2)^2 \right]    +  \rho^2 e^{-\frac{(R-\overline
  x)^2}{(\Sigma^2+\sigma^2)}}   \frac{\sigma^2
\Sigma^4}{(\sigma^2+\Sigma^2)^2}  }
{  \left[   (1-\rho)
e^{-\frac{R^2}{2\Sigma^2}} + \rho \frac{\Sigma}{\sqrt{\Sigma^2+\sigma^2}}
e^{-\frac{(R-\overline x)^2}{2(\Sigma^2+\sigma^2)}} \right]^2} \, .\label{f_c}
\eea
See also appendix \ref{appendix:mixture} where we give the form of
$f_a$ and $f_c$ for the signal model consisting of  mixture of
Gaussians.


The most likely values of parameters $\rho,\overline x,\sigma,\Delta$  can
be obtained via maximizing the partition function. Within the belief
propagation approach this is
equivalent to maximizing the Bethe free entropy $F(\rho,\overline
x,\sigma,\Delta) \equiv \log Z(\rho,\overline x,\sigma,\Delta)$ expressed as
\cite{MezardMontanari09} \be F(\rho,\overline x,\sigma,\Delta) = \sum_\mu
\log{ Z^\mu} + \sum_i \log{Z^i} - \sum_{(\mu i)} \log{Z^{\mu
    i}}\, ,\label{Bethe} \ee where \bea
Z^i&=& \int {\rm d} x_i \prod_{\mu} m_{\mu \to i}(x_i) \left[ (1-\rho)
  \delta(x_i) + \frac{\rho}{\sqrt{2\pi}\sigma}
  e^{-\frac{(x_i-\overline x)^2}{2\sigma^2}} \right]\, ,\\
Z^{\mu i}&=& \int {\rm d}x_i m_{\mu \to i}(x_i) m_{i \to \mu}(x_i)\,
. \\
Z^\mu&=& \int \prod_i {\rm d}x_i \prod_i m_{i\to \mu}(x_i)
\frac{1}{\sqrt{2\pi \Delta_\mu}} e^{-\frac{(y_\mu - \sum_i F_{\mu
      i}x_i)^2}{2\Delta_\mu}} = \frac{1}{\sqrt{2\pi(\Delta+ V_{\mu}  )}}
e^{-\frac{(y_\mu-\omega_\mu)^2}{2(\Delta+V_\mu)}  } \, .
\eea

The stationarity condition of Bethe free entropy (\ref{Bethe})
with respect to $\rho$ leads to
\be
\rho = \frac{\sum_i
  \frac{1/\sigma^2+1/\Sigma_i^2}{R_i/\Sigma_i^2 +\overline x/\sigma^2}a_i}{\sum_i \left[ 1
    -\rho + \frac{\rho}{\sigma(1/\sigma^2+ 1/\Sigma_i^2)^{\frac{1}{2}}}
    e^{\frac{(R_i/\Sigma_i^2+\overline x/\sigma^2)^2}{2(1/\sigma^2+ 1/\Sigma_i^2)}
      -\frac{\overline x^2}{2\sigma^2} }
  \right]^{-1}} \label{learn_rho}\, .
\ee
Stationarity with respect to $\overline x$ and $\sigma$ gives
\bea
\overline x & = & \frac{\sum_i a_i}{\rho \sum_i \left[\rho + (1 -
    \rho) \sigma(1/\sigma^2+ 1/\Sigma_i^2)^{\frac{1}{2}}
    e^{-\frac{(R_i/\Sigma_i^2+\overline x/\sigma^2)^2}{2(1/\sigma^2+ 1/\Sigma_i^2)}
      +\frac{\overline x^2}{2\sigma^2} }
  \right]^{-1}}\, , \label{learn_xbar}\\
\sigma^2 & = & \frac{\sum_i (v_i + a_i^2)}{\rho \sum_i \left[\rho + (1
    - \rho) \sigma(1/\sigma^2+ 1/\Sigma_i^2)^{\frac{1}{2}}
    e^{-\frac{(R_i/\Sigma_i^2+\overline x/\sigma^2)^2}{2(1/\sigma^2+ 1/\Sigma_i^2)}
      +\frac{\overline x^2}{2\sigma^2} } \right]^{-1}} - \overline
x^2\, . \label{learn_mv}
\eea
For simplicity, we consider that the noise is homogeneous,
i.e., $\Delta_{\mu}=\Delta$, for all $\mu$. The noise level $\Delta$ may be
unknown, in which case one can learn it, like the other parameters, by
maximizing the free entropy. The resulting condition for learning
of the noise variance $\Delta$ reads:
\be
\Delta =
\frac{\sum_{\mu}\frac{(y_{\mu}-\omega_{\mu})^2}{(1+\frac{1}{\Delta}
    V_{\mu})^2}}{\sum_{\mu}\frac{1}{1+\frac{1}{\Delta}V_{\mu}}}
\, ,
\label{delta_learn}
 \ee
where $\omega_{\mu}$ and $V_{\mu}$ are defined in
Eq.~(\ref{OG}).

Note that instead of using the steepest gradient descent in the Bethe
free energy for the mean and
variance (i.e. Eqs.~(\ref{learn_xbar}-\ref{learn_mv})) one can also use
simpler expressions
that are satisfied in the Bayes-optimal setting. In
particular
\bea \overline x & = & \frac{\sum_i a_i}{N \rho}, \label{learn_xbar_alternative}\\
\sigma^2 & = & \frac{\sum_i (v_i + a_i^2)}{\rho N} - {\overline x}^2. \label{learn_mv_alternative} \eea
In our numerical implementations we use these
simplified conditions.
In the case where the matrix $\bF$ is
random with iid elements of zero mean and variance $1/N$, we can
also use for learning the variance: $\sum_{\mu=1}^M y^2_\mu/N =\alpha \rho (\sigma^2 + \overline x^2)$.

Eqs.~(\ref{learn_rho}) and (\ref{learn_xbar}, \ref{learn_mv}) or
(\ref{learn_xbar_alternative}, \ref{learn_mv_alternative}) can be used
for iterative learning of the parameters, in the spirit of expectation maximization.
Eqs.~(\ref{A_mu}, \ref{B_mu},
\ref{BP_a_closed}, \ref{BP_v_closed}, \ref{learn_rho}, \ref{learn_xbar_alternative}, \ref{learn_mv_alternative})
altogether lead to the Expectation Maximization Belief Propagation
(EM-BP) algorithm that we have first presented in
\cite{KrzakalaPRX2012}. In EM-BP one update of the BP messages is
followed by an update of the parameters and this is repeated till
convergence (of both BP messages and the parameters). In our
implementations we initialize the parameters as follows
\be
 \rho^{t=0}=\alpha/10\, , \quad \quad \overline x^{t=0}=0\, , \quad \quad
\sigma^2_{t=0}= 1 \, .
\ee
In case the variance of the signal is not at all close to one, the sum rule  $\frac{1}{M}\sum_\mu
y_\mu^2= \frac{1}{M} \sum_{\mu,i} F_{\mu  i}^2 s_i^2$ suggests a more
sensible initialization $\sigma^2_{t=0}=\sum_\mu y^2_\mu / (M N {\rm var}F \rho^{t=0})$.
A new guess of
parameters is obtained using
Eqs.~(\ref{learn_rho}, \ref{learn_xbar_alternative},
\ref{learn_mv_alternative}) except if the variance becomes negative,
then the new variance is set to zero, or if the new value of $\rho$
becomes larger than $\alpha$, in which case $\alpha$ is taken as the new
value for $\rho$. To obtain an updated guess for the parameters we also use ``damping''. The
updated guess is obtained as $1/2$ times the old value plus $1/2$ times
the newly computed value. Empirically this speeds up the convergence and prevents some numerical
instabilities. If needed, such damping is also
used to improve convergence for the BP messages themselves.


\section{Asymptotic analysis: State evolution and replicas} \label{replica_homogeneous}

Belief propagation is an efficient heuristic algorithm that is in some
cases (such as the present one) amenable to asymptotic ($N\to \infty$)
analytical analysis. This statistical analysis of BP iterations is known as the ``cavity
method'' (in statistical physics) \cite{MezardParisi87b,MezardMontanari09}, the ``density
evolution'' in coding \cite{RichardsonUrbanke08}, and the ``state evolution'' in the context of
CS
\cite{DonohoMaleki09,BayatiMontanari10}. The corresponding equations can also be derived
using the replica method, that provides an exact asymptotic analysis of both
the BP performance and the performance of an optimal (perhaps
exponentially costly) reconstruction algorithm. In this section we
first concentrate (parts A to D) on the case of 'homogeneous' measurement matrices
with iid entries. We derive the density evolution equations in part
A, and we detail the replica approach in part B. Part C shows the
simplifications that takes place in the Bayes-optimal case where the
signal model gives the correct statistical properties of the
underlying signal, and part D generalizes the density evolution
equations to the case where one uses the learning procedure for the
parameters of the signal model. Part E gives the density evolution
equations in the more general case of block measurement matrices.

\subsection{Density evolution of the message passing}
\label{sec:DE}
We derive the density evolution equations in the case where the
measurement matrix $F$ has random entries that are iid, with mean 0
and variance $1/N$, and we assume that  the parameters of the signal
model are fixed.


The density evolution (or cavity method) uses a statistical analysis
of the BP messages at iteration $t$, in the large $N$ limit, in order to derive their distribution
at iteration $t+1$.
It turns out that these distributions are simply expressed in terms of
two parameters:
\begin{eqnarray}
V^t &\equiv &\frac{1}{N} \sum_{i=1}^N  v_i^t \label{BP_opv}\\
     E^t &\equiv & \frac{1}{N} \sum_{i=1}^N  (a_i^t-s_i)^2 \, . \label{BP_op}
\end{eqnarray}
We remind the reader that $s_i$ are the components of the original signal
$\bs$,  and $a_i^t$, $v_i^t$ are the mean and variance of the local beliefs defined in (\ref{a_i}), at
iteration $t$.
$V^t$ just measures the average variance of the local beliefs, and
$E^t$ is the mean-squared error achieved by BP, at a given
iteration $t$.

Using the definition of the quantity $R_i$ (\ref{UV}) and Eq.~(\ref{B_mu}), we
get
\be
      R_i^t=s_i + \frac{1}{\alpha} \left[ \sum_\mu F_{\mu i}\xi_\mu
        +\sum_\mu F_{\mu i} \sum_{j\neq i} F_{\mu j} (s_j - a_{j\to
          \mu}^t ) \right]\, ,
\ee
where $\xi_\mu$ is the measurement noise (as defined in (\ref{def})), a
centered Gaussian variable with variance $\Delta_0$.
The variable $r_i^t=\sum_\mu F_{\mu i} \xi_\mu + \sum_\mu F_{\mu i} \sum_{j\neq i} F_{\mu j} (s_j -
a_{j\to \mu}^t)$ is a random variable with respect to the distribution
of the measurement matrix elements
$F_{\mu i}$ (zero mean and $1/N$ variance matrix) and the noise
$\xi_i$. Therefore $r_i^t$ a Gaussian random variables with mean and variance
\bea
     \overline{ r^t} &=& \sum_\mu \sum_{j\neq i}  \overline{ F_{\mu i} F_{\mu j}} (s_j -
a_{j\to \mu}) = 0\, ,\\
     \overline{(r^t)^2} &=& \sum_{\mu} \xi_{\mu}^2 F^2_{\mu i}+
\sum_{\mu} \sum_{j\neq i}   \overline{
       F_{\mu i}^2 F^2_{\mu j} } (s_j - a_{j\to \mu})^2 = \alpha \Delta_0+
     \frac{1}{N^2} \sum_{\mu=1}^M \sum_{j=1}^N (s_j - a_j)^2 = \alpha
     (E+\Delta_0)\, ,
\label{secsec}
\eea
In the second inequality of (\ref{secsec}) we
neglected terms of $O(1/\sqrt{N})$.


Using the above results this leads us to the belief at iteration
$t+1$, $m_{i}^{t+1}(x_i)$, being
distributed as
\be
  m_{i}^{t+1} (x_i)  \simeq \frac{1}{\hat Z^i} [(1-\rho) \delta(x_i) + \rho
  \phi(x_i)]
e^{- \frac{\alpha\left( x_i - s_i - z \sqrt{\frac{E+\Delta_0}{\alpha}} \right)^2  }{2(\Delta+V)   } }
\label{message_crucial}
\ee
where $z$ is a random Gaussian variable with zero mean and unit
variance, and $\hat Z^i$ is a normalization constant.  Hence using
the definition of the BP order parameters given in (\ref{BP_op}) we
get for a signal
with iid elements
\bea
     V ^{t+1} &=&  \int {\rm d} s \,
     [(1-\rho_0)\delta(s)+\rho_0\phi_0(s)] \int {\cal D}z   \,
     f_c\left(\frac{\Delta+V^t }{\alpha},  s+z \sqrt{\frac{E^t+\Delta_0}{\alpha}} \right)
   \, , \label{eq_V_gen}\\
E ^{t+1}  &=&   \int {\rm d} s\,
[(1-\rho_0)\delta(s)+\rho_0\phi_0(s)] \int {\cal D}z \,  \left[
  f_a\left(\frac{\Delta+V^t}{\alpha},  s+z \sqrt{\frac{E^t+\Delta_0}{\alpha}} \right)
  -s \right]^2 \, ,  \label{eq_E_gen}
\eea
where ${\cal D}z = {\rm d}z\,  e^{-z^2/2}/{\sqrt{2\pi}}$ is a Gaussian
integration measure.
For the special case of a Gauss-Bernoulli signal model, i.e. when the
function $\phi$ is Gaussian with mean $\overline x$
and variance $\sigma^2$, the functions $f_a(\Sigma^2,R)$ and $f_c(\Sigma^2,R)$ are expressed
explicitly in Eqs.~(\ref{f_a}-\ref{f_c}).

Equations (\ref{eq_V_gen}-\ref{eq_E_gen}) are the density evolution
equations. They describe how the mean-squared error $E$ and
the variance order parameter $V$ evolve during the iterations of the
BP algorithm.
Note that the density evolution equations are the same for the message passing and for the
TAP equations as indeed factors of $O(1/N)$ are neglected in the
density evolution. If the messages are initialized as in (\ref{ainit}-\ref{omegainit}),
the initial conditions of the density evolution equations are:
\bea
    E^{t=0} &=& \rho_0 \overline{s^2} - 2 \rho \rho_0 \overline s \int
    {\rm d}x \, x \phi(x) + \rho^2 \left[ \int
    {\rm d}x \, x \phi(x) \right]^2\, , \label{init_E} \\
   V^{t=0} &=& \rho \int
    {\rm d}x \, x^2 \phi(x)  -   \rho^2 \left[ \int
    {\rm d}x \, x \phi(x) \right]^2 \label{init_V}\, .
\eea

\begin{figure}[!ht]
\includegraphics[width=0.495\linewidth]{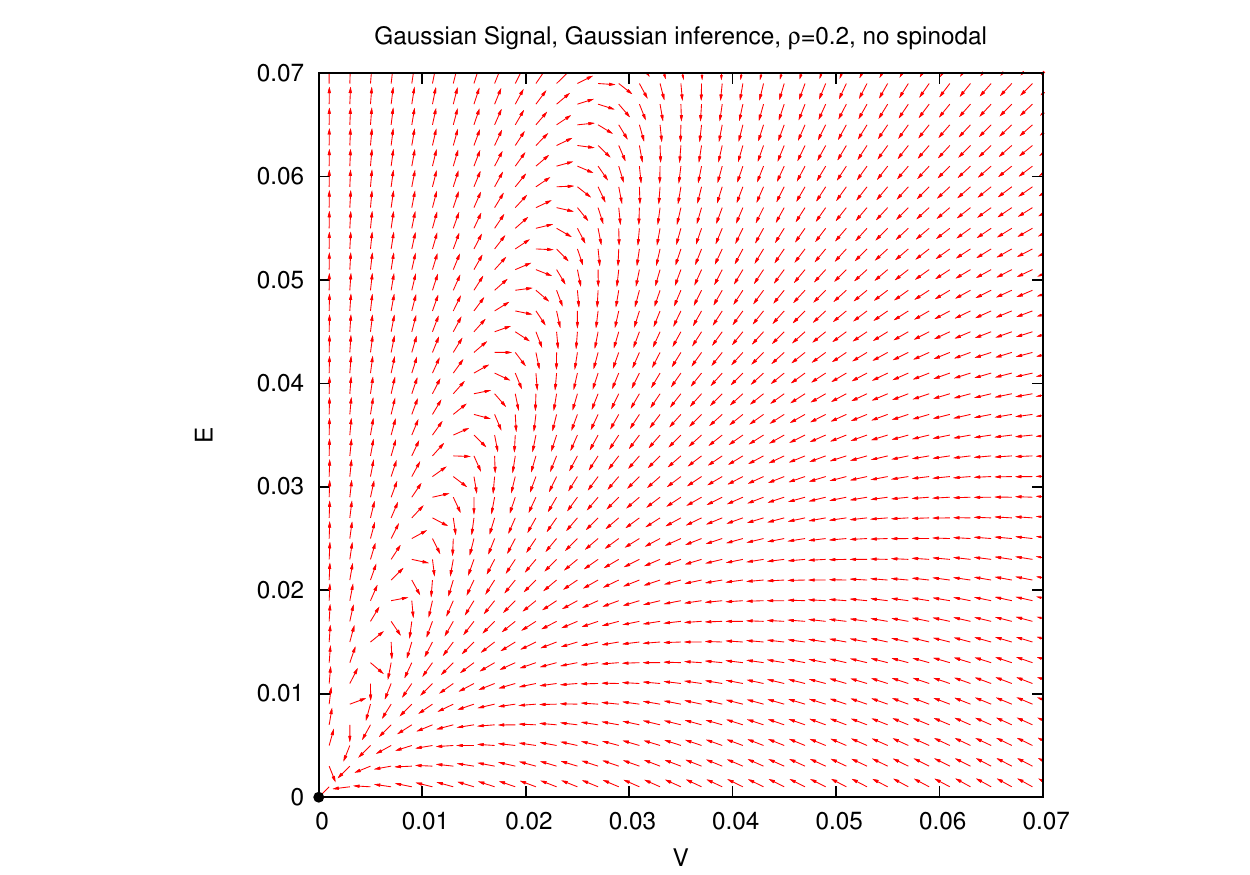}
\includegraphics[width=0.495\linewidth]{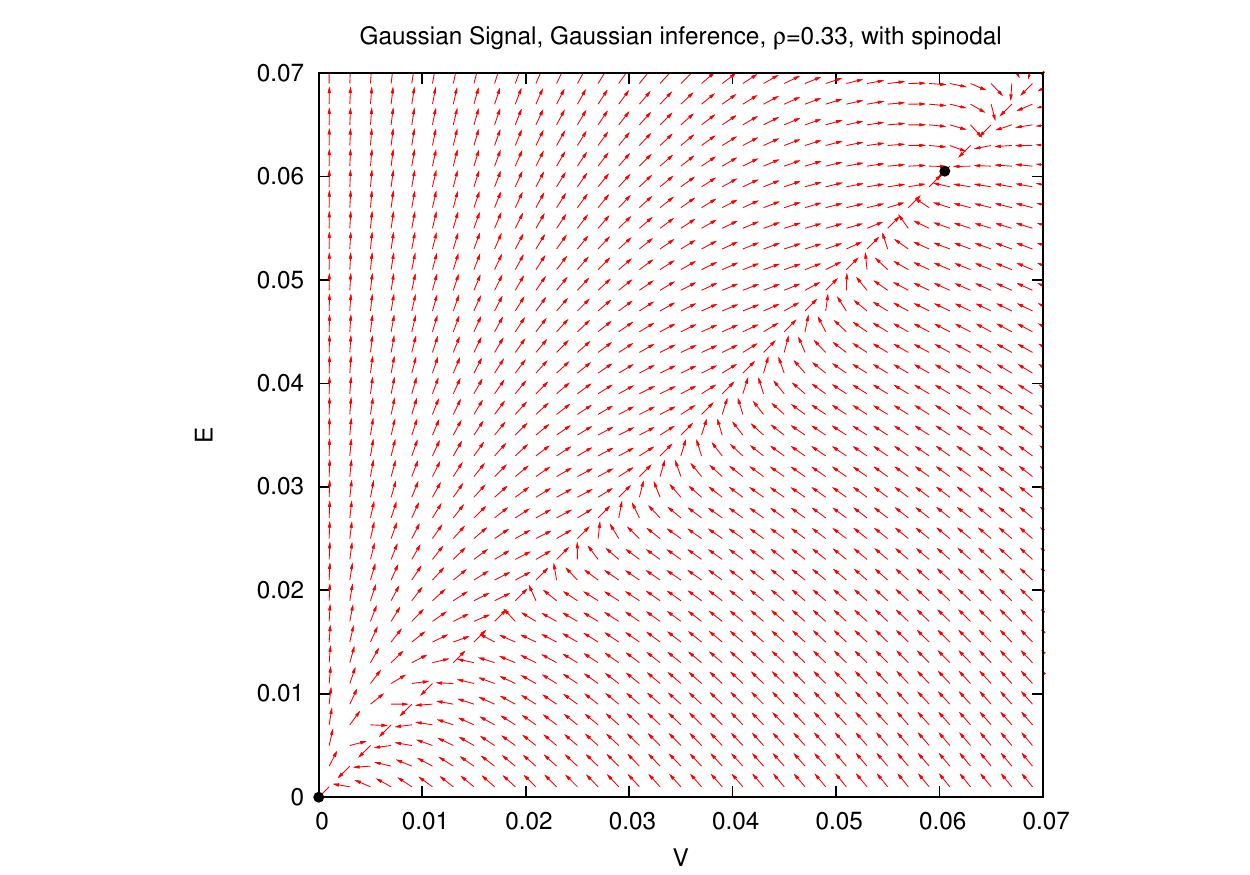}
\includegraphics[width=0.495\linewidth]{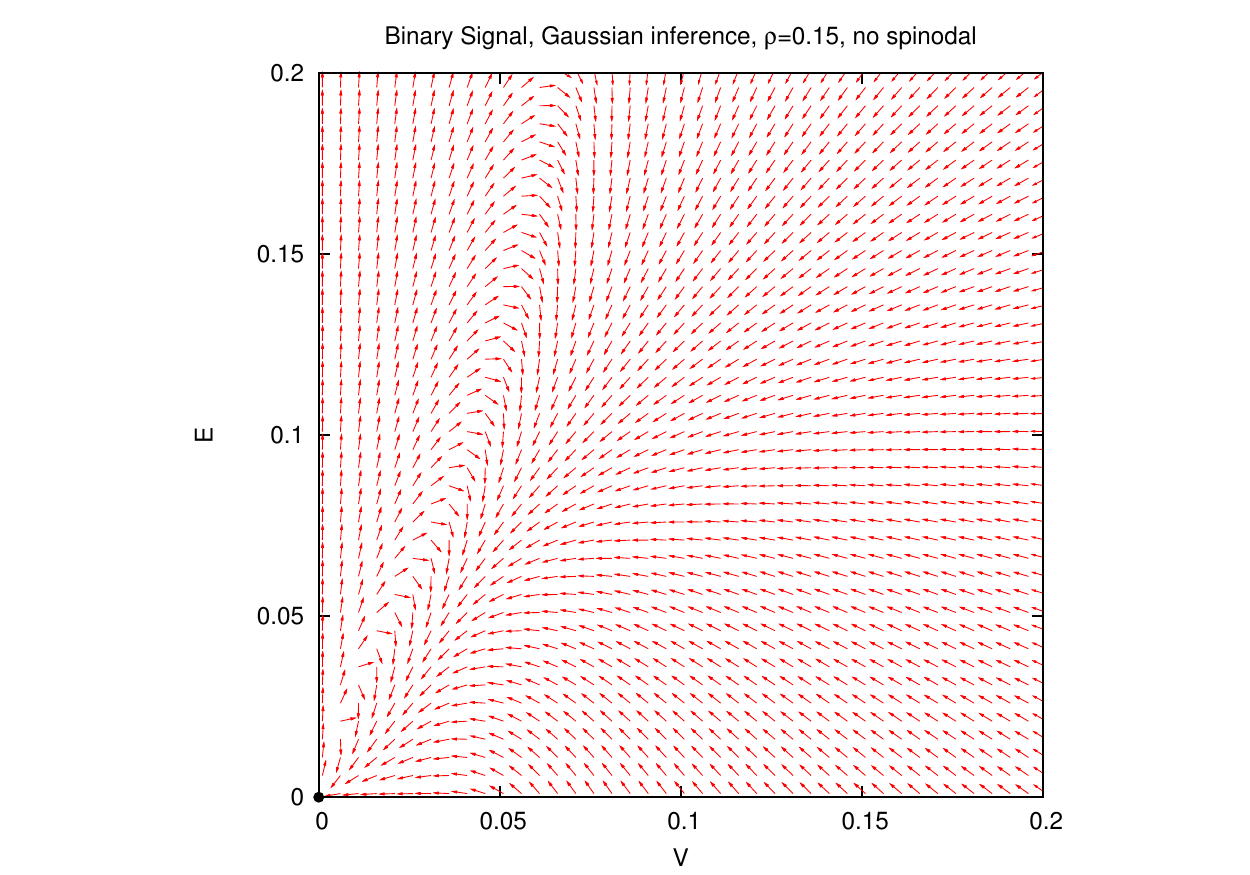}
\includegraphics[width=0.495\linewidth]{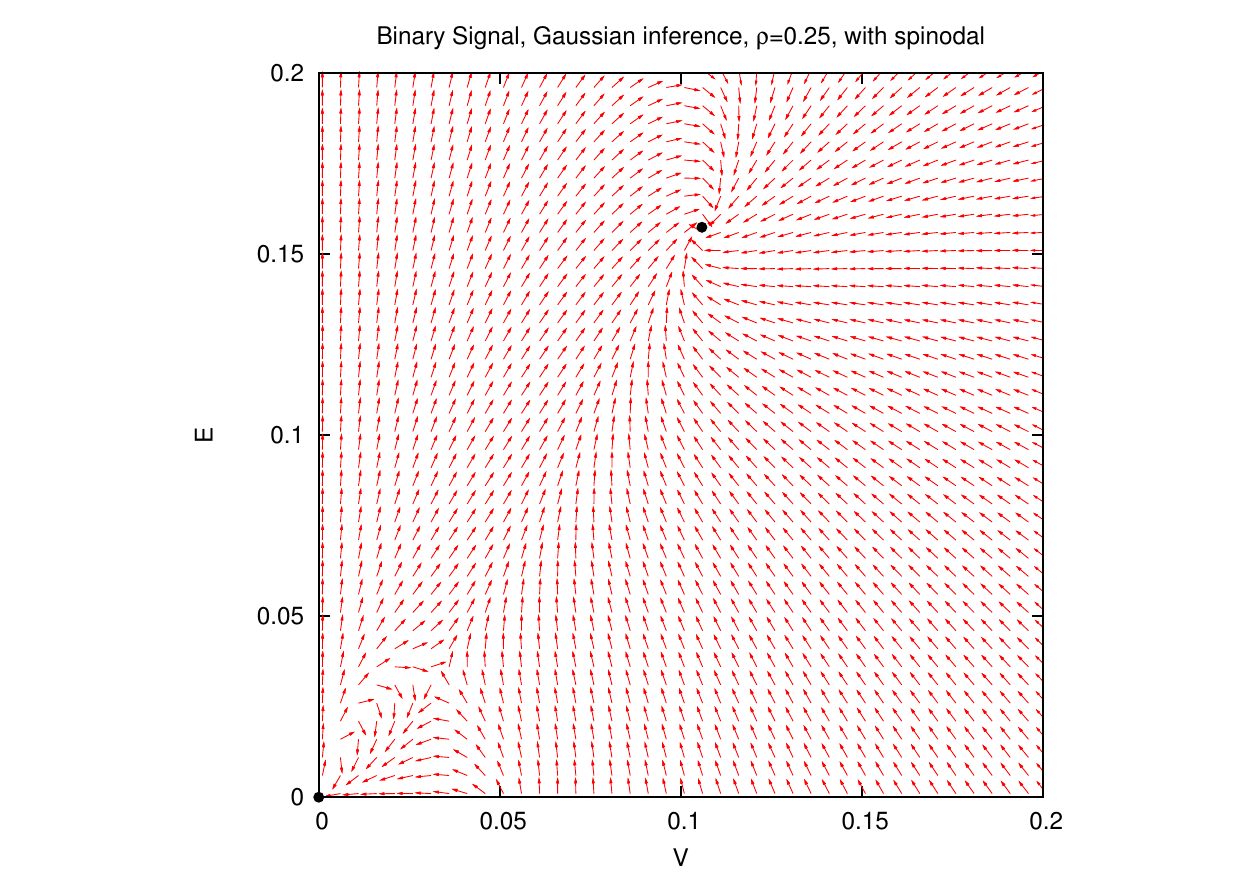}
  \caption{\label{fig_map} (color online) Examples of the BP density
    evolution, $y$-axes is the mean-squared error of the current signal
    estimate $E=q-2m+\rho_0\overline{s^2}$, the $x$-axes is the
    average variance $V = Q-q$. Each arrow is a normalized vector
    $(V^{(t+1)}-V^{(t)},E^{(t+1)}-E^{(t)})$. The
    signal model $\phi(x)$ is Gaussian with zero mean and unit
    variance, the signal distribution $\phi_0(x)$ is Gaussian on the
    top and $\{\pm 1\}$ on the bottom. The measurements are noiseless. On the left we show an example
    for relatively large measurement rate where there is a unique
    fixed point $E\to 0, V\to 0$. On the right there is another fixed
    point $E>0, V>0$ which is the attractive one for ``uninformed'' initial
    conditions. Notice that on the top plots the line $V=E$ is
    stable: this is thanks to the Nishimori condition when
    the signal is described by the correct model ($\rho_0=\rho$ and
    $\phi_0=\phi$). In that case one can work
    in the $V=E$ sub-space.}
\end{figure}

Fig.~\ref{fig_map} shows several examples of this mapping for
the noiseless case $\Delta=\Delta_0=0$. We plot the evolution of the normalized vector $(V^{(t+1)}-V^{(t)}, E^{(t+1)}-E^{(t)})$.
For a relatively high measurement density $\alpha$,  there is unique
fixed point $E=V=0$ corresponding to an exact reconstruction of the signal.  When $\alpha$ is below some critical point,
another attractive fixed point $E>0,V>0$ appears.

\subsection{Replica analysis}
\label{replicas}

The density evolution presented in the previous section can also be
derived independently using  the replica method~\cite{MezardParisi87b}.  The main advantage is that the replica
computations give a physical meaning to all the
fixed points of Eqs.~(\ref{eq_V_gen}-\ref{eq_E_gen}), even to those that are not reached by iterating the
BP algorithm.

The thermodynamic properties of a disordered system given by the Hamiltonian defined in Eq.~(\ref{Hamiltonian}) are
characterized by the average free entropy $\mathbb{E}_{{\bF}, {\bs},{\bf{\xi}}}
(\log{Z})$, where $Z$ is the partition function defined in
(\ref{p_hat}), $\bs$ is the original signal and
$\bf{\xi}=\{\xi_{\mu}\}_{\mu=1}^{M}$ are the measurement noise
with zero mean and variance $\Delta_0$ for $\mu=1,2,\ldots,M$.
The free entropy is evaluated via the replica trick as \be \Phi \equiv
\frac{1}{N} \mathbb{E}_{{\bF}, {\bs},{\bf{\xi}}}(\log{Z}) =\frac{1}{N} \lim_{n\to 0}
\frac{\mathbb{E}_{{\bF}, {\bs},{\bf{\xi}}}(Z^n)-1}{n} \, .  \ee
Introducing $n$ replicas, we get
\begin{equation}
\expect_{\bF,\bs,\bf{\xi}}(Z^n)= \int \prod_{i,a}{\rm d}x_i^a
\prod_{i,a} \left[ (1-\rho) \delta(x_i^a) + \rho \phi(x_i^a) \right]
\prod_{\mu} \expect_{\bF,\bs,\bf{\xi}}\frac{1}{\sqrt{2\pi \Delta}}
e^{-\frac{1}{2\Delta}\sum_{a=1}^n\((\sum_{i=1}^NF_{\mu i}s_i
  +\xi_{\mu}- \sum_{i=1}^N F_{\mu i}x_i^a\))^2}  \, , \label{ave_Zn}
\end{equation}
where $a,b,\dots$ denote the replica indices, $\Delta$ is the assumed measurement noise and generally $\Delta\neq \Delta_0$.

In the case where the matrix $\bF$ has
iid elements with zero mean and variance $1/N$, we introduce the order parameters as follows
\begin{eqnarray}
m^a & = & \frac{1}{N}\sum_{i=1}^N x_i^as_i\, , \quad \quad
a=1,2,\dots, n\, ,\\
Q^a & = & \frac{1}{N}\sum_{i=1}^N(x_i^a)^2, \quad \quad
a=1,2,\dots,n\, ,\\
q^{ab} & = & \frac{1}{N}\sum_{i=1}^Nx_i^ax_i^b\, , \quad \quad   a<b\, .
\end{eqnarray}
We use a common trick of rewriting the identity
\begin{eqnarray}
\nonumber
  1 & = &
  \int \prod_a {\rm d}\hat{Q}_a {\rm{d}}Q_a   {\rm{d}}\hat{m}_a {\rm{d}}m_a
  \int \prod_{a<b}  {\rm{d}}\hat{q}_{ab} {\rm{d}}q_{ab}
  e^{\sum_a \hat{Q}_a[\frac N2 {Q}_a - \frac 12  \sum_j (x_j^a)^2]-\sum_{a<b}  \hat{q}_{ab} \((N q_{ab} - \sum_j x_j^a x_j^b\))-\sum_a  \hat{m}_{a} \((N m_{a} - \sum_j x_j^a x_j^0\))}\, .
\end{eqnarray}

When averaging $Z^n$,  we first need to evaluate the quantity
\begin{equation}
X_{\mu}=\expect_{\bF,\bf{\xi}}\[[ e^{-\frac{1}{2\Delta}\sum_{a=1}^n\((\sum_i F_{\mu i}s_i+\xi_{\mu}-\sum_i F_{\mu i}x_i^a\))^2}\]]\label{Xmu}
\end{equation}
at fixed signal $\bs$ and configuration $\bx$.
In order to evaluate $X_{\mu}$ we first need to define
$v_{\mu}^a=\sum_{i=1}^N F_{\mu i} ( x_i^0 - x_i^a)+\xi_\mu$ with
$a=\{0,1,\ldots, n\}$, and where $0$ corresponds to the index of the
signal: $x_i^0=s_i$. The ${v^a}$ obeys a joint Gaussian distribution
with covariance
\bea \expect_{\bF,\bf{\xi}} \[[ \((v_{\mu}^a\))^2\]] &=&
\expect_{\bF,\bf{\xi}} \sum_i F^2_{\mu i} \(( x_i^0 - x_i^a \))^2 + \Delta_0 =
\frac 1N \sum_i \(( x_i^0 - x_i^a \))^2+\Delta_0
= Q^a-2m^a+\rho \overline{s^2} + \Delta_0\\
\expect_{\bF,\bf{\xi}} \[[ v_{\mu}^a v_{\mu}^b\]] &=&
\expect_{\bF,\bf{\xi}} \sum_i F_{\mu i}^2 \(( x_i^0 - x_i^a \)) \((
x_i^0 - x_i^b \))+\Delta_0 = q^{ab}-(m^a+m^b)+\rho \overline{s^2} + \Delta_0\eea
We shall use the so-called replica symmetric (RS) Ansatz. This is consistent
with using Belief Propagation, and it is known to be
correct for the optimal Bayesian inference (i.e. when the signal model
correspond to the empirical signal distribution) \cite{NishimoriBook,MezardMontanari09}. In this Ansatz the replicas are
considered as equivalent, therefore:
\be
m^a=m, \quad  q^{ab}=q, \quad Q^{a}=Q \, .\\
\ee
Going back to $X_{\mu}$, we now have
\be X_{\mu} =\expect_{\bv} \[[ e^{-\frac{1}{2\Delta }
  \sum_{a=1}^n\((v_\mu^a\))^2}\]] \ee
with
\be
P(\bv) = \frac 1{\sqrt{(2\pi)^n {\text{det}}(G)}}e^{-\frac 12 \sum_{a,b} v_a
  (G^{-1})_{ab} v_b}\, ,
\ee
where (under the RS hypothesis) the covariance matrix reads
\begin{eqnarray}
G_{aa}=E_{\bv}(v_{\mu}^av_{\mu}^a) & = &
Q+\rho\overline{s^2}-2m+\Delta_0, \quad a=1,2, \ldots, n\, , \\
G_{ab}=E_{\bv}(v_{\mu}^av_{\mu}^b) & = &
q+\rho\overline{s^2}-2m+\Delta_0, \quad a<b\, .
\end{eqnarray}
Computing explicitly $X_{\mu}$, one now finds
\be
X_{\mu}  =  \frac 1{\sqrt{(2\pi)^n {\text{det}}(G)}}
\int D\bv \, e^{-\frac{1}{2}  \sum_{a,b} v^a \[[ (G^{-1})_{ab} + \frac  1{\Delta} \delta_{a,b} \]] v^b}
=\frac{\int D\bv
  e^{-\frac{1}{2}\bv ^T(G^{-1}+\frac{\one}{\Delta})\bv}}{\int
  D\bv e^{-\frac{1}{2} \bv ^TG^{-1}\bv}}
 =  \frac{1}{\sqrt{\text{det}(\one+\frac{G}{\Delta})}} \, .
\ee
We now  compute this determinant. We have
\be
G = (q+\rho\overline{s^2}-2m + \Delta_0)  \amalg + (Q-q) \one\, ,
\ee
where $\amalg$ stands for the $n \times n$ matrix with elements all equal
to one. The eigenvectors of $G$ are
(a) one eigenvector $(1,1,\dots,1)$ with an eigenvalue $Q-q+n(q-2m+\rho \overline{s^2}+\Delta_0)$,
and (b) $n-1$ eigenvectors of the type $(0,0,1,-1,0,\dots,0)$ with
eigenvalues $Q-q$. Therefore
%
\be
\text{det}(\one+\frac{G}{\Delta})  =\[[ 1+ \frac 1{\Delta} \((Q-q + n
(q-2m+\rho  \overline{s^2} + \Delta_0)
\)) \]] \[[1+\frac{Q-q}{\Delta}\]]^{n-1}
\ee
To conclude the computation of $X_\mu$ we get
\be
\lim_{n \to 0} X_\mu
= e^{-\frac{n}{2}\left[\frac{q-2m+\rho\overline{s^2}+\Delta_0}{Q-q+\Delta}+\text{log}(1+\frac{Q-q}{\Delta})\right]} \, .
\ee

We thus obtain
\begin{eqnarray}
 \expect_{\bF,\bs,\bf{\xi}}Z^n & = &
  \int \prod_a {\rm d}\hat{Q}_a {\rm{d}}Q_a  {\rm{d}}\hat{m}_a
  {\rm{d}}m_a
  \int  \prod_{ab} {\rm{d}}\hat{q}_{ab} {\rm{d}}q_{ab}
e^{N    \[[\frac{1}{2} \sum_a \hat{Q}_aQ_a - \sum_{a<b}
  \hat{q}_{ab}q_{ab}-\sum_a\hat{m}_am_a\]]}
\prod_{\mu} \frac{X_{\mu}}{\sqrt{2 \pi \Delta}} \\
\nonumber
&\times& \left\{
\int dx_0 \[[(1-\rho_0) \delta(x_0) + \rho_0 \phi_0(x_0)\]]
\prod_a dx_a  \[[(1-\rho) \delta(x_a) + \rho \phi(x_a)\]]
e^{-\frac 12 \sum_a \hat{Q}_a x_a^2 + \frac 12 \sum_{a\neq b} x_a
  x_b \hat{q}_{ab} + \sum_a \hat{m}_a x_a x_0}\right\}^N
\end{eqnarray}
Let us call $Y$ the expression in the $\{{.\}}$ in the last
equation. Introducing the following transformation into the last
equation
\begin{equation}
e^{\frac{1}{2}\hat{q}_p\sum_{a\neq b}x^ax^b}=\int Dz \, e^{z\sqrt{\hat{q}_p}\sum_{a=1}^nx_a}e^{-\frac{\hat{q}_p}{2}\sum_{a=1}^n(x^a)^2}
\end{equation}
where ${\cal D}z$ is a Gaussian integration measure with zero mean and
variance one, we obtain under the RS hypothesis
\be
Y=\int dx_0 \[[(1-\rho_0) \delta(x_0) + \rho_0 \phi_0(x_0)\]]
\int Dz \, \left\{
\int dx  \[[(1-\rho) \delta(x) + \rho \phi(x)\]]
e^{-\frac{\hat Q + \hat q}{2} x^2 + \hat m x x_0 + z\sqrt{\hat{q}} x}\right\}^n
\ee
In the $n \to 0$ limit, one can write that $f(z)^n=1+n \log f(z)$ and
thus $\int Dz f(z)^n=1+n\int Dz \log f(z) \approx e^{n\int Dz \log
  f(z)}$.
Grouping all terms together we finally get
\be
  \expect_{\bF,\bs,\bf{\xi}}Z^n =  \int {\rm d}\hat{Q} \, {\rm{d}}Q \,
  {\rm{d}}\hat{q} \,  {\rm{d}}q \,  {\rm{d}}\hat{m} \, {\rm{d}}m\,
e^{nN    \Phi(Q,q,m,\hat Q,\hat q,\hat m)  }
\ee
where $\Phi$ is the replica free energy function
\bea
     && \Phi(Q,q,m,\hat Q,\hat q,\hat m) = -\frac{\alpha}{2}
     \frac{q-2m+\rho_0 \overline {s^2}+\Delta_0}{\Delta +Q-q} -
     \frac{\alpha}{2}  \log{(\Delta+Q-q)} + \frac{Q\hat Q}{2} - m\hat
     m + \frac{q \hat q}{2} \nonumber \\ && + \int {\rm d}s
     \left[(1-\rho_0)\delta(s) + \rho_0 \phi_0(s) \right] \int {\cal
       D}z \log{\left\{  \int {\rm d}x \, e^{-\frac{\hat Q+\hat q}{2}x^2 + \hat m x s + z \sqrt{\hat q}x} \left[  (1-\rho)\delta(x) +\rho\phi(x) \right] \right\}}\, .\label{free_rep}
\eea
We remind that ${\cal D}z$ is a Gaussian integration measure with zero mean and
variance one, $\rho_0$ is the density of the signal,
and $\phi_0(s)$ is the distribution of the signal components and $\overline{s^2 }= \int ds
\, s^2 \, \phi_0(s)$ is its second moment, $\Delta_0$ is the true variance of the
measurement noise.

The physical meaning of the order parameters is
\be
 Q = \frac{1}{N} \sum_i \langle x_i^2 \rangle \, , \quad  q = \frac{1}{N} \sum_i \langle x_i \rangle^2 \, , \quad
 m = \frac{1}{N} \sum_i s_i \langle x_i \rangle \, ,\label{def_order}
\ee
in which the average is with respect to the measure $\hat P$
(\ref{p_hat}),
whereas the other three $\hat m$, $\hat q$, $\hat Q$ are auxiliary
parameters.
Using the saddle point method and  performing  derivatives with
respect to $m$, $q$, $Q-q$, $\hat m$,
$\hat q$, and $\hat Q + \hat q$ we obtain the self-consistent
equations
\bea
     \hat m &=& \hat Q + \hat q = \frac{\alpha}{\Delta + Q-q}\, , \quad \quad  \hat q =
   \frac{\alpha(q-2m+\rho_0  \overline {s^2} +\Delta_0)}{(\Delta +Q
     -q)^2}\, ,  \label{hats_gen}  \\
  m &=& \rho_0\int {\rm d} s \, s\,  \phi_0(s)  \int {\cal D}z\,
  f_a\left(\frac{1}{\hat m}, s+z \frac{\sqrt{\hat q}}{\hat m}\right)\, , \label{eq_m_gen}\\
     Q-q &=&  \int {\rm d} s \,
     [(1-\rho_0)\delta(s)+\rho_0\phi_0(s)] \int {\cal D}z   \, f_c\left(\frac{1}{\hat m}, s+z \frac{\sqrt{\hat q}}{\hat m}\right)
   \, , \label{eq_Q_gen}\\
q  &=&   \int {\rm d} s\,
[(1-\rho_0)\delta(s)+\rho_0\phi_0(s)] \int {\cal D}z \, f_a^2\left(\frac{1}{\hat m}, s+z \frac{\sqrt{\hat q}}{\hat m}\right)  \, .  \label{eq_q_gen}
\eea
From the definition of the order parameters (\ref{def_order}) we
obtain
\be
   E=q-2m+\rho_0\overline{s^2}\, , \quad \quad V=Q-q\, .
\ee
It is easily seen that the set of stationary point equations
(\ref{hats_gen}-\ref{eq_q_gen}) exactly reproduces the fixed point
condition of  the density evolution
equations (\ref{eq_V_gen}-\ref{eq_E_gen}):
BP fixed points are stationary points of the free entropy
(\ref{free_rep}).

The
uniform sampling from the measure $\hat P$, Eq.~(\ref{p_hat}), is
described by the global maximum of $\Phi$.
We can use equation
(\ref{free_rep})  in order to confirm (non-rigorously) our previous result about
the optimality of the probabilistic approach for any $\phi(x)$ with
 a support that contains that of the signal $\phi_0$,  and finite
 second moment. Indeed the free entropy $\Phi$, evaluated close
 to the
the signal i.e. when $Q=q=m=\rho_0 \overline{s^2}$,
diverges as  $-(\alpha-\rho_0)\log(\Delta+Q-q)/2$. Therefore in the
noiseless limit $\Delta\to 0$, $\Phi$ diverges when $E,V\to 0$,
whenever $\alpha>\rho_0$.

It is useful to compute the free entropy
restricted to configurations $\bx$ at a fixed squared distance $D$ from the
signal, $D=\sum_i(x_i-s_i)^2/N$. When sampling from the
probability $\hat P =P(\bx|\bF,\by)$, in the limit of large $N$, the
probability that the reconstructed signal $\bx$ is at a squared distance
$D=\sum_i(x_i-s_i)^2/N$ from the original signal $\bs$ is proportional
to $e^{N\Phi (D)}$ where $\Phi(D)$ is the free
entropy restricted to squared distance $D$.  In order to compute $\Phi(D)$ we need to
evaluate the following saddle point
\be
\Phi(D) = {\rm
  SP}_{Q,q,\hat Q,\hat q,\hat m} \Phi(Q,q,(Q-D+\rho_0\langle s^2
\rangle)/2,\hat Q,\hat q,\hat m) \, , \label{free_D}
\ee
which can be
done using  Eqs.~(\ref{eq_m_gen}-\ref{eq_q_gen}) and $\hat q =
\alpha(q-2m+\rho_0 \langle s^2\rangle)/(Q-q)^2$, and $\hat m = \hat Q
+ \hat q$. The resulting free entropy $\Phi(D)$ is a useful quantity
to visualize when the BP reconstruction fails. It will be shown and analyzed in
the next section.

Let us, at this point, underline the difference between
distance $D=\sum_i\langle (x_i-s_i)^2\rangle /N = Q - 2m + \rho_0 \overline{s^2}$ and
the mean-squared error $E=\sum_i(\langle x_i\rangle -s_i)^2/N = q - 2m + \rho_0 \overline{s^2}$.
Clearly $D=E+V$, and one should not confuse the
two definitions.

\subsection{Analysis of Bayes-optimal inference}
\label{sec:Bayes}

So far we were discussing the general case when the signal is created
using density $\rho_0$ and empirical distribution of the non-zero
elements $\phi_0$, and the belief propagation reconstruction algorithm
is used with a signal model with density $\rho\neq \rho_0$ and entry
distribution $\phi\neq \phi_0$. As we explained in section
(\ref{optimal_B}) the Bayes-optimal inference corresponds to the case
when the statistical properties of the signal, and the distribution of
the measurement noise are known. Then one can use a signal model with
\be \rho = \rho_0\, , \quad \quad \phi(x) = \phi_0(x)\, , \quad \quad
\Delta= \Delta_0\, . \label{Bayes_case} \ee In such a case exact
sampling from the measure $\hat P$ (\ref{p_hat}) corresponds to the
information-theoretic optimal way of reconstructing the signal. This
means that the predictions obtained in this case represent the best
possible reconstruction performances {\it regardless of the algorithm
  used}.

The replica symmetric computation presented in the previous section becomes
exact in this case, for reasons similar to those known in mean field
spin glasses on the 'Nishimori line' \cite{Iba99,NishimoriBook,MezardMontanari09}.
Hence in this Bayes-optimal case the above replica calculation can be used to study the information-theoretic
limits for reconstruction in CS. This is equivalent to
what was rigorously established by \cite{WuVerdu11,WuVerdu11b}.

The density evolution and the free entropy can be simplified greatly
in the Bayes-optimal case, since the Nishimori condition (\ref{Nishi_gen})
gives the following equalities:
\be
q=m \, , \quad \quad
Q=\rho\overline{s^2}\, , \quad \quad E=V\, .
\ee
%
%
Hence in the Bayes-optimal case the density evolution is characterized
by a single parameter, the mean-squared error $E=\rho\overline{s^2}-m$. Note that the mean-squared distance from the
signal to a configuration sampled from the distribution $\hat P$ is $D=E+V=2E$. The density evolution equations
(\ref{eq_V_gen}-\ref{eq_E_gen}) or (\ref{hats_gen}-\ref{eq_q_gen})
reduce to:
\be
  E^{t+1} =  \rho\overline{s^2}-\rho \int {\rm d} s \, s\,  \phi(s)
  \int {\cal D}z   f_a\left(\frac {\Delta+E^t}{\alpha}, s+z\frac{
      \sqrt{\Delta+E^t} }{\sqrt{\alpha}} \right)\, .\label{E_NL}
\ee
(we remind that the function $f_a$ is defined in (\ref{f_a_gen})).
The initial condition of Eq.~(\ref{init_E})  is $E^{t=0}=\rho
\overline{s^2} -\rho^2 {\overline s}^2$.

The free entropy also becomes a function of the single variable $E$:
\bea
    \Phi_{\rm NL}(E) &=& -\frac{\alpha}{2} - \frac{\alpha}{2}
     \log{(\Delta+E)} - \frac{\alpha(\rho\overline{s^2}+E)}{2(\Delta+E)}
     \nonumber \\ &+& \int {\rm d}s  \left[(1-\rho)\delta(s) + \rho \phi(s)
     \right] \int
     {\cal D}z  \log{\left\{  \int {\rm d}x \,
         e^{\frac{\alpha}{\Delta+E} x
           (s-\frac{x}{2}) + z x \frac{\sqrt{\alpha}}{\sqrt{\Delta+E}}} \left[  (1-\rho)\delta(x) +\rho\phi(x) \right] \right\}}\, .\label{free_rep_NL}
\eea
When the signal distribution is known, the value of the MSE $E$ at the global maximum of this free
entropy provides the Bayes optimal reconstruction of the signal,
i.e. the lowest achievable MSE given the knowledge of the measurement
vector $\by$ and the measurement matrix $\bF$.  As we will see, depending on
parameters $\alpha$, $\rho$ and $\phi(x)$, the BP
algorithm where the MSE evolves according to (\ref{E_NL}) will either
find this global maximum or it will get blocked in a local suboptimal maximum.

For completeness let us give the explicit form of the free entropy
(\ref{free_rep_NL}) for a Gauss-Bernoulli signal where $\phi_0$ has zero mean and unit
variance:
\bea \Phi_{NL}(E) &=& -\frac{\alpha}{2}
\left[\log{(\Delta+E)} + \frac{\Delta}{\Delta+E}\right] + (1-\rho) \frac {\alpha}{2(\alpha+\Delta+E)}
\\
&+& (1-\rho) \int {\cal D}z \log{\left[ (1-\rho) e^{-\frac{z^2 \alpha
      }{2(\alpha +\Delta+ E)}}+ \frac{\rho
      \sqrt{\Delta+E}}{\sqrt{\Delta+E+\alpha}} \right]} + \rho \int
{\cal D}z \log{\left[ (1-\rho) e^{\frac{-z^2 \alpha
      }{2(\Delta+E)}}+\frac{\rho
      \sqrt{\Delta+E}}{\sqrt{\Delta+E+\alpha}}
  \right]}\label{Phi_E}\, . \eea
In this case, the condition of stationarity of the free entropy,
giving also the fixed-point condition of density evolution, takes the
simple form:
\be
E = \rho - \frac{\rho^2}{\alpha+\Delta+E} \int {\cal D}z \frac{ z^2
}{\rho + (1-\rho) \frac{\sqrt{\alpha+\Delta+E}}{\sqrt{\Delta+E}}
  e^{-\frac{z^2}{2}\frac{\alpha}{\Delta+E} }}\, .
\ee

\subsection{Density evolution with parameter learning}
\label{DE_learning}

We study here the general case where the signal is created
using a density $\rho_0$ and empirical distribution of the non-zero
elements $\phi_0$, and the belief propagation reconstruction algorithm
is used with a different  signal model, with density $\rho\neq \rho_0$ and
distribution of the non-zero
elements $\phi\neq \phi_0$. In this case, expectation maximization can be used to
learn the parameters, as described in Sec.~\ref{BP_learning}.
This modified BP procedure, including parameter learning, can also be
studied with density evolution. We describe here the case that we use
in our implementation, namely a model signal which is
Gauss-Bernoulli, where $\phi$ is Gaussian with mean $\overline x$ and variance $\sigma^2$.
The learning conditions (\ref{learn_rho}-\ref{learn_mv}) give the
evolution of the parameters:
\bea
\rho^{(t+1)} &=&\rho^{(t)}
\frac{ \int {\rm d}s
\left[(1-\rho_0)\delta(s) +\rho_0 \phi_0(s) \right]
\int {\cal D}z \frac{
g(\Sigma^2,s+zU)
}
{
1-\rho^{(t)}+\rho^{(t)} g(\Sigma^2,s+zU)
}
} {
\int {\rm d}s
\left[(1-\rho_0)\delta(s) +\rho_0 \phi_0(s) \right]
\int {\cal D}z \frac{
1
}
{
1-\rho^{(t)}+\rho^{(t)} g(\Sigma^2,s+zU)
}
} \, ,\label{eq:learn_analytic1}
\\
\ox^{(t+1)} &=&
\frac{
{\int {\rm d}s
\left[(1-\rho_0)\delta(s) +\rho_0 \phi_0(s) \right]
\int \cal D}z   f_a(\Sigma^2,s+zU)
}{ \rho^{(t)}
\int {\rm d}s
\left[(1-\rho_0)\delta(s) +\rho_0 \phi_0(s) \right]
\int {\cal D}z \frac{
g(\Sigma^2,s+zU)
}
{
1-\rho^{(t)}+\rho^{(t)} g(\Sigma^2,s+zU)
}
} \, ,\label{eq:learn_analytic2}
\\
(\sigma^2)^{(t+1)}&=&
\frac{
 \int {\rm d}s
\left[(1-\rho_0)\delta(s) +\rho_0 \phi_0(s) \right]
\int {\cal D}z [f^2_a(\Sigma^2,s+zU)+f_c(\Sigma^2,s+zU)]
}{\rho^{(t)}
{ \int {\rm d}s
\left[(1-\rho_0)\delta(s) +\rho_0 \phi_0(s) \right]
\int \cal D}z\frac{
g(\Sigma^2,s+zU)
}
{
1-\rho^{(t)}+\rho^{(t)} g(\Sigma^2,s+zU)
}
}  - \left[\ox^{(t+1)}\right]^2\, ,\label{eq:learn_analytic3}
\eea
where the function $g$ is defined as
\be
   g(\Sigma^2,R) =  \, \frac{\Sigma}{\sqrt{\Sigma^2+\sigma^2}}e^{\frac{(R/\Sigma^2+\overline x/\sigma^2)^2}{2(1/\Sigma^2+1/\sigma^2)}-\frac{\overline x^2}{2\sigma^2}} .
\ee
And we use
\be
    \Sigma^2 = \frac{\Delta + V^t}{\alpha}\, , \quad \quad U \equiv
    \sqrt{\frac{\Delta_0+E^t}{\alpha}}\, .
\ee
The density evolution for the simplified learning
(\ref{learn_xbar_alternative}, \ref{learn_mv_alternative}) reads
\bea
\ox^{(t+1)} &=&
\frac{1}{\rho^{(t)}}\int {\rm d}s
\left[(1-\rho_0)\delta(s) +\rho_0 \phi_0(s) \right]
\int {\cal D}z \,   f_a(\Sigma^2,s+zU)  \, ,\label{eq:learn_analytic2_bis}
\\
(\sigma^2)^{(t+1)}&=& \frac{1}{\rho^{(t)}}
 \int {\rm d}s
\left[(1-\rho_0)\delta(s) +\rho_0 \phi_0(s) \right]
\int {\cal D}z \, \left[f^2_a(\Sigma^2,s+zU)+f_c(\Sigma^2,s+zU)\right]
- \left[\ox^{(t+1)}\right]^2\, .
\label{eq:learn_analytic3_bis}
\eea

The density evolution equations now provide a mapping
\be
\left(E^{(t+1)},V^{(t+1)},\rho^{(t+1)},\overline x^{(t+1)},\sigma^{(t+1)}\right)= f
\left(E^{(t)},V^{(t)},\rho^{(t)},\overline
  x^{(t)},\sigma^{(t)}\ \right)
\ee  obtained by complementing the
  previous equations on $V$, and $E$
  (\ref{eq_V_gen}-\ref{eq_E_gen})  with
  the learning update equations
  (\ref{eq:learn_analytic1}, \ref{eq:learn_analytic2}, \ref{eq:learn_analytic3}).
In our implementation we initialize $\rho^{t=0}=\alpha/10$, $\overline
x^{t=0}=0$, and $\sigma^2_{t=0}= 1$.

When a measurement noise is present the variance of the noise can be
learned using Eq.~(\ref{delta_learn}) which in the density evolution becomes
\be
        \Delta^{(t)} = \frac{\Delta_0+E^t}{1+\frac{V}{\Delta^{(t)}}}\, .
\ee

\subsection{Density evolution for block matrices}
\label{replica_1D}


In the case of the block measurement matrices defined in
Sec.~\ref{block_matrices}, one can easily generalize the above
derivation of the density evolution and of the replica analysis. We
just give the results here. For details of the derivation see appendix \ref{appendix:free_entropy}.

 The order parameters are now
\be
Q_p \equiv \frac{1}{N_p} \sum_{i\in B_p} \langle x_i^2 \rangle  \, ,
\quad
q_p \equiv \frac{1}{N_p} \sum_{i\in B_p} \langle x_i \rangle^2\, ,\quad
m_p \equiv \frac{1}{N_p} \sum_{i\in B_p} s_i \langle x_i\rangle
\ee
 in each block $p\in\{1,\dots,L_c\}$. The free entropy analogous
to that in Eq.~(\ref{free_rep}) becomes
\bea
&&
\Phi(\{Q_p\}_{p=1}^{L_c},\{q_p\}_{p=1}^{L_c},\{m_p\}_{p=1}^{L_c},\{\hat
Q_p\}_{p=1}^{L_c},\{\hat q_p\}_{p=1}^{L_c},\{\hat m_p\}_{p=1}^{L_c}) = \nonumber
\\ && -\frac{1}{2}\sum_{q=1}^{L_r} n_1 \alpha_{q1} \left[ \frac{\tilde
    q_q-2\tilde m_q+\tilde \rho_q+\Delta_0}{\tilde
    Q_q-\tilde q_q+\Delta} + \log{(\Delta+\tilde Q_q-\tilde q_q)}
\right] + \sum_{p=1}^{L_c} n_p\left( \frac{Q_p\hat Q_p}{2} - m_p\hat m_p +
  \frac{q_p \hat q_p}{2} \right)\nonumber \\ && + \sum_{p=1}^{L_c}n_p\int {\rm d}s \left[(1-\rho_0)\delta(s) + \rho_0 \phi_0(s)
\right] \int
{\cal D}z \log{\left\{ \int {\rm d}x \, e^{-\frac{\hat Q_p+\hat
        q_p}{2}x^2 + x(\hat m_p s + z \sqrt{\hat q_p})} \left[
      (1-\rho)\delta(x) +\rho \phi(x) \right] \right\}} \,
, \label{free_seeded}
\eea
where we introduced
\be
\tilde{\rho}_q=\rho_0 \overline{s^2}\sum_{p=1}^{L_c} J_{qp} n_p,\quad
\tilde{m}_q=\sum_{p=1}^{L_c} J_{qp}n_p m_p, \quad \tilde{q}_q=\sum_{p=1}^{L_c}
J_{qp} n_p q_p,\quad \tilde{Q}_q=\sum_{p=1}^{L_c} J_{qp}n_p Q_p\,
. \label{def_tilde}
\ee
The equations corresponding to the stationarity condition for this
free entropy  read:
\bea
    \hat q_p &=& n_p\sum_{q=1}^{L_r}  \frac{\alpha_{qp} J_{qp} (\tilde q_q-2\tilde m_q+\tilde \rho_q+\Delta_0)}{(\tilde Q_q-\tilde q_q+\Delta)^2}\, , \quad \quad\label{hqp1}\\
     \hat m_p &=& n_p\sum_{q=1}^{L_r}\frac{\alpha_{qp} J_{qp}}{\tilde
       Q_q-\tilde q_q+\Delta}\,
     , \label{hmp1}\\
\hat Q_p&=&\hat m_p-\hat q_p\label{hQp1}\\
      m_p &=& \rho_0 \int {\rm d} s \, s \, \phi_0(s) \int {\cal D}z  f_a\left(\frac{1}{\hat m_p},s+z\frac{\sqrt{\hat q_p}}{\hat m_p}\right)\, , \label{eq_m_GB}\\
     Q_p-q_p &=& \int {\rm d} s [(1-\rho_0)\delta(s)+\rho_0\phi_0(s)]
     \int {\cal D}z\,   f_c\left(\frac{1}{\hat m},s+z\frac{\sqrt{\hat q_p}}{\hat m_p}\right)
   \, , \label{eq_Q_GB-}\\
 q_p  &=&  \int {\rm d} s [(1-\rho_0)\delta(s)+\rho_0\phi_0(s)] \int
 {\cal D}z \,   f_a^2\left(\frac{1}{\hat m_p},s+z\frac{\sqrt{\hat q_p}}{\hat m_p}\right)  \ .  \label{eq_q_GB}
  \eea

When interpreted as a mapping (given the order parameters
$Q_p,q_p,m_p$ at time $t$, one computes $\hat Q_p$, $\hat q_p$, $\hat
m_p$ form (\ref{hqp1}-\ref{hQp1}), and then finds the new order
parameters $Q_p,q_p,m_p$ at time $t+1$ using
(\ref{eq_m_GB}-\ref{eq_q_GB})), these equations are exactly the
density evolution equations for the case of block matrices.
These equations can be written
in term of only $2 L_c$ order parameters, the mean-squared error $E_p=q_p-2m_p+\rho_0\overline
{s^2}$ and the variance $V_p=Q_p-q_p$ in each block $p\in\{1,\dots,L_c\}$.
The explicit form of the density evolution equations in terms of these
$2 L_c$ order parameters is:
\bea
  E_p^{(t+1)}  &=& \int {\rm d}s
      \left[(1-\rho_0)\delta(s)+\rho_0 \phi_0(s) \right] \int
      {\cal D}z  \left[
 f_a\left(\frac{1}{\hat m_p},s+z\frac{\sqrt{\hat q_p}}{\hat m_p}\right)-s\right]^2\, ,\\
V_p^{(t+1)}  & =&\int {\rm d}s \left[(1-\rho_0)\delta(s)+\rho_0
     \phi_0(s) \right] \int {\cal D}z   f_c\left(\frac{1}{\hat m_p},s+z\frac{\sqrt{\hat q_p}}{\hat m_p}\right)\, ,
\eea
where:
\bea
\hat m_p&=& n_p \sum_{q=1}^{L_r}\frac{\alpha_{qp} J_{qp}}{\Delta +\sum_{r=1}^{L_c}
   J_{q r}n_r V_r^{(t)}}\ , \label{hatm}\\
\hat q_p&=& n_p\sum_{q=1}^{L_r}  \left\{ \frac{\alpha_{qp} J_{qp}}{\left[\Delta +\sum_{r=1}^{L_c}
   J_{q r}n_r V_r^{(t)}\right]^2}\; \left[ \Delta_0 + \sum_{s=1}^{L_c}
 J_{qs} n_s E_s^{(t)} \right]
\right\}\, .\label{hatq}
\eea

If one uses block measurement matrices together with expectation-maximization learning of the parameters, for a
Gauss Bernoulli signal model, the density evolution equations for the
parameters are:
\bea
\rho^{(t+1)} &=&\rho^{(t)}
\left(
\frac{1}{{L_c}}\sum_{p=1}^{L_c}\int {\cal D}z \int {\rm d}s
\left[(1-\rho_0)\delta(s) +\rho_0 \phi_0(s) \right]
\frac{
g\left(\frac{1}{\hat m_p},s+z\frac{\sqrt{\hat q_p}}{\hat m_p}\right)
}
{
1-\rho+\rho g\left(\frac{1}{\hat m_p},s+z\frac{\sqrt{\hat q_p}}{\hat m_p}\right)
}
\right) \\
&&
\left(
\frac{1}{{L_c}}\sum_{p=1}^{L_c}\int {\cal D}z \int {\rm d}s
\left[(1-\rho_0)\delta(s) +\rho_0 \phi_0(s) \right]
\frac{
1
}
{
1-\rho+\rho g\left(\frac{1}{\hat m_p},s+z\frac{\sqrt{\hat q_p}}{\hat m_p}\right)
}
\right)^{-1}\, ,
\\
\ox^{(t+1)} &=&
\frac{1}{\rho}
\left(
\frac{1}{{L_c}}\sum_{p=1}^{L_c}\int {\cal D}z \int {\rm d}s
\left[(1-\rho_0)\delta(s) +\rho_0 \phi_0(s) \right]
f_a\left(\frac{1}{\hat m_p},s+z\frac{\sqrt{\hat q_p}}{\hat m_p}\right)
\right) \nonumber \\
&&
\left(
\frac{1}{{L_c}}\sum_{p=1}^{L_c}\int {\cal D}z \int {\rm d}s
\left[(1-\rho_0)\delta(s) +\rho_0 \phi_0(s) \right]
\frac{
g\left(\frac{1}{\hat m_p},s+z\frac{\sqrt{\hat q_p}}{\hat m_p}\right)
}
{
1-\rho+\rho g\left(\frac{1}{\hat m_p},s+z\frac{\sqrt{\hat q_p}}{\hat m_p}\right)
}
\right)^{-1}\, ,
\\
(\sigma^2)^{(t+1)}&=&
\frac{1}{\rho}
\left(
\frac{1}{{L_c}}\sum_{p=1}^{L_c}\int {\cal D}z \int {\rm d}s
\left[(1-\rho_0)\delta(s) +\rho_0 \phi_0(s) \right]
[f^2_a\left(\frac{1}{\hat m_p},s+z\frac{\sqrt{\hat q_p}}{\hat m_p}\right)+f_c\left(\frac{1}{\hat m_p},s+z\frac{\sqrt{\hat q_p}}{\hat m_p}\right)]\right) \nonumber \\
&&
\left(
\frac{1}{{L_c}}\sum_{p=1}^{L_c}\int {\cal D}z \int {\rm d}s
\left[(1-\rho_0)\delta(s) +\rho_0 \phi_0(s) \right]
\frac{
g\left(\frac{1}{\hat m_p},s+z\frac{\sqrt{\hat q_p}}{\hat m_p}\right)
}
{
1-\rho+\rho g\left(\frac{1}{\hat m_p},s+z\frac{\sqrt{\hat q_p}}{\hat m_p}\right)
}
\right)^{-1} -\left[\ox^{(t+1)}\right]^2\, .
\eea

As in the homogeneous case, the density evolution equation of the
block measurement matrices simplify in the optimal Bayesian approach, when the correct distribution of the
signal and its density are known $\rho_0=\rho$, $\phi_0=\phi$. In this
case, the Nishimori conditions $m_p=q_p$
and $Q_p=\rho \overline{s^2}$ hold,  hence $E_p=V_p$ holds for every
block $p=1,\dots,L_c$. This leads to a
single set of closed density evolution equations for the vector $E_p$, $p=1,\dots,L_c$,
that reads
\bea
 E_p^{(t+1)}  &=& \int {\rm d}s
      \left[(1-\rho)\delta(s)+\rho\phi(s) \right] \int
      {\cal D}z  \left[
 f_a\left(\frac{1}{\hat m_p},s+z\frac{1}{\sqrt{\hat m_p}}\right)-s\right]^2\, ,\\
\hat m_p&=&  \sum_{q=1}^{L_r}n_p\frac{\alpha_{qp} J_{qp}}{\Delta +\sum_{r=1}^{L_c}
   J_{q r}n_r E_r^{(t)}}\ .
\eea

In the case where $\phi_0$ is a centered Gaussian with
unit variance, we get explicitly:
\be
E_p^{(t+1)} =\rho-\frac{\rho^2 \hat m_p}{\hat m_p +1} \int {\cal D}z \frac{ z^2
}{\rho + (1-\rho)e^{-\frac{z^2\hat m_p}{2}} \sqrt{\hat m_p +1} }\, .
\ee


\newpage
\section{The phase diagrams}

In this section we turn the equations from the previous section into phase diagrams to
display the performance of belief propagation in CS
reconstruction. We first discuss the noiseless case, with random
homogeneous measurement matrices, this is a benchmark case that has
been widely used to demonstrate the power of
the $\ell_1$ reconstruction. We use measurement matrices with iid entries with
zero mean and variance $1/N$ (we remind that our approach is
independent of the distribution of the iid matrix elements and depends
only on their mean and variance).  Finally we discuss the phase
diagram for noisy measurements, that present several interesting
features.


\subsection{Noiseless measurements and the optimal Bayes case}
\label{Res:Bayes}

\begin{figure}[!ht]
  \begin{center}
      \includegraphics[width=0.4\linewidth]{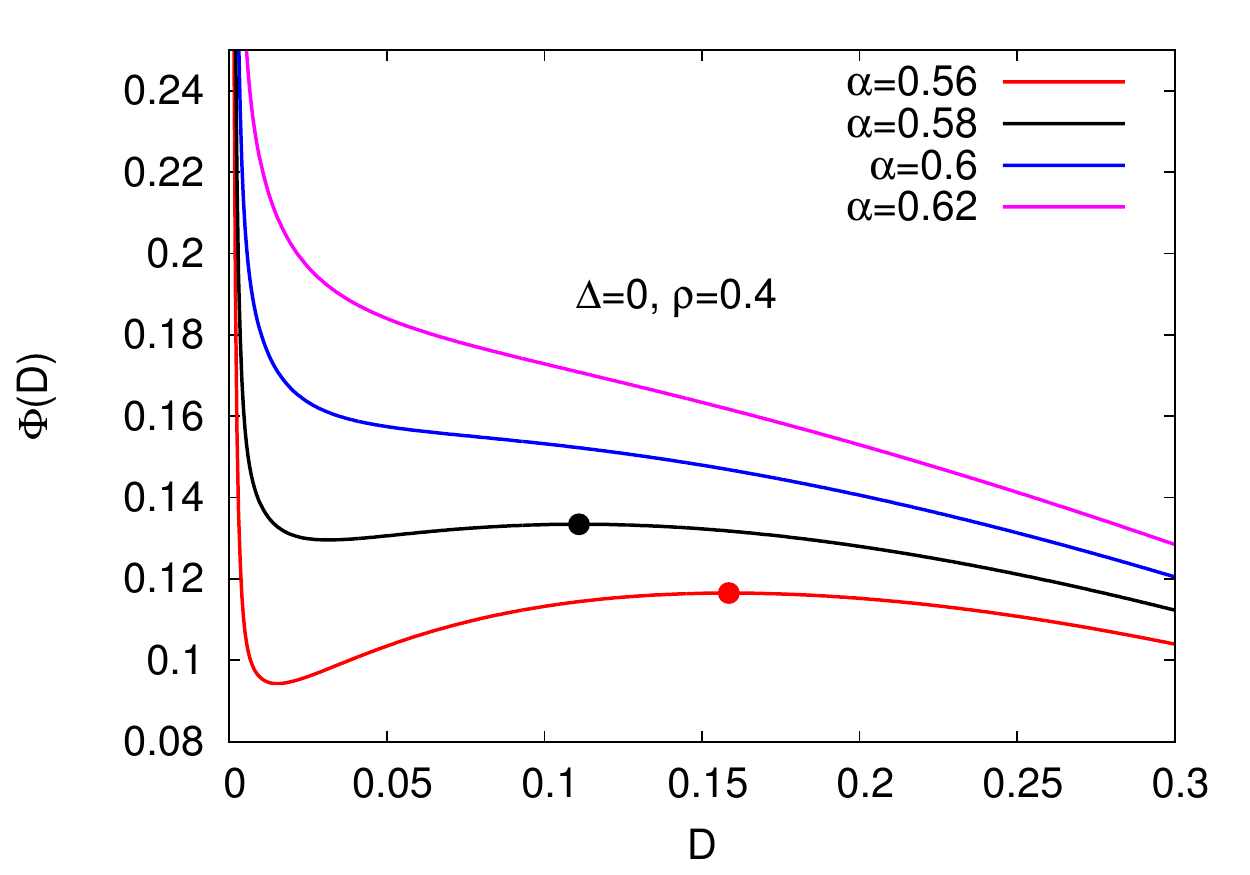}
      \includegraphics[width=0.4\linewidth]{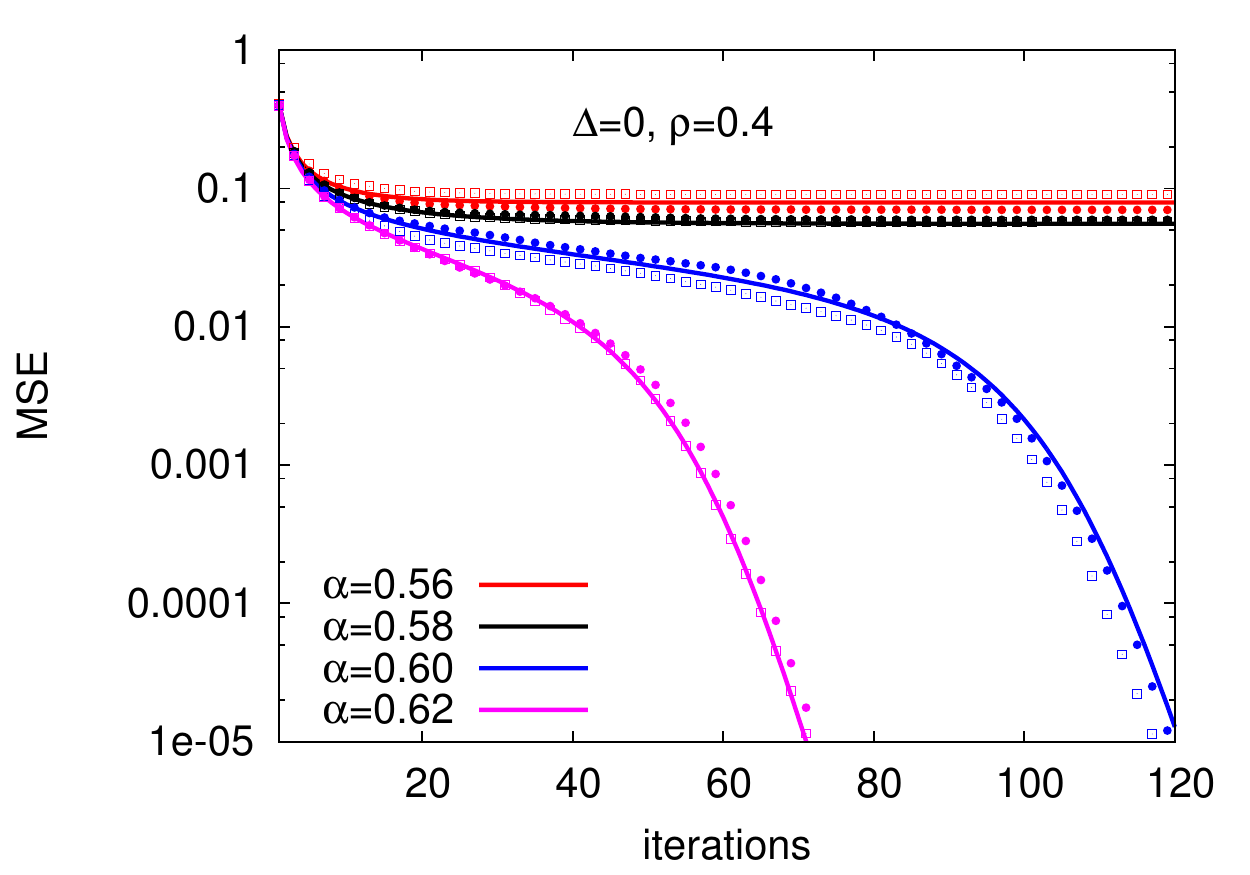}
      \caption{Left: The free entropy, $\Phi(D)$, is plotted as a function
        of $D=\big \langle \sum_i(x_i-s_i)^2/N \big \rangle$ for
        $\rho_0=0.4$ and several measurement rates $\alpha$ in the
        Bayesian approach (when both the signal and the signal model are
        described by a Gauss-Bernoulli distribution).  The evolution
        of the BP algorithm is basically a steepest ascent in
        $\Phi(D)$ starting from a large value of $D$. Such ascent goes to the
        global maximum at $D=0$ for large value of $\alpha$ but is
        blocked in the local maximum that appears for
        $\alpha<\alpha_{\rm BP}(\rho_0=0.4)\approx 0.59$. For
        $\alpha<\rho_0$, the global maximum is not at $D=0$ and exact inference
        is impossible. Right: Using the same conditions as for the
        left figure, we show the evolution of the MSE measured experimentally during the
        iterations of BP for a signal of size $N=15000$ (data points)
        compared to the theory using density evolution (line). For the two
        lower measurement rates, where $\alpha<0.59$, the MSE saturates
        at a finite value. For
        the two higher ones it goes to zero. The full circles are
        for measurement matrices with iid Gaussian elements, the empty
        squares for matrices with iid $\pm 1$ elements.  We see small finite
        size corrections, but otherwise there is excellent agreement between the
        two cases, as expected from the theory which states that only
        the mean and variance of the distribution of each matrix
        element matters.}
    \label{fig:FreeEntropy}
  \end{center}
\end{figure}

In Fig.~\ref{fig:FreeEntropy} we show the free entropy density
at fixed squared distance, $\Phi(D)$, for the Bayes-optimal case
in which both $\phi_0$ and $\phi$ are Gaussian with zero mean
and unit variance. The elements of the $M\times N$ measurement matrix
$\bF$ are independent random variables with zero mean and variance
$1/N$.

The free entropy $\Phi(D)$  is computed using
Eq.~(\ref{free_rep}) and (\ref{free_D}) which was derived using the
replica method.  The dynamics of the message passing algorithm (without learning)
is a gradient dynamics leading to a maximum of the free-entropy
$\Phi(D)$ starting from high distance $D$. As expected, we see in
Fig.~\ref{fig:FreeEntropy} that $\Phi(D)$ has a global maximum at $D=0$ if and
only if $\alpha >\rho_0$, which confirms that the Bayesian optimal inference
is in principle able to reach the theoretical limit
$\alpha=\rho_0$ for exact reconstruction. The left-hand side of the
figure shows the existence of a critical
measurement rate $\alpha_{\rm BP}(\rho_0)>\rho_0$, below which a secondary local maximum of $\Phi(D)$
appears at $D>0$.  When this secondary maximum exists, the BP algorithm converges instead to
it, and does not reach exact reconstruction.  The
threshold $\alpha_{\rm BP}(\rho_0)$ is obtained analytically as the smallest
value of $\alpha$ such that $\Phi(D)$ is monotonic.
The behavior of $\Phi(D)$ is
typical of a first order transition. The equilibrium transition
appears at a number of measurement per unknown $\alpha=\rho_0$, which
is the point where the global maximum of $\Phi(D)$ switches
discontinuously from being
at $D=0$ (when $\alpha>\rho_0$) to a value $D>0$. In this sense the
value $\alpha=\alpha_{\rm BP}(\rho_0)$ appears like a spinodal point:
it is the point below which the global maximum of $\Phi(D)$ is no
longer reached by the dynamics. Instead, in the regime below the
spinodal ($\alpha<\alpha_{\rm BP}(\rho_0$), the dynamical evolution is
attracted to a metastable non-optimal state with $D>0$.

On the right-hand side of Fig.~\ref{fig:FreeEntropy}, we show the
evolution of the MSE as predicted by the density evolution equations,
as well as the MSE measured using the BP
algorithm for a system with size $N=15000$. Below the spinodal point
$\alpha_{\rm BP}(\rho_0)$ the MSE does not converge to zero, because
the system is trapped in a metastable state.

\begin{figure}[!ht]
  \begin{center}
     \includegraphics[width=0.48\linewidth]{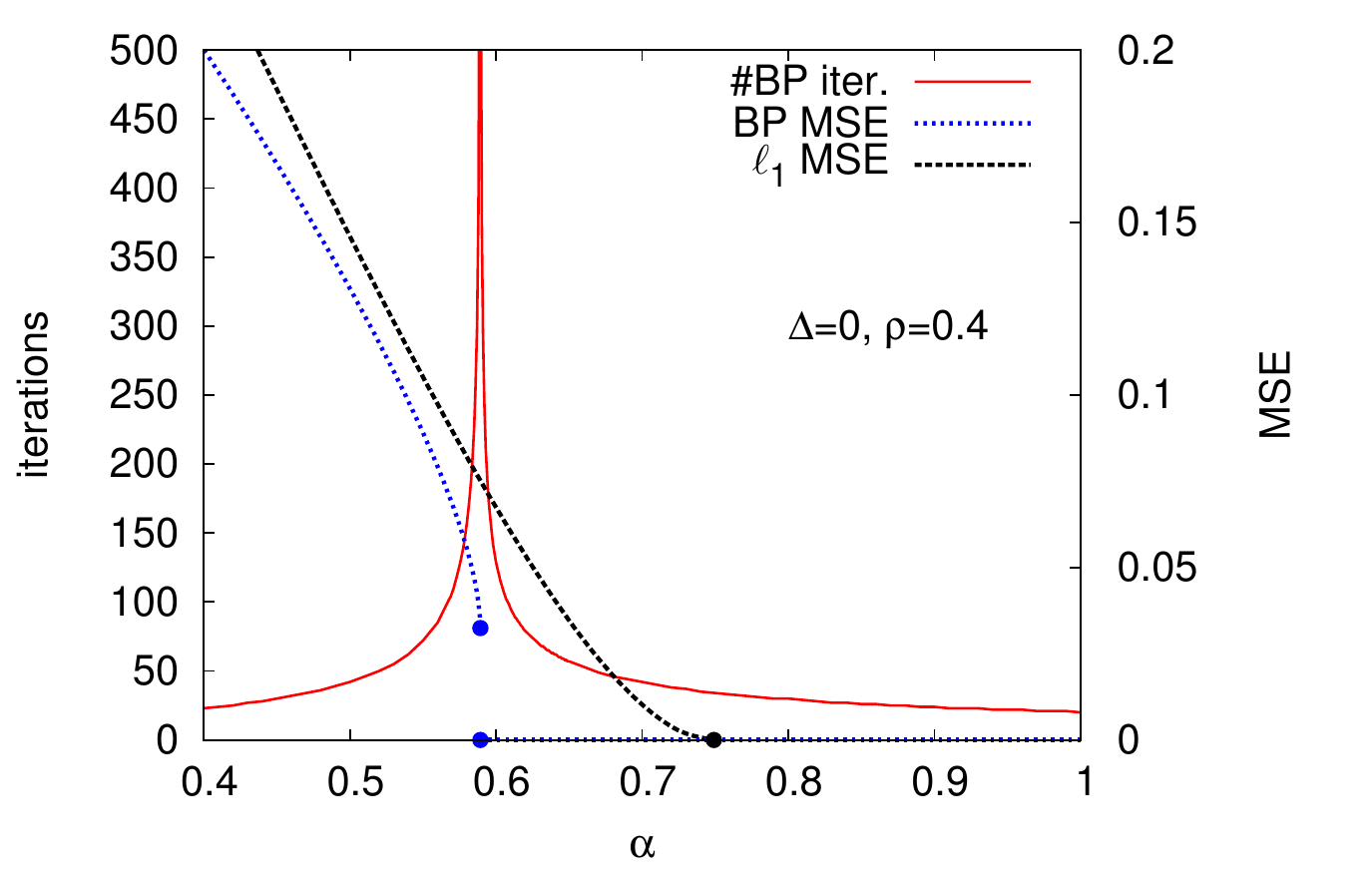}
     \caption{The full (red) line (left $y$-axis) is the convergence time of
       the BP algorithm, defined as the number of iterations needed
       such that the
       MSE obtained by the algorithm at a given iteration does not
       change more than by $10^{-7}$ in the next iteration.
       The data are obtained with the density evolution for a signal
       with density $\rho_0 = 0.4$, where the non-zero elements of the signal are
       Gaussian with zero mean and unit variance. Reconstruction is
       done in the Bayes-optimal case.   The BP convergence time diverges as
       $\alpha \to \alpha_{\rm BP}$. The dotted lines (right $y$-axis) give the mean-squared error
       achieved by the BP algorithm (blue) and by the $\ell_1$
       minimization (black)
       for reconstruction of the same signal. Exact reconstruction is
       in principle possible in the whole region $\alpha
       > \rho_0$. The reconstruction with BP is exact for $\alpha>\alpha_{\rm
         BP}(\rho_0=0.4)\approx0.59$, whereas the $\ell_1$-reconstruction is
       exact only for $\alpha \gtrsim 0.75$. Note also in the regime
       $\alpha<\alpha_{\rm BP}$
       where BP does not reconstruct  exactly the signal, the MSE
       achieved by BP is always smaller than the one of $\ell_1$.}
    \label{fig:MSQ_BEPL}
  \end{center}
\end{figure}

 The spinodal transition is the physical reason that limits the
 performance of the BP algorithm.
To illustrate this statement, we plot in
Fig.~\ref{fig:MSQ_BEPL} the BP convergence time as a function the
measurement rate $\alpha$. As expected, the convergence time diverges around the spinodal transition
$\alpha_{\rm BP}$. In the same Fig.~\ref{fig:MSQ_BEPL} we also plot
the MSE achieved by the BP reconstruction algorithm compared to the MSE
achieved by the $\ell_1$ minimization reconstruction for the same
signal and the same measurement matrix as before. We remind that here we are still in
the favorable case when the signal model was equal to the signal
distribution $\rho=\rho_0$, $\phi(x)=\phi_0(x)$.

Notice that the $\ell_1$ transition at $\alpha_{\ell_1}$ is continuous (second order), whereas the spinodal
transition is discontinuous (first order).
The transition at $\alpha_{\rm BP}$ is called a
spinodal transition in the mean field theory of first order phase
transitions. It is similar to the one found
in the cooling of liquids which go into a super-cooled glassy state
instead of crystallizing, and appears in the decoding of error
correcting codes \cite{RichardsonUrbanke08,MezardMontanari09} as well.
This difference might seem formal, but it is absolutely essential for what concerns the possibility
of achieving the theoretically optimal reconstruction with the use of
seeding measurement matrices (as discussed in the next section).

\begin{figure}[!ht]
  \begin{center}
		\includegraphics[width=0.48\linewidth]{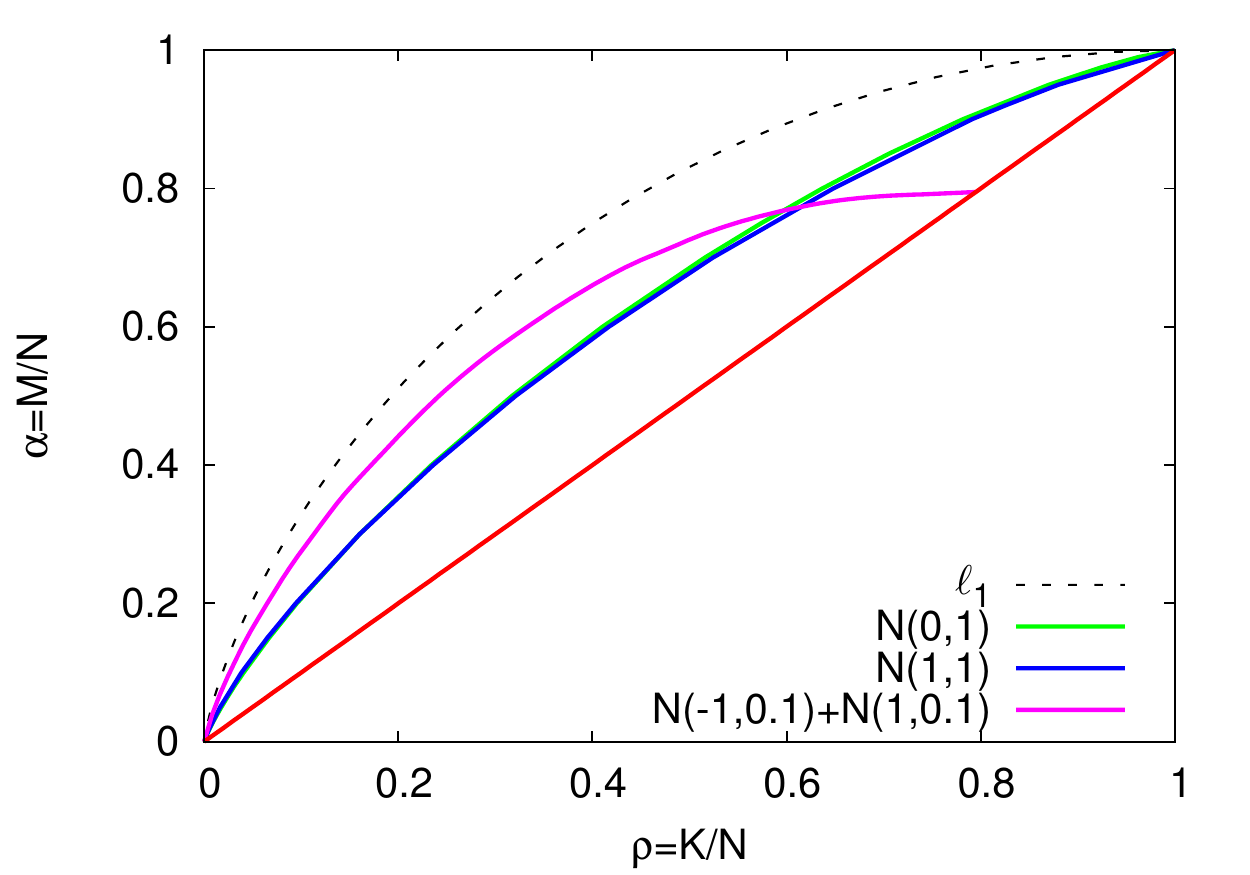}
                \includegraphics[width=0.48\linewidth]{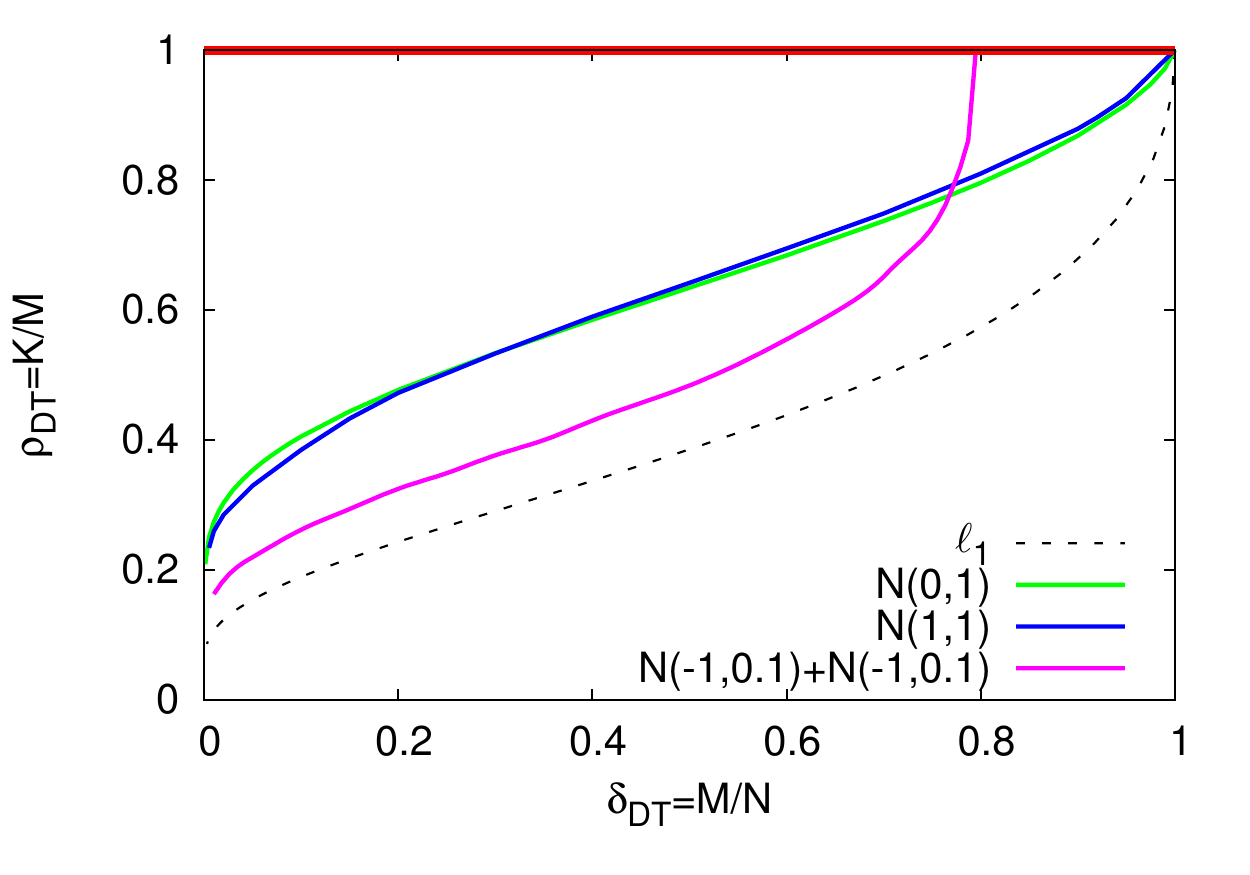}
       \caption{Phase diagram for the BP reconstruction in the optimal Bayesian
                 case when the signal model is
                  matching the empirical distribution of signal
                  elements, i.e. $\phi(x)=\phi_0(x)$. The elements of the $M\times N$
                  measurement matrix $\bF$ are iid variables with zero mean and variance $1/N$.
                The spinodal transition $\alpha_{\rm
                    BP}(\rho_0)$ is computed with the asymptotic replica
                  analysis and plotted for the following signal
                  distributions: $\phi(x)={\cal N}(0,1)$ (green),
                  $\phi(x)={\cal N}(1,1)$ (blue)
                  $\phi(x)=[{\cal N}(-1,0.1)+{\cal N}(1,0.1)]/2$
                  (magenta, equations needed to obtain this curve are
                  summarized in appendix \ref{appendix:mixture}).
Note that for some signals, e.g. the third case, there is a region of
signal densities
(here $\rho_0 \gtrsim 0.8$) for which the BP reconstruction is possible down to the optimal
subsampling rates $\alpha=\rho_0$.
              The data are compared to the Donoho-Tanner phase
              transition $\alpha_{\ell_1}(\rho_0)$ (dashed) for $\ell_1$ reconstruction that does not
              depend on the signal distribution, and to the
              theoretical limit for exact reconstruction $\alpha=\rho_0$
              (red). The left hand side represents the undersampling
              rate $\alpha$ as a function of the signal density
              $\rho_0$. The right hand side shows the same data in the
              Donoho-Tanner notation, i.e. the number of nonzero
              elements in the signal per measurement
              is plotted as a function for the undersampling rate.
  \label{fig:phase_diag_Bayes}}
\end{center}
\end{figure}

In Fig.~\ref{fig:phase_diag_Bayes} we show how  the critical value
$\alpha_{\rm BP}$ depends on the signal density $\rho$ and on the type
of the signal, for several Gauss-Bernoulli signals. In this figure we still assume that the signal
distribution is known, and hence $\rho_0=\rho$ and $\phi_0=\phi$. We compare to the
Donoho-Tanner phase transition   $\alpha_{\ell_1}$  that gives the limit for exact
reconstruction with the $\ell_1$ minimization
\cite{DonohoTanner2009,DonohoMaleki09,KabashimaWadayama09}, and to the
information-theoretical
 limit for exact reconstruction $\alpha=\rho$.

Note that for some signals, e.g. the mixture of Gaussians $\Phi(x)=
[{\cal N}(-1,0.1)+{\cal N}(1,0.1)]/2$, there is a region of signal  densities
(here $\rho_0 \gtrsim 0.8$) for which the BP reconstruction is possible down to the optimal
subsampling rates $\alpha=\rho_0$.

\subsection{Noiseless measurements and the mismatching signal model}
\label{Res:learning}

In this section we show the performance of BP
reconstruction and the corresponding phase diagrams in the general
case when the density of the signal and the
distribution of the non-zero signal elements is not known
\be
    \rho\neq \rho_0\, , \quad \quad \phi(x) \neq \phi_0(x) \, .\label{no_Nish}
\ee
All the results we show are for the Gauss-Bernoulli model of the
signal, i.e. $\phi(x)=e^{-(x_i-\overline x)^2/(2\sigma^2)}/(\sqrt{2\pi}\sigma )$.
As we argued in Sec.~\ref{Optimality}, for noiseless measurements the
probabilistic reconstruction for CS is optimal as long
as $\alpha>\rho_0$ even if the signal model is not the correct one,
as in (\ref{no_Nish}). This property can also be seen by analyzing the
replica calculation of the free entropy (\ref{free_rep}) that close to exact
reconstruction ($Q \to \rho_0 \overline{s^2}$, $q \to \rho_0
\overline{s^2}$, $m \to \rho_0 \overline{s^2}$) behaves as  $\Phi \to -
(\alpha-\rho_0)\log{(Q-q)}/2$. Unfortunately, in general, BP encounters a spinodal line (barrier) as in the
case discussed in the previous section. The position of this line
(phase transition) depends on both the signal
model $\phi(x)$ and the signal distribution $\phi_0(x)$.

\begin{figure}[!ht]
  \begin{center}
		\includegraphics[width=0.48\linewidth]{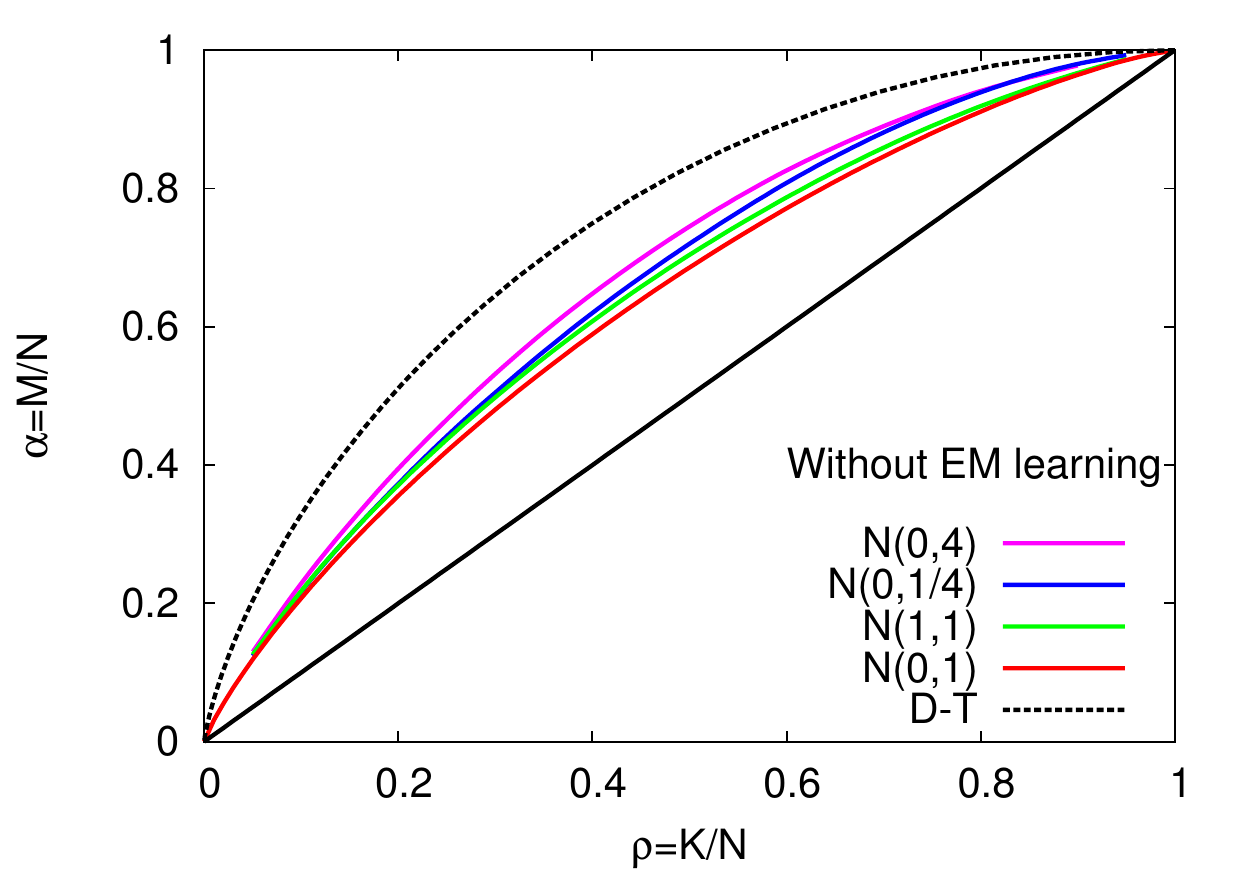}
		\includegraphics[width=0.48\linewidth]{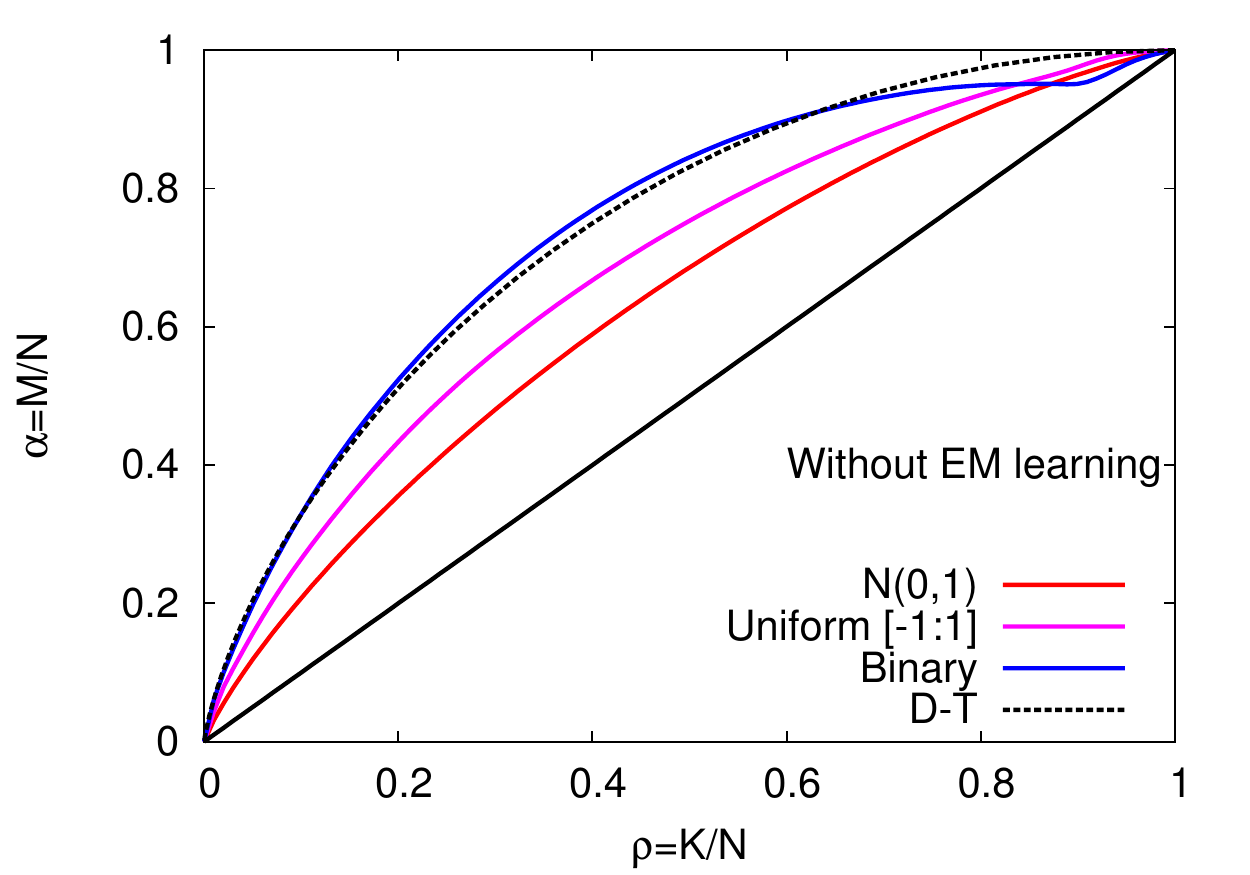}
       \caption{
     \label{fig:phase_diag_nolearn} Phase diagram for the
     reconstruction with BP when the signal model is not
     matching the empirical distribution of signal elements. The signal model is
     Gauss-Bernoulli with zero mean and unit variance. The measurement
     matrix is the homogeneous one with Gaussian  iid entries. In this plot we assume that the signal density is
     known $\rho=\rho_0$. Different
     curves correspond to different distributions $\phi_0$ of the
     signal. The dashed line gives the Donoho-Tanner transition line
     for $\ell_1$ reconstruction,
     which is independent of the signal distribution. }
\end{center}
\end{figure}

In Fig.~\ref{fig:phase_diag_nolearn} we show the phase diagram for
Gauss-Bernoulli signal model, i.e. the distribution of components being
\be
   P(x) = (1-\rho) \delta(x) + \rho \frac{1}{\sqrt{2\pi}} e^{-\frac{x^2}{2}}
\ee
and various signal components distributions $P_0(x)= (1-\rho_0) \delta(x) +
\rho_0 \phi_0(x)$. Here we assume $\rho=\rho_0$. We see that the performance of BP mostly slightly decreases. For some signal distributions
(e.g. the binary case $\phi_0(x)=[\delta(x-1)+\delta(x+1)]/2$) there
is a  narrow region of parameters in which the $\ell_1$-reconstruction
becomes better than the probabilistic-BP approach.

\begin{figure}[!ht]
  \begin{center}
		\includegraphics[width=0.48\linewidth]{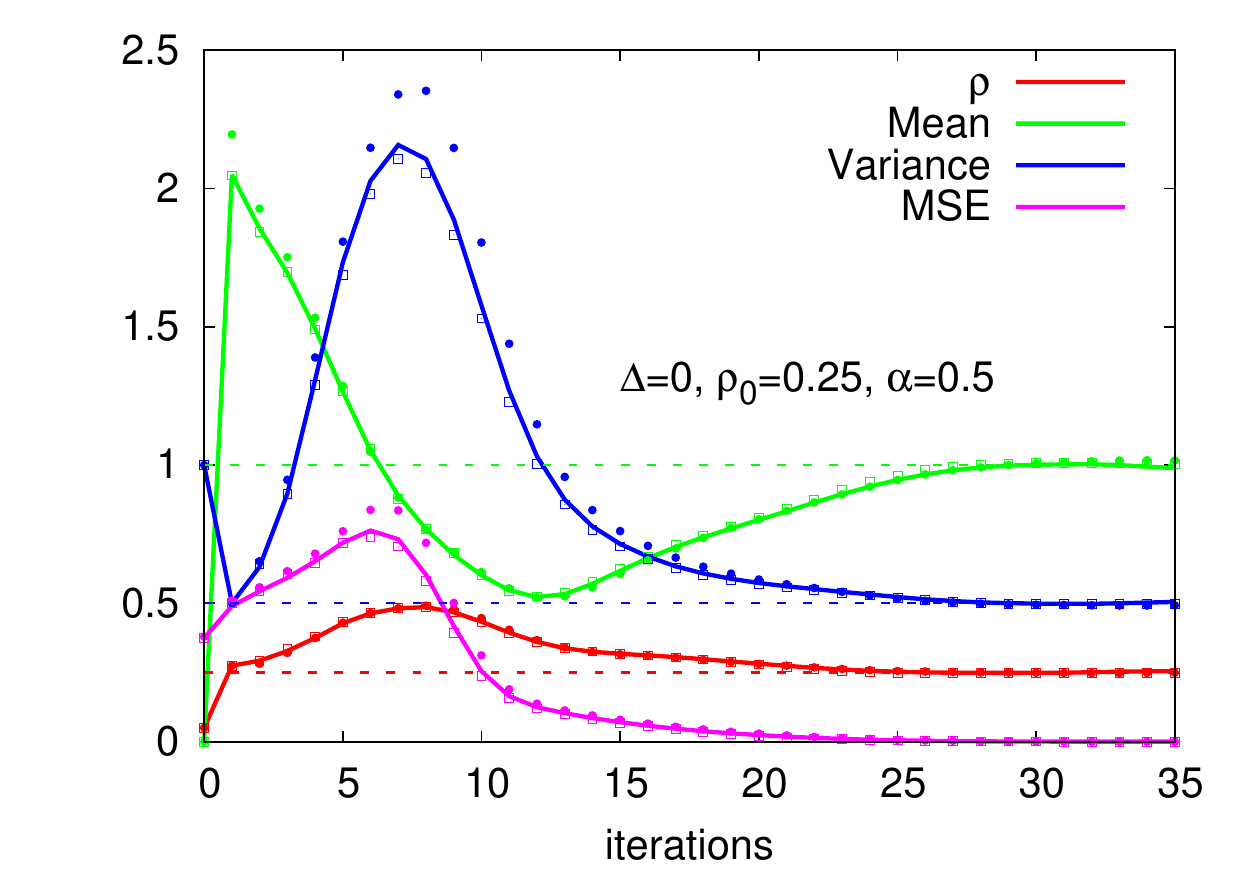}
                \includegraphics[width=0.48\linewidth]{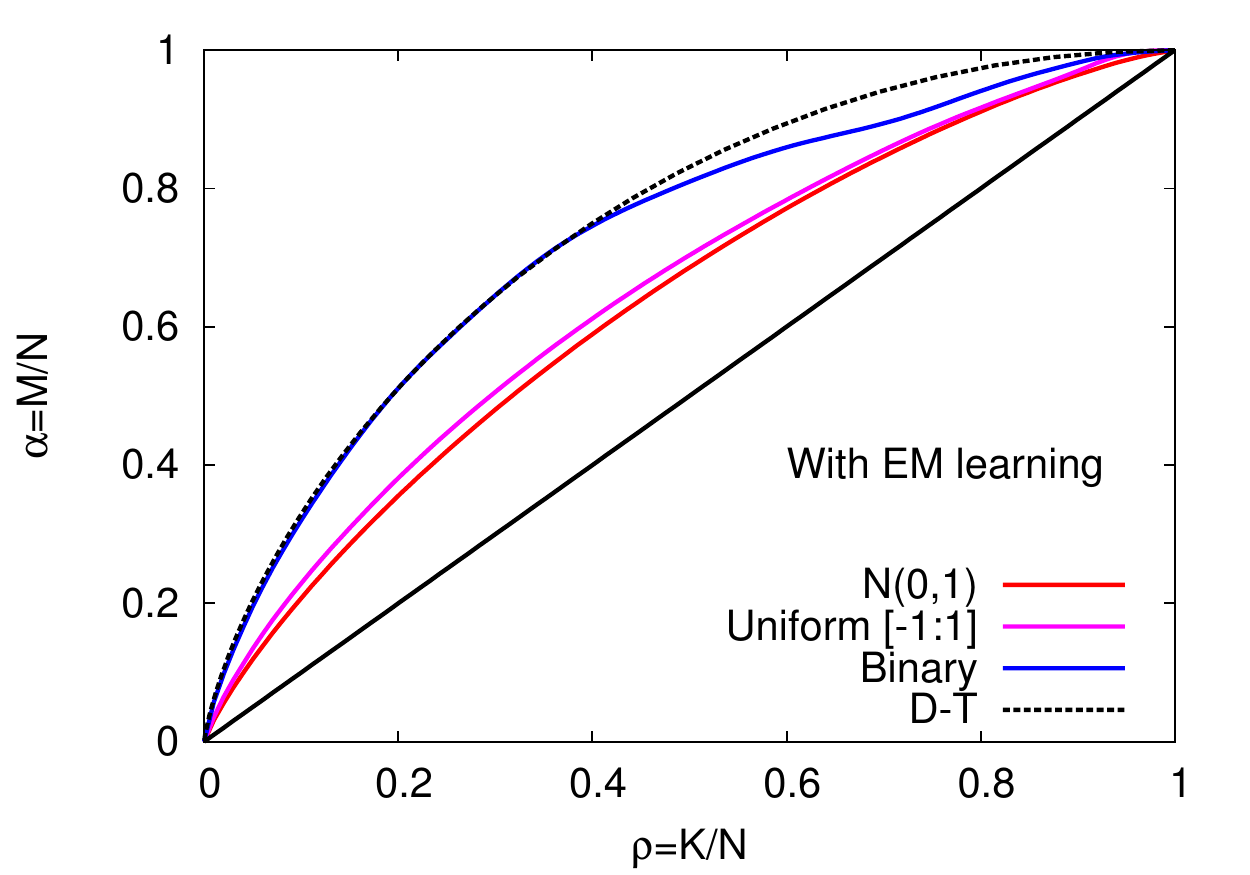}
       \caption{
     \label{fig:evolution_learn}
Left: Learning of parameters for noiseless measurements. The signal is
Gauss-Bernoulli with density $\rho_0=0.25$, mean $\overline s=1$ and variance
$\overline{s^2} -{\overline s}^2=0.5$. The measurement density is
$\alpha=0.5$.  The EM-BP algorithm is
initialized with $\rho=0.05$, $\overline x=0$, $\sigma^2=1$. In the figure we plot the evolution of the
parameters and of the mean-squared error $E$. The full
line is the analytic prediction using density evolution, the data points is the EM-BP
algorithm on an instance of $N=12 000$, the full points are for a
measurement matrix with Gaussian elements, the empty points for a
matrix with elements $\pm 1/N$.
Right: Phase diagram for the EM-BP reconstruction, that is when the
signal model is not matching the empirical distribution of signal
elements, i.e. $\phi(x)\neq\phi_0(x)$. Different
     curves correspond to different distributions $\phi_0$ of the
     signal. The dashed line gives the Donoho-Tanner transition line
     for $\ell_1$ reconstruction,
     which is independent of the signal distribution.
}
\end{center}
\end{figure}

In case the signal distribution and its sparsity are not known, the performance of BP can
be improved by including the expectation maximization learning. We
call this generalization  EM-BP. In this paper we study the
performance of EM-BP in the case where the signal model is Gauss-Bernoulli
\be
   P(x) = (1-\rho) \delta(x) + \rho \frac{1}{\sigma\sqrt{2\pi}}
   e^{-\frac{(x-\overline x)^2}{2\sigma^2}} .
\ee
Expectation maximization is used to learn the three parameters $\rho,
\overline x$ and $\sigma$.
In EM-BP we do one
update of BP messages followed by one update of the parameters. New
values of parameters are computed using Eqs.~(\ref{learn_rho}, \ref{learn_xbar_alternative},
\ref{learn_mv_alternative}). BP message are then updated again using
parameter values $\rho= [\rho_{\rm old} + \min(\rho_{\rm new},\alpha)]/2$, $\overline x =
(\overline x_{\rm old} + \overline x_{\rm new})/2$, $\sigma^2 =
[\sigma^2_{\rm old} + \max(\sigma^2_{\rm new},0)]/2$. And this is
repeated till convergence. The evolution of parameters under learning
is illustrated in the left part of Fig.~\ref{fig:evolution_learn}.

We observe that for the Gaussian-distributed signal elements (left
part of Fig.~\ref{fig:phase_diag_nolearn}) the correct mean and variance are always
learned (even in the region where exact reconstruction is not
possible). In this case the spinodal line is always the same as in the
case when the signal distribution was known, see Fig.~\ref{fig:phase_diag_Bayes}.
For signals with non-Gaussian distribution of elements, right
part of Fig.~\ref{fig:phase_diag_nolearn}, the spinodal line changes slightly,
the lines with learning are shown in the right part of Fig.~\ref{fig:evolution_learn}.
We conclude that EM-BP improves on pure BP and on
$\ell_1$-reconstruction in many cases and  hence it can be useful in
 practical situations. Of course if one has some knowledge of the signal
distribution it is helpful to further include it in the signal
model.

\subsection{Phase diagram for noisy measurements}
\label{Sec:Noisy}

In this section we discuss compressed sensing with noisy measurements,
$\Delta>0$. We first describe the performance of the BP algorithm and
the corresponding phase diagrams in the Bayes optimal case when the
signal model corresponds to the signal distribution. In a second part
we then discuss the general noisy case with non-matching signal model
and learning.

\begin{figure}[!ht]
(a)
\includegraphics[width=0.45\linewidth]{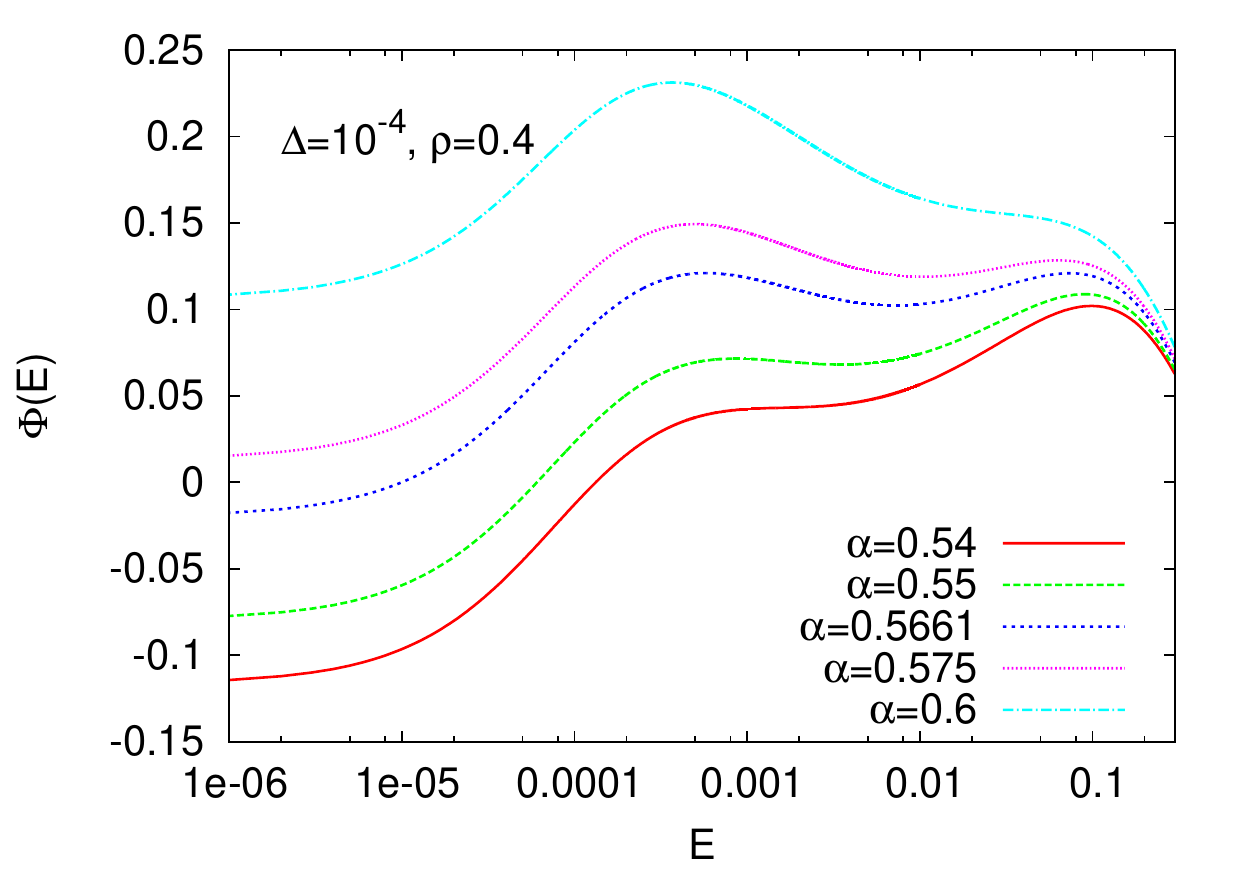}
(b)
\includegraphics[width=0.45\linewidth]{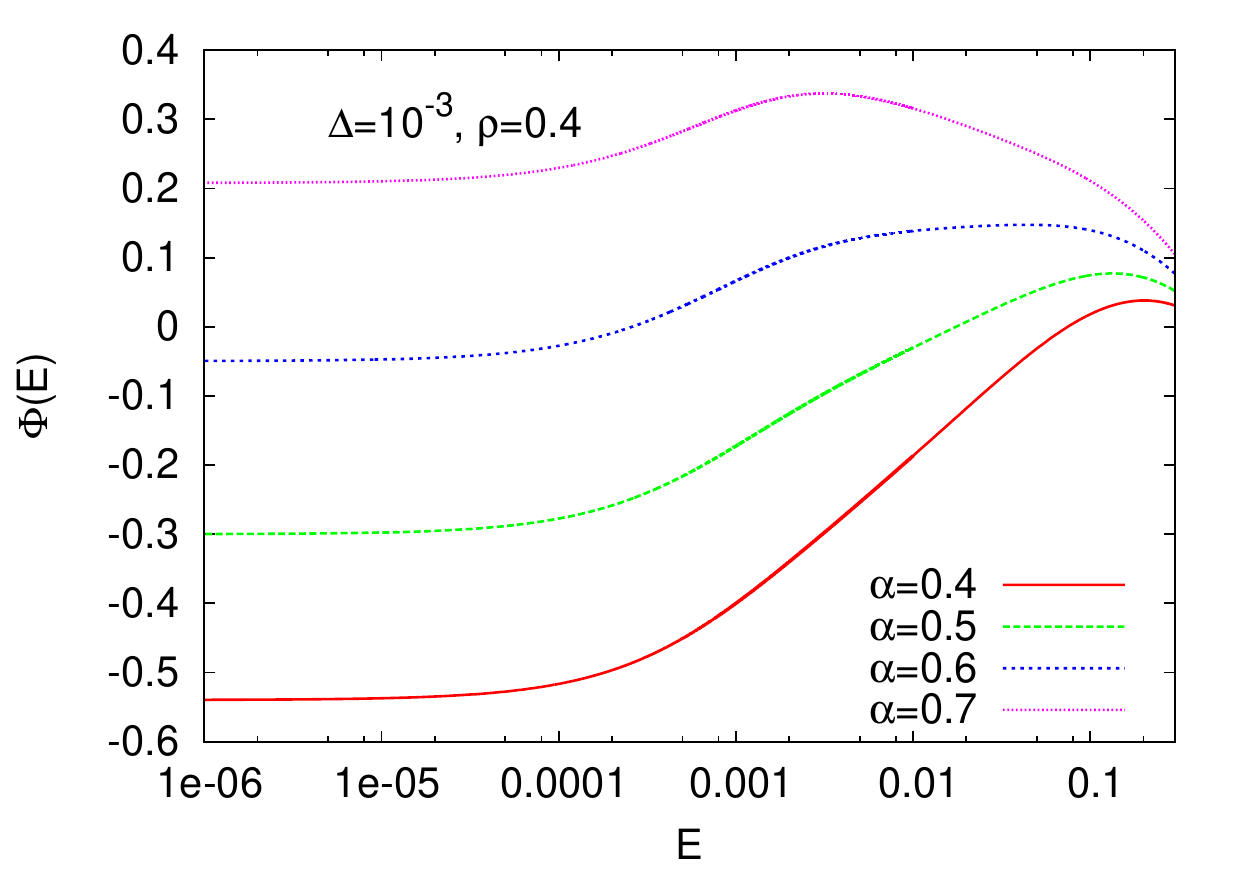}
\caption{\label{fig_noise1} (color online) The free entropy $\Phi(E)$ in
  presence of noise as a function of the
  MSE. (a) $\rho=0.4$ and $\Delta=10^{-4}$, there is a first
  order phase transition and two local maxima do co-exist for region
  of subsampling rates $\alpha_d>
\alpha>\alpha_s$. (b) for larger noise, $\Delta=10^{-3}$, there
  is always only one maxima, in this case the EM-BP approach is always
  optimal, although the mean-squared error may be quite large.}
\end{figure}

In Fig.~\ref{fig_noise1} we plot the free entropy $\Phi(E)$, obtained
from Eq.~(\ref{Phi_E}), as
a function of the mean-squared error $E$, for signal with nonzero
elements being iid Gaussian variables with zero mean and unit
variance, and a matching signal model. The main difference with the
noiseless case, Fig.~\ref{fig:FreeEntropy}, is that the global maximum of the free
entropy, that described the optimal achievable mean-squared error, is
at non-zero values of the MSE. This indeed reflects the fact that with
noisy measurements exact reconstruction is no longer possible.

Let us investigate whether BP algorithm finds a configuration
with the best achievable MSE or not. Again, BP is basically performing
steepest ascent in the free entropy starting from a large value of
MSE. Depending on the value of the signal density
$\rho$ and the measurement noise variance $\Delta$, we see two kinds
of behavior as a function the subsampling rate $\alpha$. For some
values of $\rho$, $\Delta$, see Fig.~\ref{fig_noise1} (b), the global maximum of $\Phi(E)$ is the only
maximum for all $\alpha$, and in that case BP will converge to it. For other
values of $\rho$, $\Delta$, see Fig.~\ref{fig_noise1} (a), the
situation is similar to the noiseless case:
\begin{itemize}
\item For $\alpha>\alpha_d$ the free entropy has a single maximum at a
  small value of MSE comparable to $\Delta$.
\item For $\alpha_d> \alpha>\alpha_c$ the free entropy has two maxima,
  the one at lower MSE being the global one.
\item For $\alpha_c> \alpha>\alpha_s$ the free entropy has two maxima,
  the one at higher MSE being the global one.
\item For $\alpha<\alpha_s$ the free entropy has a single maximum at a
  value of MSE much larger than $\Delta$.
\end{itemize}

\begin{figure}[!ht]
(a)
\includegraphics[width=0.45\linewidth]{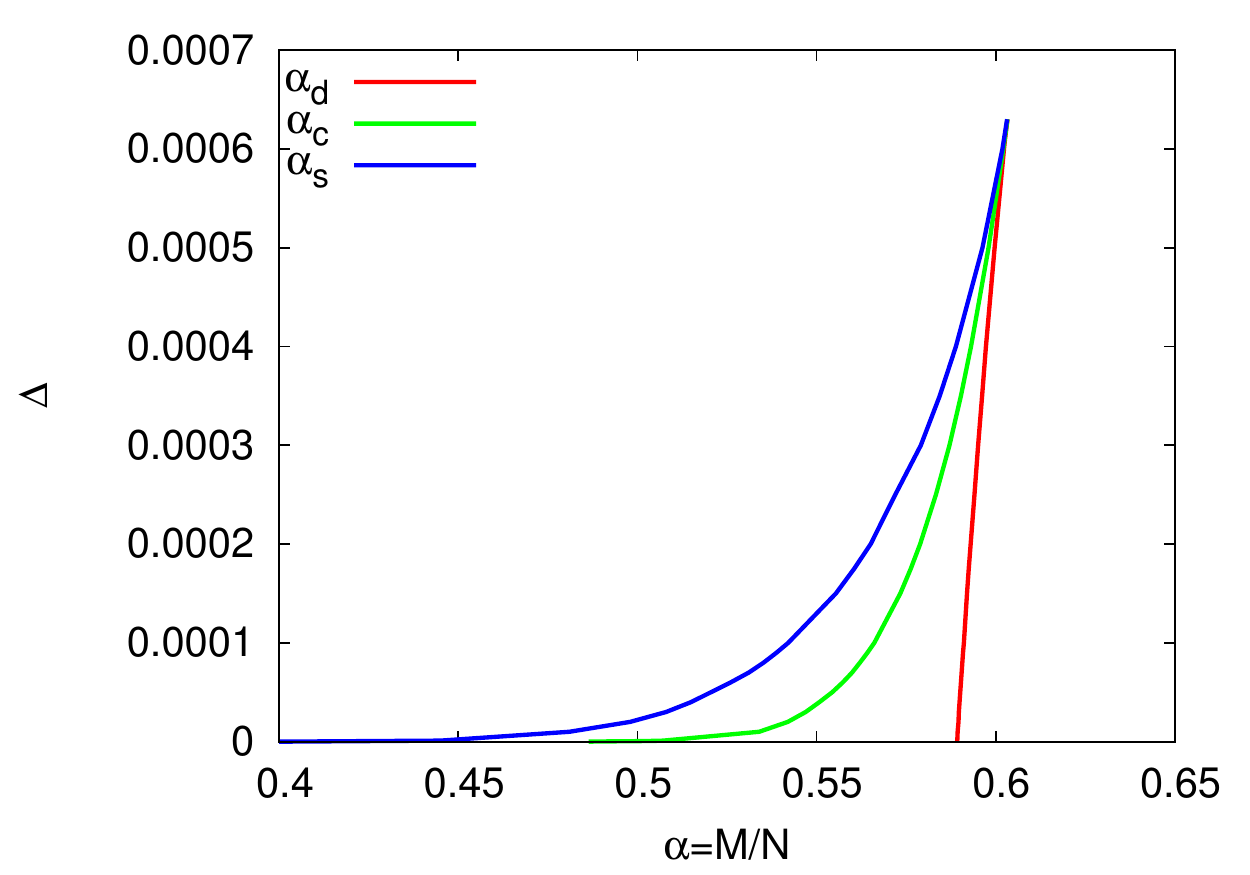}
(b)
\includegraphics[width=0.45\linewidth]{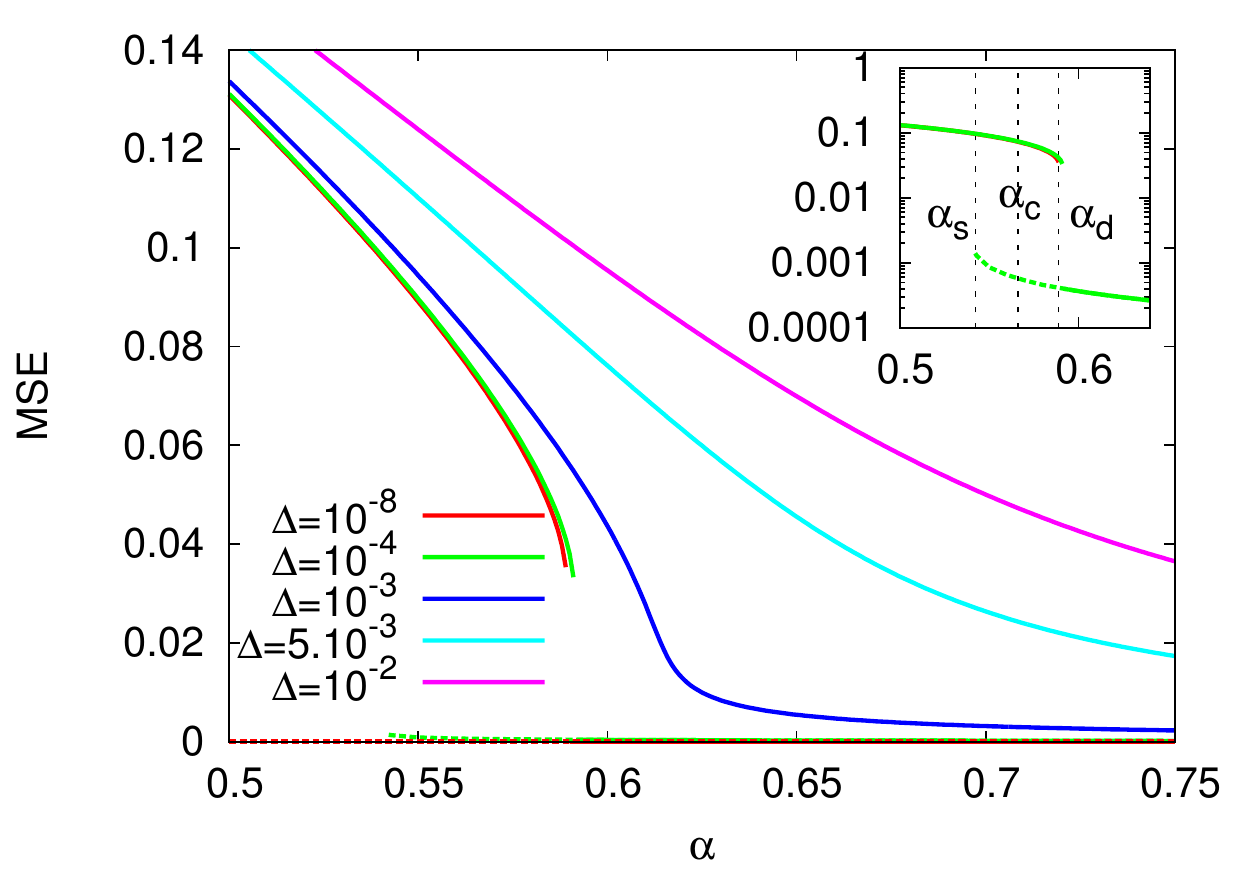}
\caption{\label{fig_noise2} (color online) (a) The three phase
  transition lines in CS with noisy measurements for
  Gauss-Bernoulli signal and matching signal
  model with density $\rho=0.4$. The blue line is the spinodal
  line $\alpha_s$, red line is the dynamical line $\alpha_d$, and green is
  the critical line $\alpha_c$. For larger noise there is no such
  sharp threshold.  A perfect sampling
  algorithm changes its behavior abruptly at $\alpha_c$ (the green line), where the
  quality of reconstructed signal would jump discontinuously from high
  MSE to low MSE. The BP algorithm (with the uninformed initialization) always converges to the local
  maxima of the free entropy corresponding to the largest MSE, hence
  its MSE jumps from a relative low value to a high value at
  $\alpha_d$ (the red line). BP is hence suboptimal for $\alpha_d>
\alpha>\alpha_c$.
(b) The MSE achieved by BP for several noise strengths. In the inset is
the case of $\Delta=10^{-4}$ with the three phase transitions
depicted. For $\alpha_d>\alpha>\alpha_c$ the best achievable MSE
corresponds to the lower part of the curve, whereas BP reconstruction
achieves the MSE corresponding the the upper part of the curve.
Note that in this case the MSE achieved by
$\ell_1$ reconstruction would be much larger (nonzero for $\alpha \gtrsim 0.75$
even for the noiseless case, see Fig.~\ref{fig:MSQ_BEPL}).}
\end{figure}

The above result means that for a region of subsampling rates $\alpha_d>
\alpha>\alpha_c$ the BP algorithm is sub-optimal, as it converges to
much higher MSE than the MSE corresponding to the optimal Bayes
inference (global maximum of the free entropy). In
the left part of Fig.~\ref{fig_noise2} we plot the dependence of $\alpha_d$,
$\alpha_c$, and $\alpha_s$ on the noise variance. In the right part we
plot the MSE achieved by BP as a function of the subsampling rate. In
cases where BP is suboptimal (for the two lowest
noise variances) we compare to the optimal MSE. The data presented in
Figs.~\ref{fig_noise1} and \ref{fig_noise2} are obtained from the
density evolution, i.e. $N\to \infty$ limit of BP behavior. The
behavior of BP for finite $N$ agrees well with these results for
systems sizes of several thousands of elements and more.

\begin{figure}[!ht]
\includegraphics[width=0.31\linewidth]{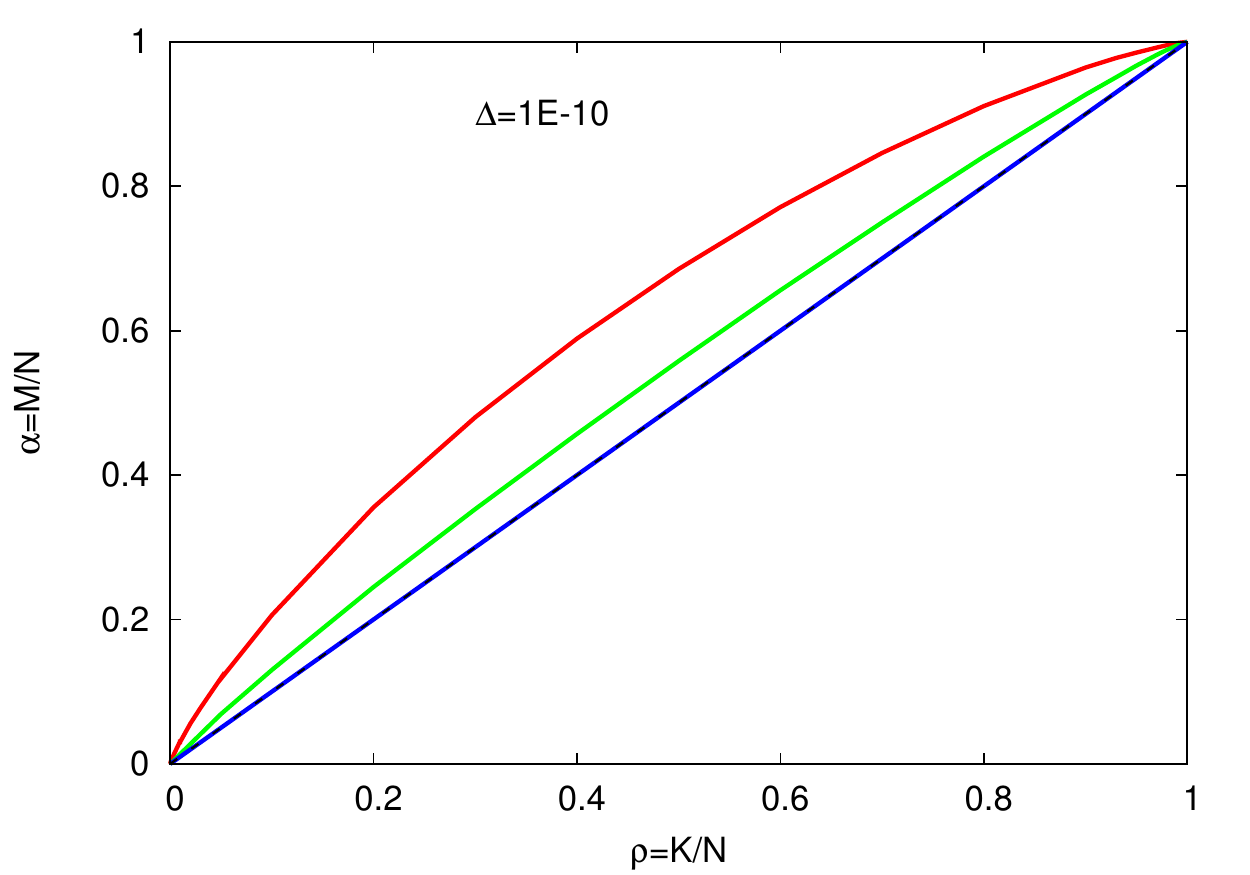}
\includegraphics[width=0.31\linewidth]{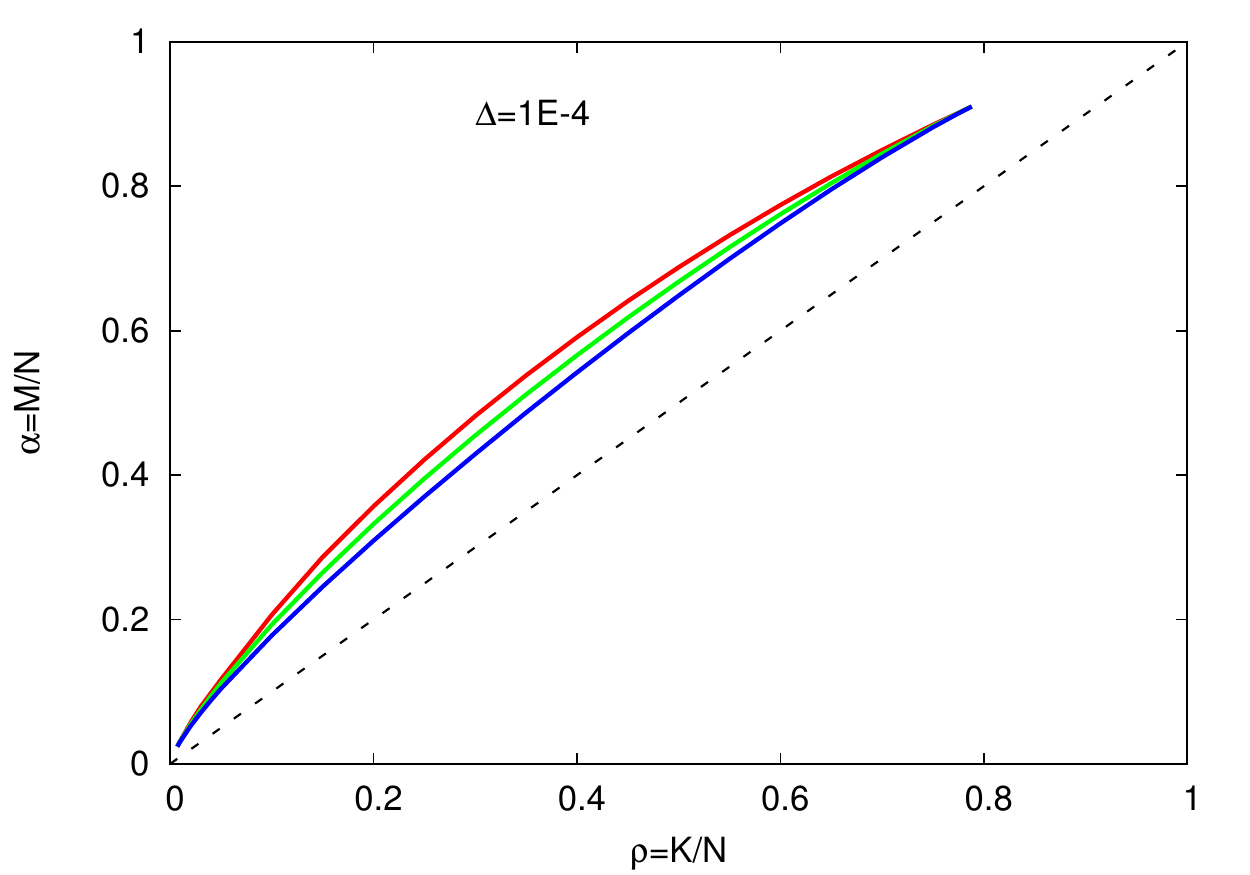}
\includegraphics[width=0.31\linewidth]{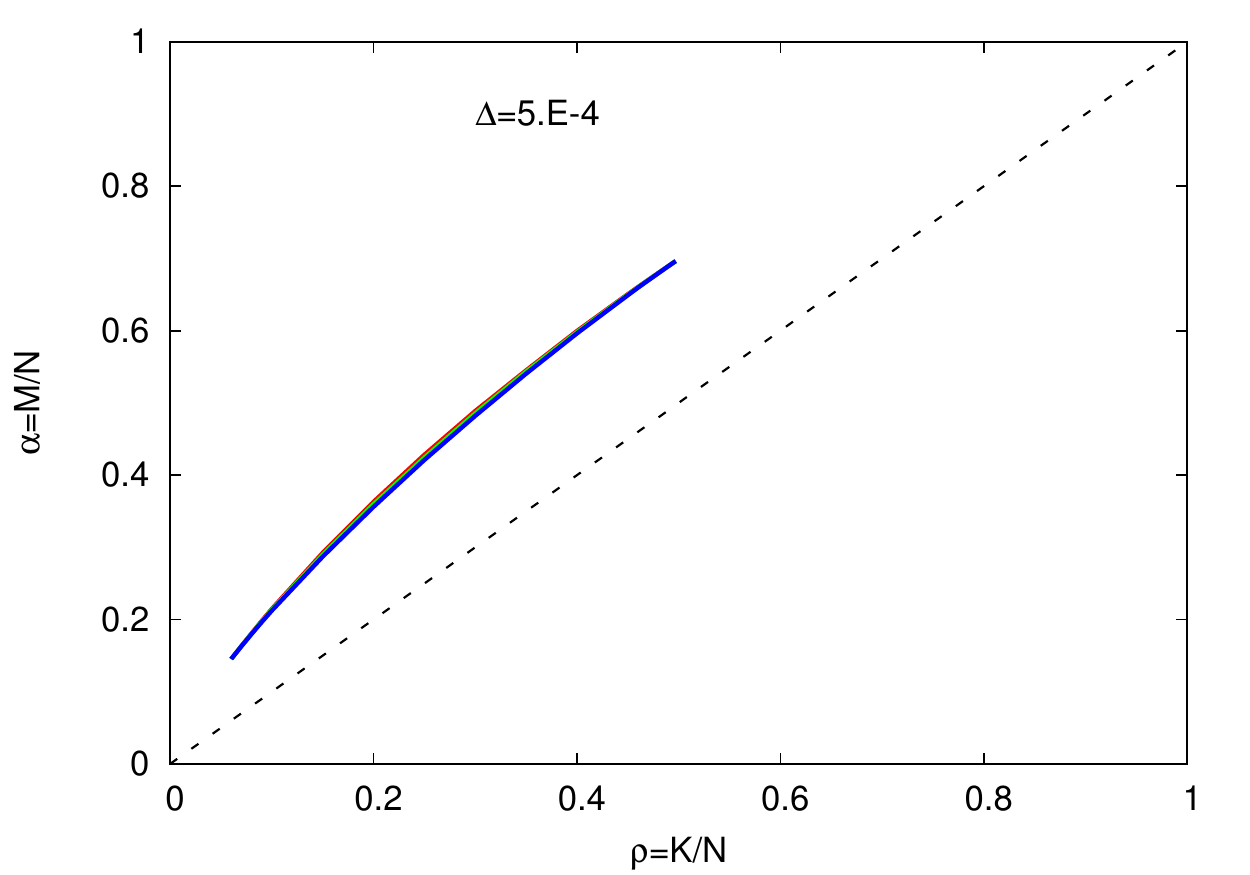}
\caption{\label{fig_noise3} The three transition lines $\alpha_d$,
  $\alpha_c$, $\alpha_s$ shown in Fig.~\ref{fig_noise2} for different
  values of the noise variance, growing from left to right: left:
  $\Delta=10^{-10}$, middle $\Delta=10^{-4}$, and right $\Delta=5\cdot
  10^{-4}$. }
\end{figure}

In Fig.~\ref{fig_noise3} we plot again the three phase transition
lines for reconstruction with measurement noise. This time we plot
the lines in the $\rho$-$\alpha$ phase diagram for several values of
the variance $\Delta$. As the noise increases the region of
  densities for which there is a sharp phase transition shrinks.  For
  large enough values of $\Delta  \gtrsim  0.00078$ there is no sharp phase transition
  for the inference of Gauss-Bernoulli signal (with matching
  Gauss-Bernoulli signal model).

\begin{figure}[ht]
a)
\includegraphics[width=0.47\linewidth]{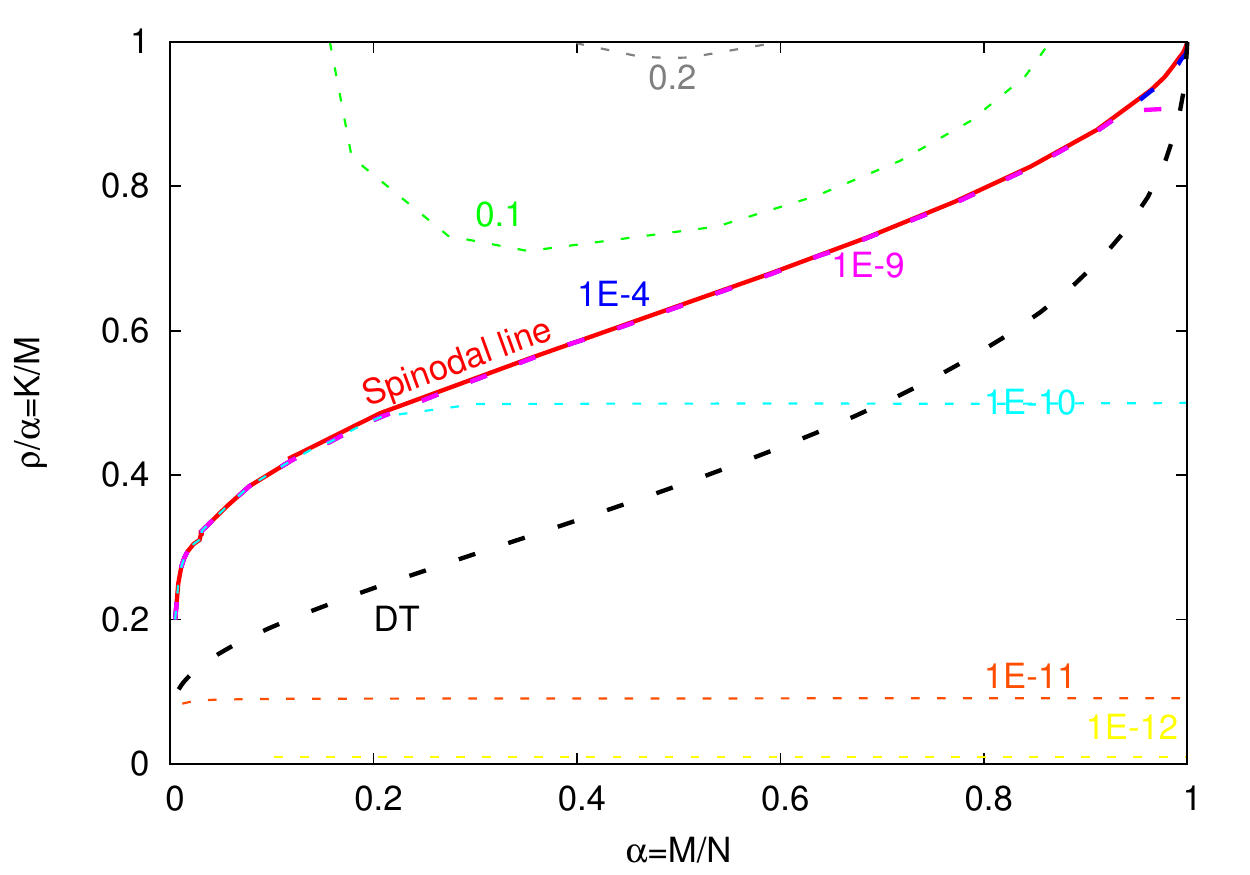}
b)
\includegraphics[width=0.47\linewidth]{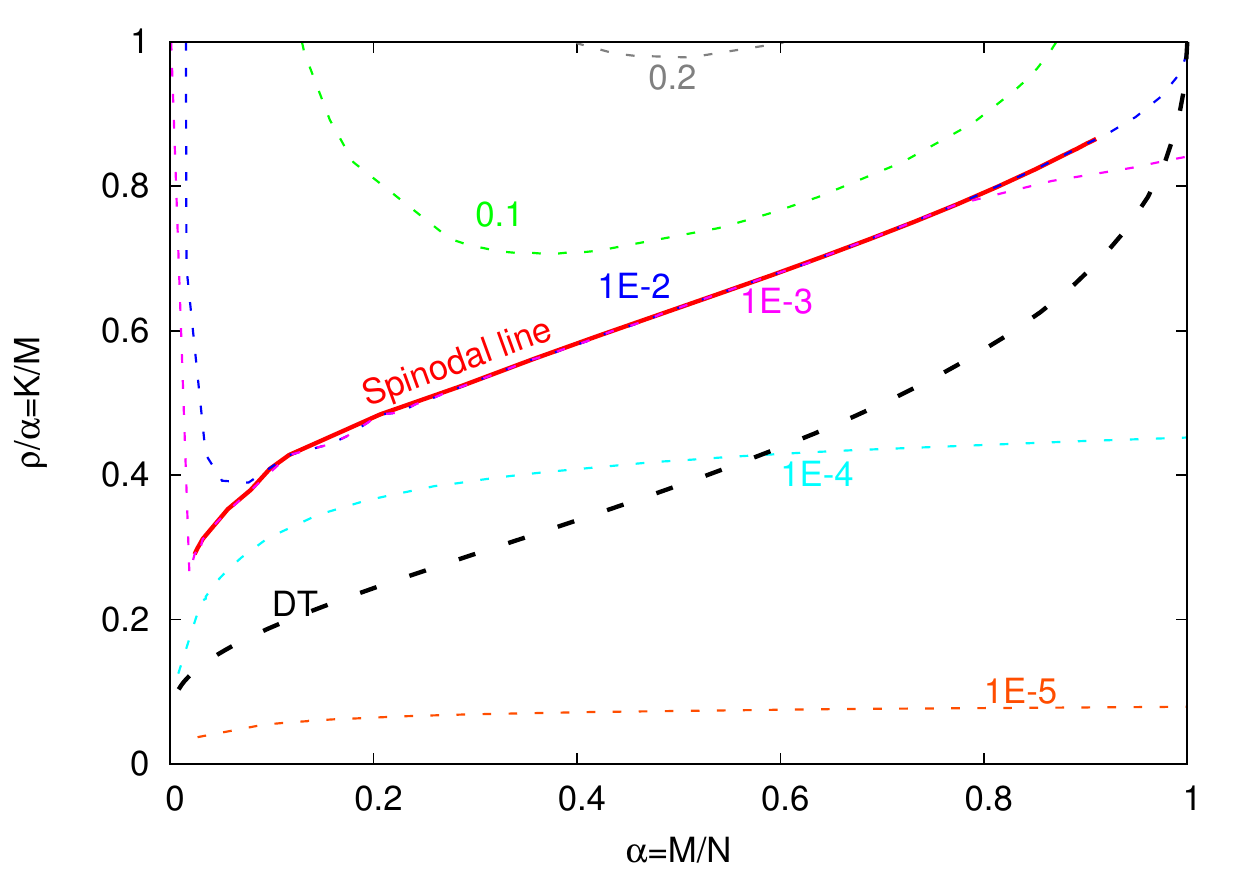}
\caption{ (color online) \label{fig_noise4} Phase diagram and level
  lines of the MSE for the BP algorithm in presence of noise in the
  Donoho-Tanner convention. Left: using a noise with variance
  $\Delta=10^{-10}$.  Right: using a noise with variance
  $\Delta=10^{-4}$. The Donoho-Tanner transition line for $\ell_1$ is
  shown for comparaison.}
\end{figure}

Another illustration of this phase diagram with noise is in
Fig.~\ref{fig_noise4} where we plot level lines following the MSE
achieved by BP reconstruction. On the line $\alpha_d$ the MSE of BP
reconstructions increases discontinuously from values comparable to
$\Delta$ to large values.

\begin{figure}[ht]
  \begin{center}
    \includegraphics[width=0.48\linewidth]{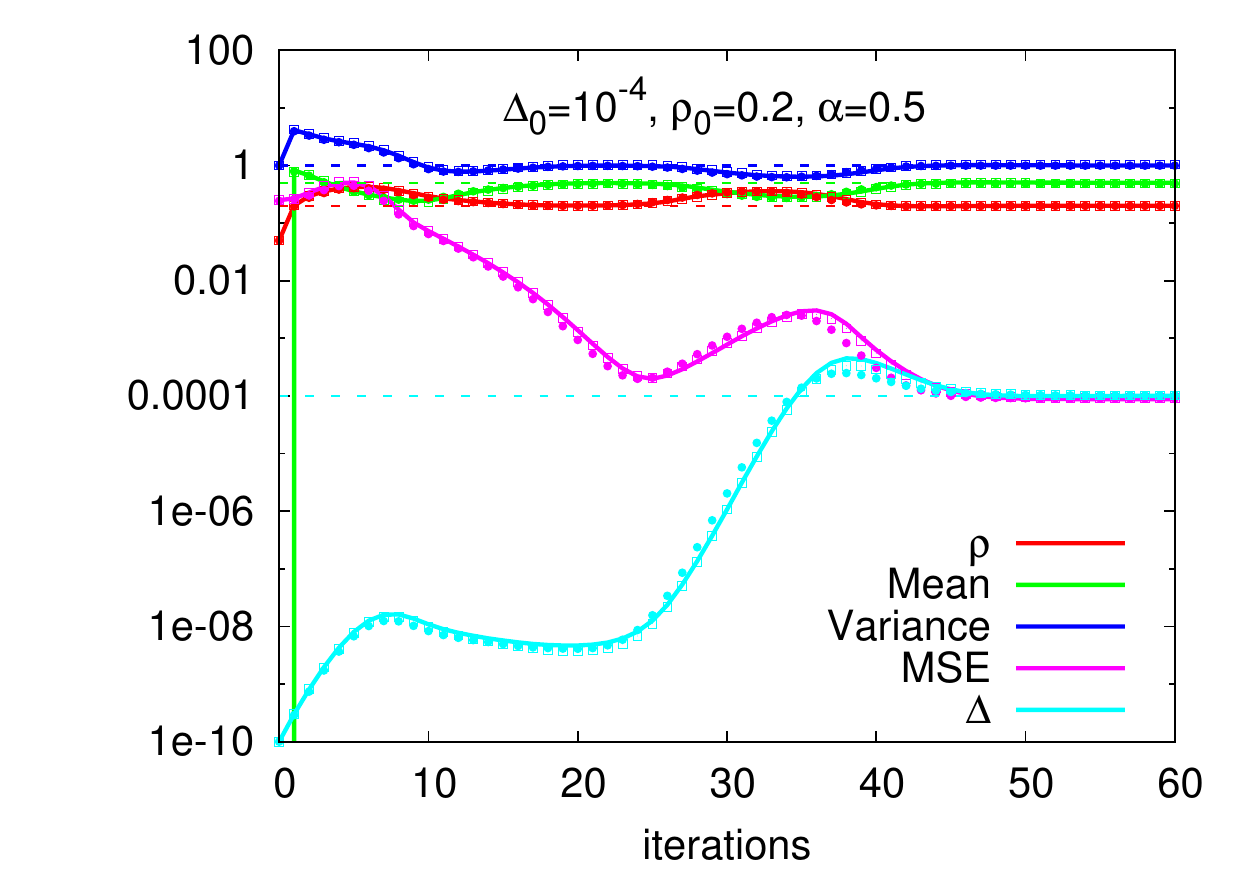}
       \caption{
         \label{fig:evolution_learn_noise} Learning of parameters for
         noisy measurements. The signal is Gauss-Bernoulli of density
         $\rho_0=0.2$, mean $\overline s=0.5$ and variance
         $\overline{s^2} -{\overline s}^2=1$. The measurement rate is
         $\alpha=0.5$ and the noise variance $\Delta_0=10^{-4}$. The
         EM-BP algorithm is initialized with $\rho=0.05$, $\overline
         x=0$, $\sigma^2=1$, $\Delta=10^{-10}$.  In the figure we plot
         the evolution of the parameters and of the mean-squared error
         $E$ for three cases. The full line is the density evolution,
         the data points is the EM-BP algorithm on an instance of
         $N=10 000$, the full points are for a measurement matrix with
         Gaussian elements, the empty points for a matrix with elements
         $\pm 1/N$.}
\end{center}
\end{figure}

Of course in practical applications the noise level $\Delta_0$ is
often not known.  In such cases learning of the noise level can be
included in the EM-BP algorithm, using noise variance update
Eq.~(\ref{delta_learn}). In Fig. \ref{fig:evolution_learn_noise} we illustrate
the evolution of parameters and the mean-squared error $E$ under such
expectation maximization learning for a Gaussian signal of density $\rho_0=0.2$, with
measurement rate $\alpha=0.5$ and noise variance $\Delta_0=10^{-4}$.

\newpage
\section{Seeding matrices: a way to achieve
  optimality}\label{Sec:Seeding}

In the previous section we exposed the reason why BP reconstruction
for homogeneous measurement matrices $\bF$ does not achieve
subsampling rates down to the information theoretical limit
$\alpha=\rho_0$. In \cite{KrzakalaPRX2012} we developed a new type of
measurement matrices ---that we coined {\it seeding matrices}--- for
CS for which the limit $\alpha=\rho_0$ is achievable
using the BP reconstruction. 
This was built on several result in the error correcting code
community\cite{FelstromZigangirov99,LentmaierFettweis10,KudekarRichardson10,KudekarRichardson12}. Here
we shall explain further our motivations for the construction of the
seeding matrices.

We shall give heuristic arguments why with these matrices it is
possible to achieve theoretically optimal reconstruction ande show,
using the replica method (or equivalently, density evolution) that
this is indeed the case. We want to point out that, while we use
mostly the replica method/density evolution formalism, some rigorous
results can be obtained. In particular, in the special Bayes optimal
case ---when the signal model corresponds to the empirical
distribution of the nonzero signal elements--- it has been now proven
rigorously in \cite{DonohoJavanmard11} that the for CS
with seeding matrices the BP reconstruction is indeed able to achieve
the information theoretical limit $\alpha =\rho_0$. Here, we shall
show, using the statistical physics tools, that seeding matrices allow
close to optimal reconstruction also when the signal distribution is
not known, which is even more appreciable.

\subsection{Why and when does seeding work?}
\label{whywhen}

As exposed in the previous section, for homogeneous measurement
matrices with iid entries, BP is able
to reconstruct the signal correctly at $\alpha>\alpha_{\rm BP}$,
bellow $\alpha_{\rm BP}$ a {\it metastable state} (i.e. a local
maximum of $\Phi(D)$ at $D>0$) appears in the measure
$P(\bx|\bF,\by)$. The iterations of the BP algorithm get ``trapped''
in this state and BP is
therefore unable to find the global maximum corresponding to the
original signal (see Fig.~\ref{fig:MSQ_BEPL}).  This is a situation
well known in physics, that is typical for a system undergoing a first order
phase transition. A familiar example of first order phase transition
being crystallization, i.e. the way a liquid changes into
a solid. In physics, systems undergoing a first order phase transition
can be divided into two groups: (a) Mean field systems, where the size
of the boundary of a sphere of a (large) finite radius drawn around
one particle (variable) is of the same order as the volume of this
sphere. (b) Finite dimensional systems where the size of the boundary
is much smaller than its volume. Typically in $d$ dimensions, a sphere
of radius $r$ has surface $s_d r^{d-1}$ and volume
$v_d r^{d}$ ($s_d$ and $v_d$ being the surface and volume of a sphere of radius one).

In mean field systems metastable states have exponentially large (in
the size of the system) living time, meaning that is would take an
exponential time to randomly find a fluctuation that would be able to
overcome the barrier between the local maximum and the global
one. Whereas in finite dimensional systems the living time of
metastable states is always constant. A simplified argument leading to
this conclusion uses the fact that maximization of the entropy is the
driving force of system dynamics. Consider the system being in the
metastable state (e.g. supercooled liquid), if a random fluctuation
appears flipping a droplet of radius $R$ into the equilibrium state
(crystal) then this causes free entropy increase of $\Delta \Phi v_d
R^d$ and decrease because of the surface terms $\Gamma s_d R^{d-1}$
for $R$ large enough $R> R^* = \Gamma s_d (d-1)/(\Delta\Phi v_d d)$
the gain is more important than the loss and such a randomly created
droplet will start to grow. The crucial point is that the critical
radius $R^*$ does not depend on the system size $N$ and hence such a
fluctuation arises with a constant probability in the finite
dimensional systems. The processus we described here is on the basis
of nucleation theory in physics that described the growth of crystal
droplets close to a first order phase transition
\cite{ReviewBinder,JCHEM}.

\begin{figure}[!ht]
  \begin{center}
    \begin{tabular}{cc}
    (i)
    \includegraphics[scale=0.35]{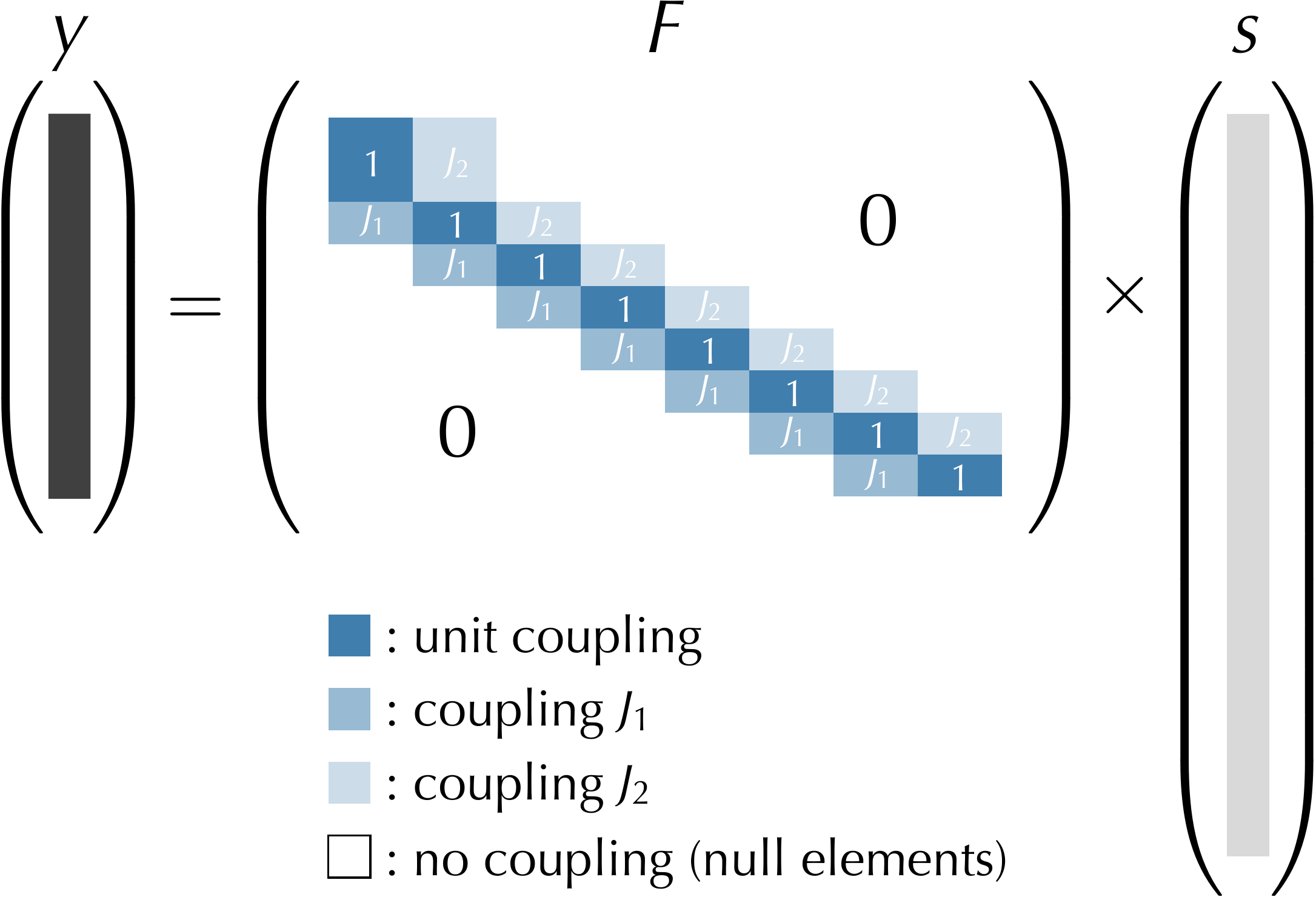}&
    \hspace{0.2cm}
    (ii)
    \includegraphics[scale=0.35]{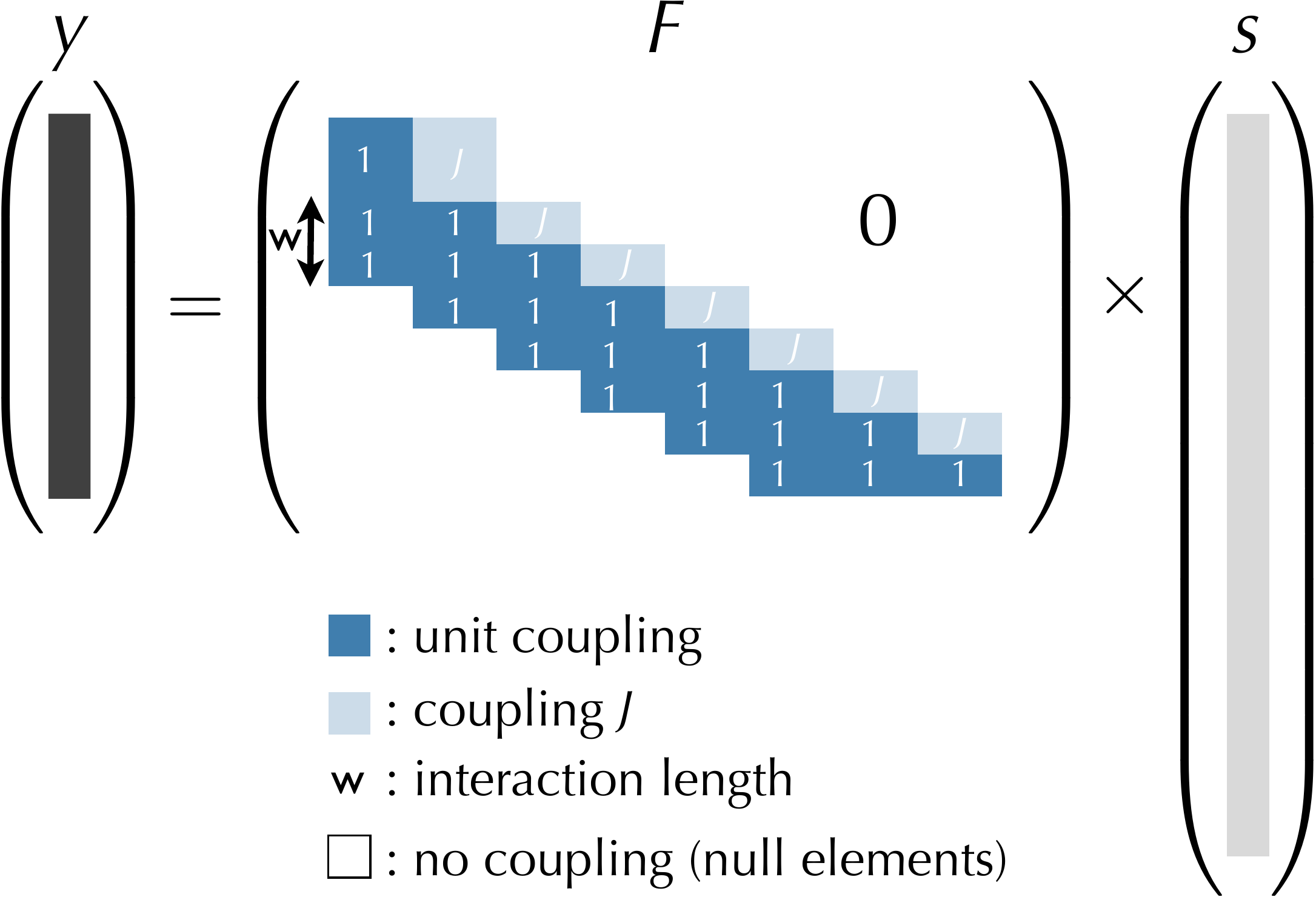}\\
\\
        \hspace{0.2cm}
   (iii)
    \includegraphics[scale=0.35]{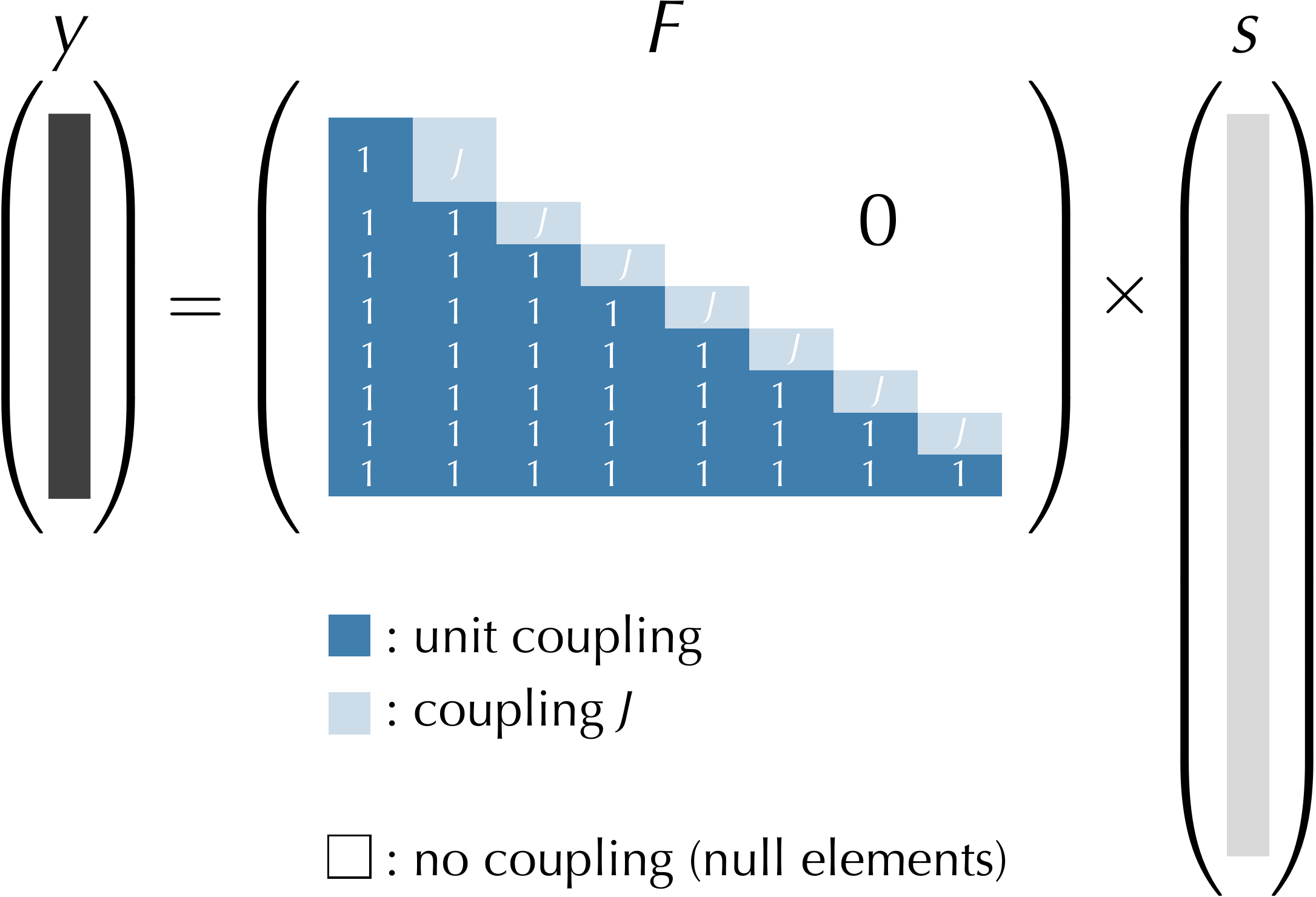}&
        \hspace{0.2cm}
    (iv)
    \includegraphics[scale=0.35]{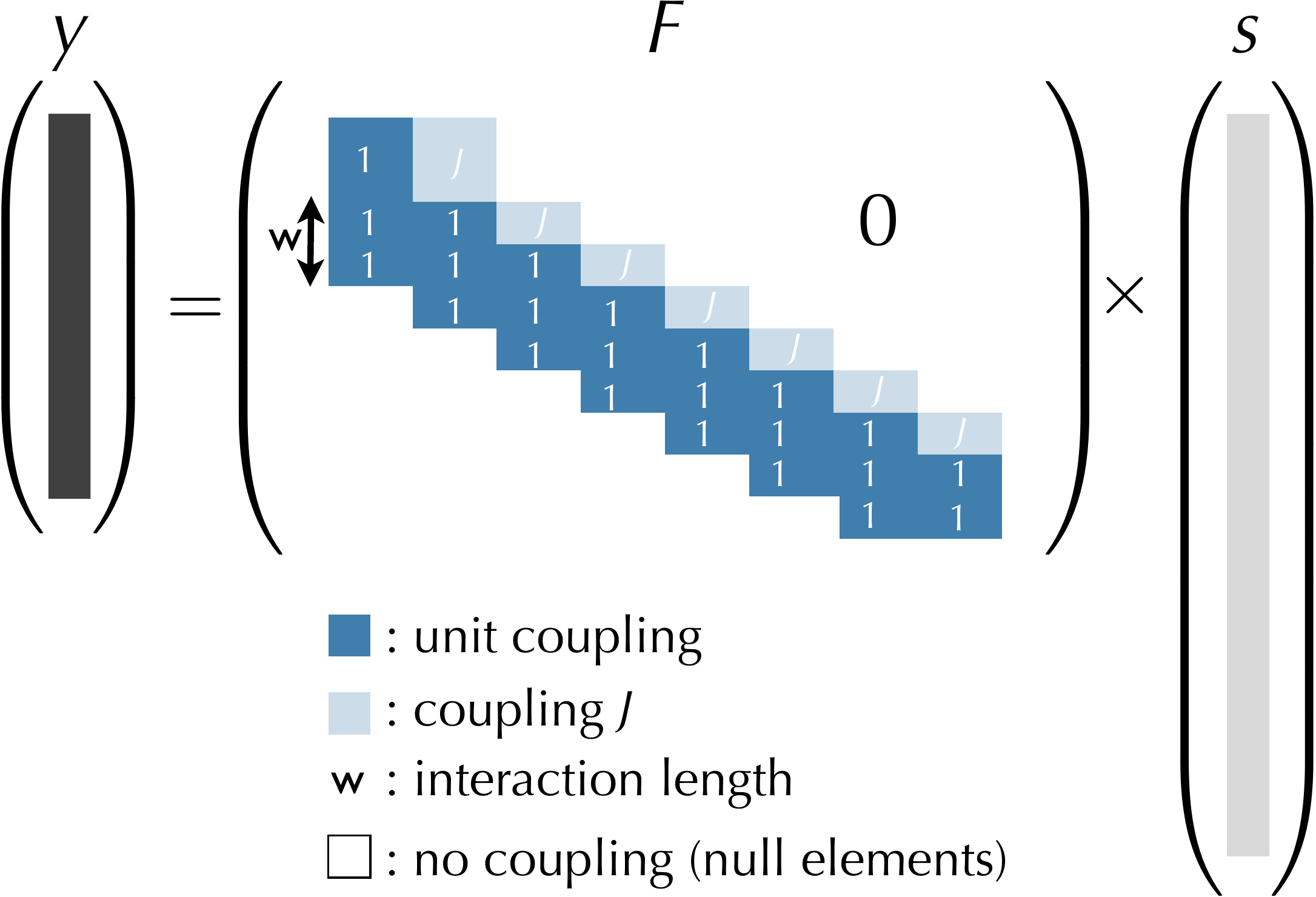}
        \hspace{0.2cm}
    \end{tabular}
    \caption{Examples of seeding measurement matrices $\bF$ for
      CS. Here $L_c=8$. (i) A band-diagonal matrix,
      already introduced in \cite{KrzakalaPRX2012}, where $L=L_c=L_r$,
      and $J_{p,q}=0$, except for $J_{p,p}=1$, $J_{p,p-1}=J_1$, and
      $J_{p-1,p}=J_2$. Good performance is typically obtained with
      large $J_1$ and small $J_2$.  (ii) Another band-diagonal matrix
      where $L=L_c=L_r$, and $J_{p,q}=0$, except for $J_{p,p}=1$,
      $J_{p-1,p}=J$, and $J_{p,p-w}=1$ with $w=1,\dots,W$. Good
      performance is typically obtained with small $J$ and $W \ge
      2$. Since all variances are lower or equal to one, this matrix
      can be realized only having elements $(0,\pm 1)$. (iii) A lower
      triangular matrix, that can be viewed as the matrices of type
      (ii) when $W = L-1$. Again, good performance is obtained with
      relatively small $J$. (iv) In some cases, we observed that the
      last block of variables was not recovered correctly. Adding a
      new line in the matrix ($L_r=L_c+1$), as in this example, cures
      the problem. All these matrices are motived by the same
      consideration: more measurement are made in the first block of
      the signal such that the information will first appear in this
      block, and then propagate into the whole vector.}
    \label{fig:1d_matrix}
  \end{center}
\end{figure}

The whole idea of seeding matrices is to mimic the process of
nucleation and crystal growth in the reconstruction of compressed
sensing signal. This idea, together with the previous work on
spatially coupled LDPC codes \cite{KudekarRichardson10}, also
motivated the design of the seeding matrix in \cite{KrzakalaPRX2012}.
There are three key ingredients that need to be present in the system
in order for the seeding to work.
\begin{itemize}
\item[(a)] The free entropy driving force. To escape
from a metastable state we need the existence of a higher maximum of
the free entropy $\Phi(D)$. This ingredient is present in the BP
reconstruction of the original signal as long as $\alpha>\rho_0$ (or
$\alpha>\alpha_c$ for the nosy case). Let
us note here that seeding does not improve performance of the $\ell_1$
reconstruction algorithms (see appendix \ref{appendix:L1_seeded}), because this ``driving force'' is missing
since the Donoho-Tanner transition is continuous (it is a second order
transition in the
physics classification).
\item[(b)] The existence of a nucleus (seed). We need a
part of the system to be already in the equilibrium state. This
ingredient can be achieved by making the measurement matrix
inhomogeneous and measuring at a much higher subsampling rate a small
subpart of the signal -- that we call a ``seed''.
\item[(c)] An interaction between the seed and the rest of the signal
  that enables the growth of the seed. In \cite{KrzakalaPRX2012} and
  \cite{DonohoJavanmard11} this was achieved via the so-called spatial
  coupling. The signal was divided into  blocks and the measurements designed in
  such a way that only several neighborhooding blocks are measured at
  a time.  Similar ideas have
been used recently in the design of sparse coding matrices for error
correcting
codes~\cite{FelstromZigangirov99,KudekarRichardson10,LentmaierFettweis10,HassaniMacris10}.
  Here we also give an example of a seeded measurement matrix
  that does not have spatially coupled structure.
\end{itemize}

In this article we present several ways how to achieve points (b) and
(c), and hence be able to do reconstruction in CS at
yet lower subsampling rates. We, however, stress that there is
relatively a lot of freedom in the construction of these matrices and
their optimization and adaptation to physically constraint
measurements is surely a promising area of future research. 

The matrix we used are presented n Fig.~\ref{fig:1d_matrix}.  These
are block-matrices defined as follows: The $N$ variables are divided
into $L_c$ groups of $N_p$, $p=1,\dots,L_c$, variables in each
group. We denote $n_p=N_p/N$. And the $M$ measurements are divided
into $L_r$ groups of $M_q$, $q=1,\dots,L_r$, measurements in each
group, we define $\alpha_{qp}=M_q/N_p$. Then the matrix $F$ is
composed of $L_r\times L_c$ blocks and the matrix elements $F_{\mu i}$
are generated independently, in such a way that if $\mu$ is in group
$q$ and $i$ in group $p$ then $F_{\mu i}$ is a random number with zero
mean and variance $J_{q,p}/N$. Thus we obtain a $L_r\times L_c$
coupling matrix $J_{q,p}$. For the asymptotic analysis we assume that
$N_p\to \infty$, for all $p=1,\dots,L_c$ and $M_q\to \infty$ for all
$q=1,\dots,L_r$.  The total subsampling rate is then
$\alpha=\sum_{q=1}^{L_r}M_q/(\sum_{p=1}^{L_c}N_p)$. The case of
homogeneous matrix can easily be recovered by setting $L_c=L_r=1$. We
define $I(\mu)$ or $I(i)$ to be the index of the block to which $\mu$
or $i$ belongs, $B_q$ is the set of indices in block $q$.

In all the examples of seeding matrices used in this article and
presented in Fig.~\ref{fig:1d_matrix}, the elements of the signal
vector are split into $L_c$ equally sized blocks ($N_p=N/L_c$). The
first block of measurements has size $M_1$ and the other $L_r-1$
measurement blocks have equal size $M_q=(M-M_1)/(L_r-1)$ for $q>1$.
In all the examples here we achieve the seeding by taking $\alpha_{\rm
  seed}=M_1L_c/N$ larger than $\alpha_{\rm BP}>\rho_0$, and
$\alpha_{\rm bulk}=M_qL_c/N$ for $q>1$ that can be approaching
$\rho_0$. The overall measurement rate is then \be \alpha=
\frac{\alpha_{\rm seed} + (L_r-1) \alpha_{\rm bulk} }{L_c}\, .\ee
Hence $\alpha \to \alpha_{\rm bulk}$ as $L_c/L_r \to 1$, and $L_r \to
\infty$. The matrix elements $F_{\mu i}$ are chosen as random i.i.d
variables with variance $J_{q,p}/N$ if variable $i$ is in the block
$p$ and measurement $\mu$ in the block $q$.

\subsection{Seeding experiments for noiseless measurements}
\label{1D_no_noise}

\begin{figure}[!ht]
  \begin{center}
    a)
    \includegraphics[scale=0.35]{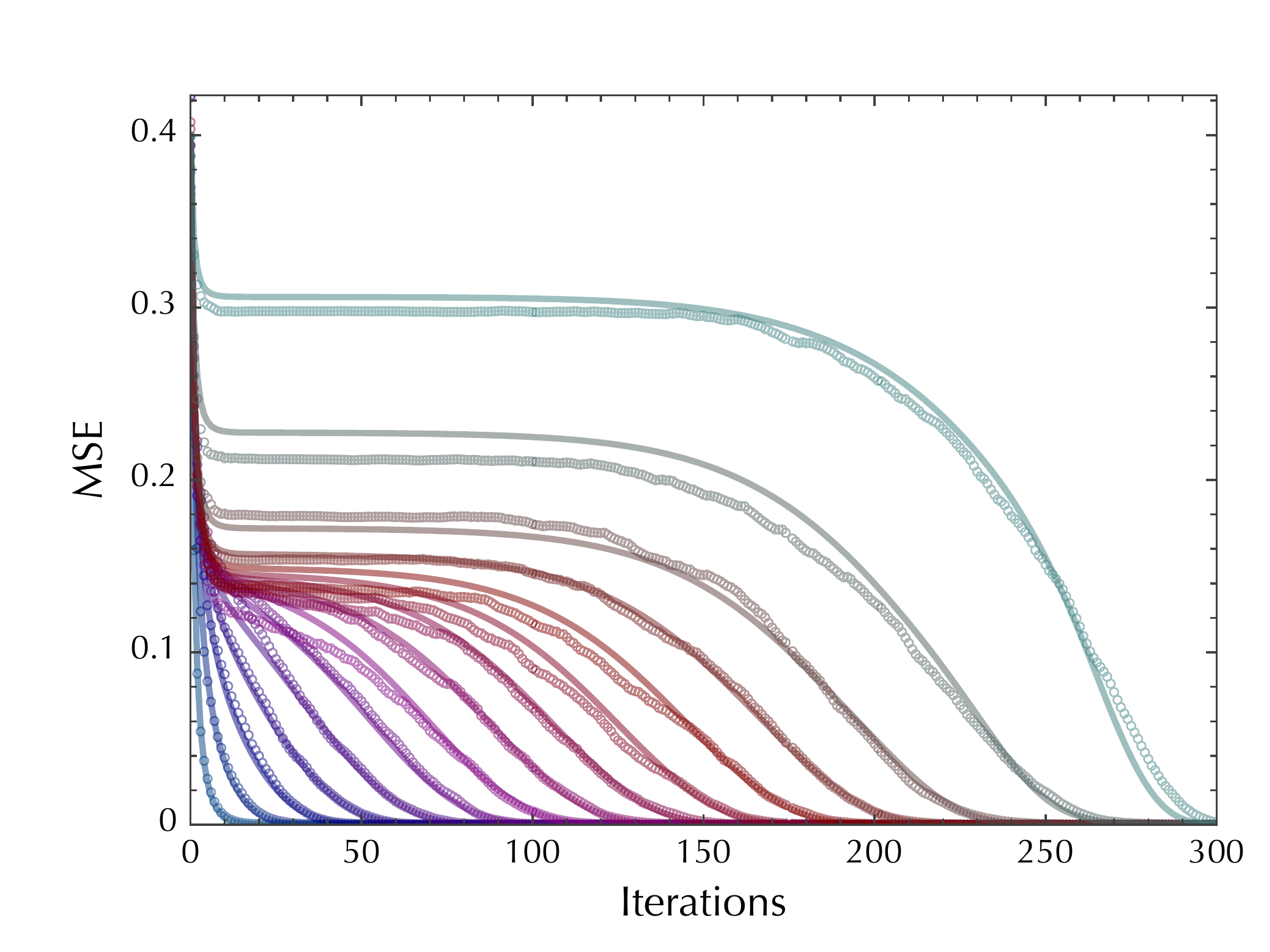}
    \hspace{0.5cm}
    b)
    \includegraphics[scale=0.35]{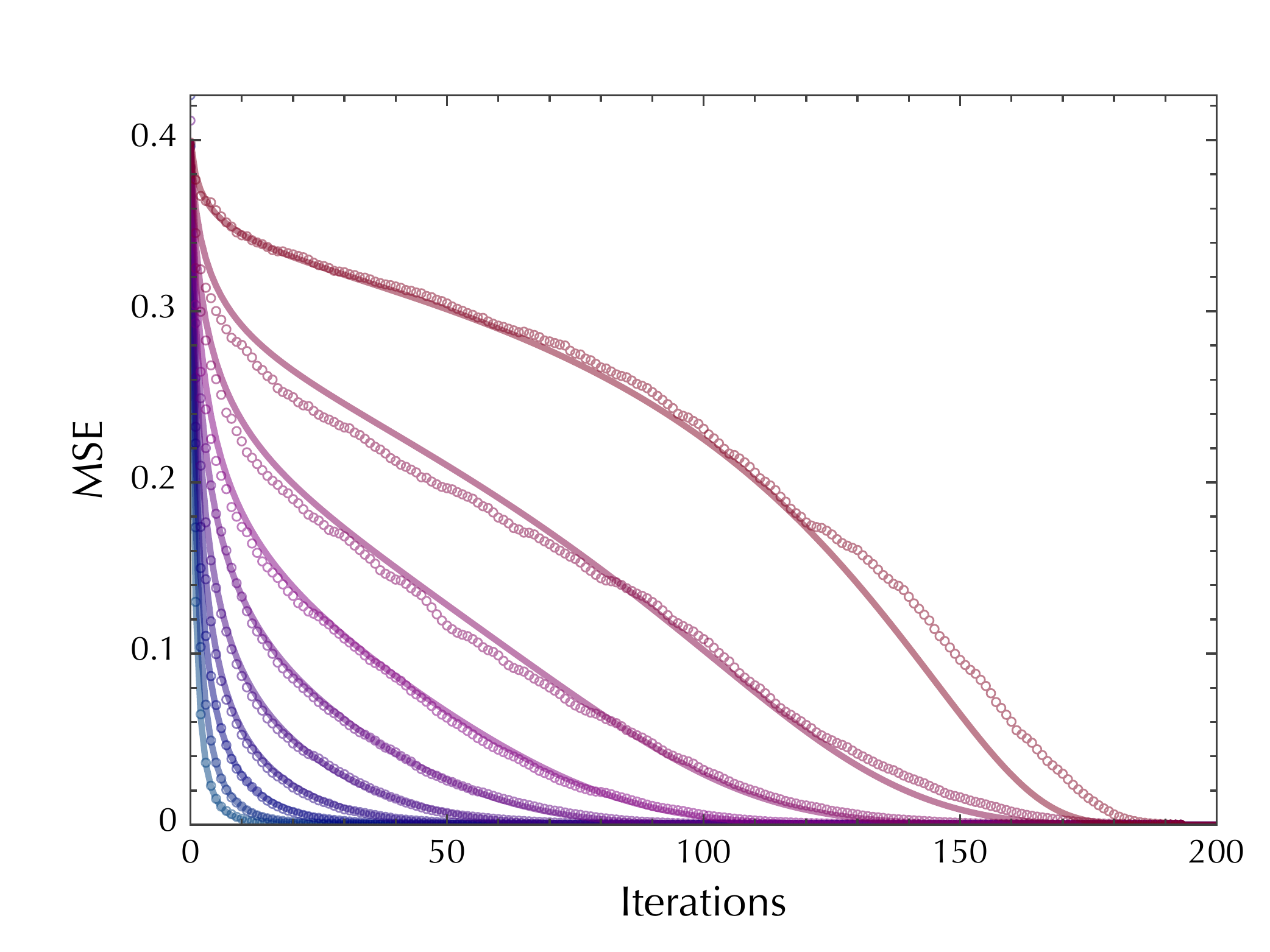}
        \hspace{0.5cm}
        \caption{Reconstruction on the signal with seeded measurement
          matrices. The mean-squared error in every block is plotted
          as a function of the iteration time. We compare the
          numerical analysis
          of BP for a signal of $N=40000$ elements with the analytic
          result obtained in the $N\to \infty$ limit using
          density evolution. The agreement is very good. The
          density of the signal is $\rho_0=0.4$, the non-zero elements
          are Gaussian with zero mean and unit variance. The
          measurement rate is $\alpha=0.5$. The two cases are: (a) The
          seeding matrix of the type (ii) from
          Fig.~\ref{fig:1d_matrix} with i.i.d. $0,\pm 1$ random
          elements, $\alpha_{\rm seed}=0.7$, $\alpha_{\rm
            bulk}=0.485$, $L=15$, $J=0.01$ and $W=2$. (b) The seeding
          matrix of the type (iii) from Fig.~\ref{fig:1d_matrix} with
          i.i.d. Gaussian random elements, $\alpha_{\rm seed}=0.68$,
          $\alpha_{\rm bulk}=0.48$, $L=10$ and $J=0.1$.  }
\label{fig:wave_1}
 \end{center}
\end{figure}

In Fig.~\ref{fig:wave_1} we demonstrate how BP reconstruction works
for seeded measurement matrices. We generated signal elements of
density $\rho_0=0.4$, the non-zero elements are Gaussian random
variables with zero mean and unit variance. We obtained $\alpha=0.5$
noiseless measurements per signal element using seeded matrices
generated as described above. We plot the mean-squared error in every
block (different lines) as a function of BP iteration time. We compare
a result from BP with its asymptotic density evolution behavior,
obtaining excellent agreement. Note that in this case, the BP
reconstruction for standard homogeneous matrices would fail. Notice
that in both cases illustrated in Fig.~\ref{fig:wave_1} the first
blocks are reconstructed fast and by interaction with the subsequent
blocks the reconstructed region is propagated to the following blocks.

\begin{figure}[!ht]
  \begin{center}
    (a)
    \includegraphics[scale=0.6]{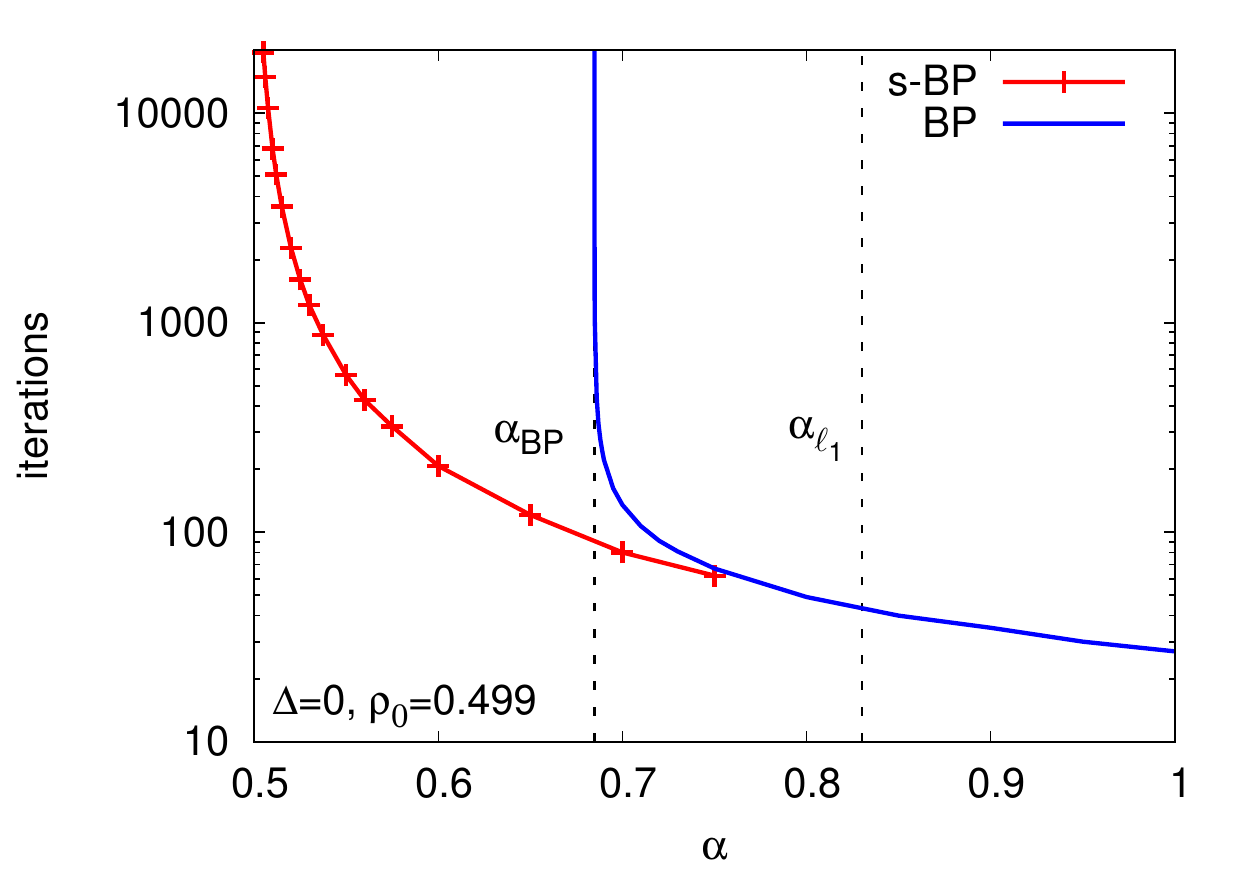}
    \hspace{0.5cm}
    (b)
    \includegraphics[scale=0.6]{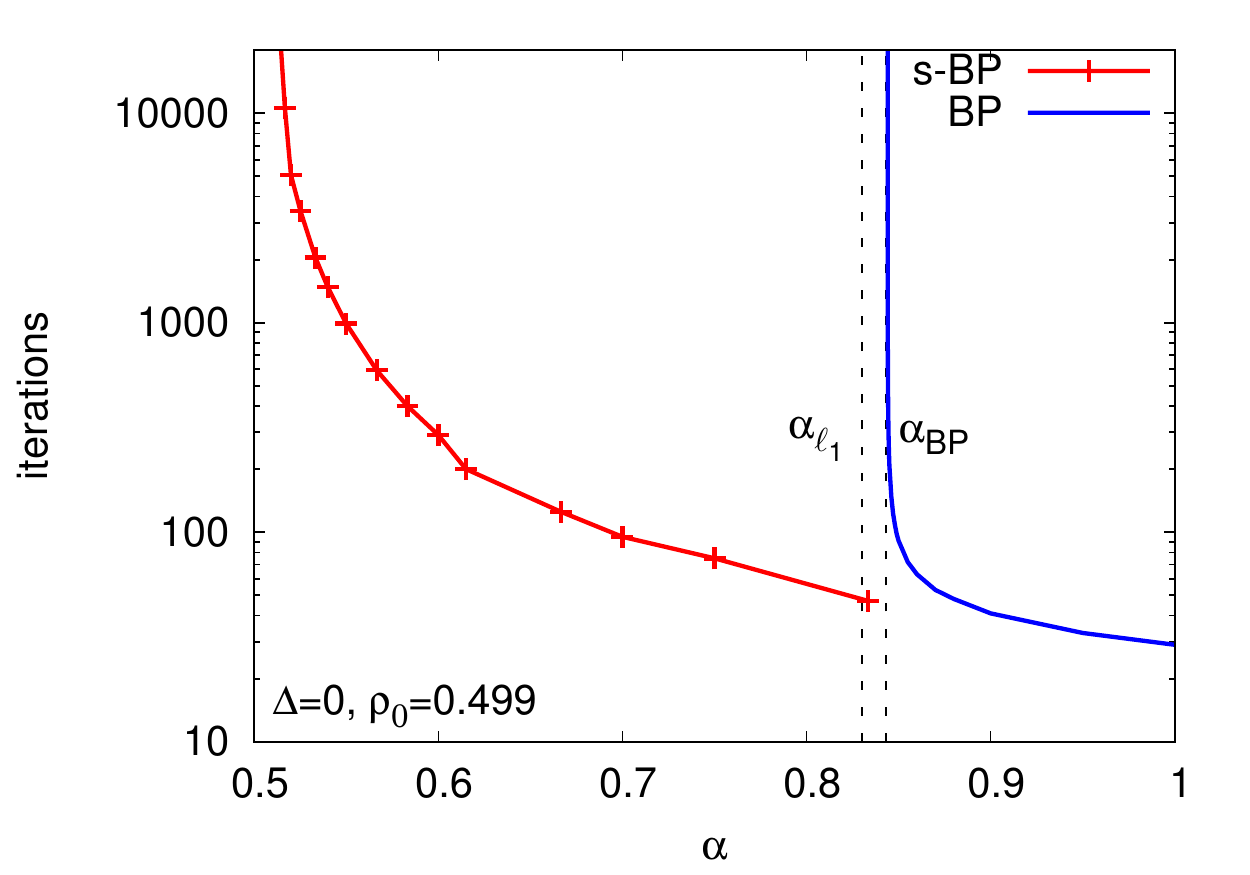}
        \hspace{0.5cm}
        \caption{Reaching the $\alpha\to \rho_0$ limit. Number of iterations needed to
          find the original signal of density $\rho_0=0.499$ for
          (a) a Gauss-Bernoulli signal and (b) a $0,\pm 1$
          signal. In both cases, we used BP with a Gauss-Bernoulli
          signal model with $\rho=\rho_0$.  The blue line shows the BP
          convergence time for homogeneous matrices, that
          diverges at the spinodal line $\alpha_{BP}$. The red line
          shows the BP reconstruction done with type (iii) seeding
          matrices: (a) using $\alpha_{\rm seed}=0.8$,
          $\alpha_{\rm bulk}=0.5$, and $J=2.10^{-3}$, (b) using
          $\alpha_{\rm seed}=1$, $\alpha_{\rm bulk}=0.5$, and $J=0.01$
          (in this case we added one block of measurements, $L_r=L+1$). As $L$
          increases in both cases, the total measurement rate $\alpha$ decreases and approaches
          $\alpha_{\rm bulk}=0.5 \approx \rho_0=0.499$. The number of
          iterations needed for
          exact reconstruction then diverges with $L$. The difference
          between the reconstruction limits of BP  and of
          $\ell_1$  is striking.}
    \label{fig:time_evolution_seeded}
  \end{center}
\end{figure}

Now that we illustrated that the BP reconstruction for large systems
indeed agrees with the asymptotic density evolution analysis we plot
in Fig.~\ref{fig:time_evolution_seeded} two examples of the number of
iterations (defined as the time when mean-squared error $E<10^{-7}$) it
takes to reconstruct exactly signal of density $\rho_0$ with
measurement rate $\alpha \to \rho_0$.

\begin{figure}[!ht]
  \begin{center}
    (a)
    \includegraphics[scale=0.6]{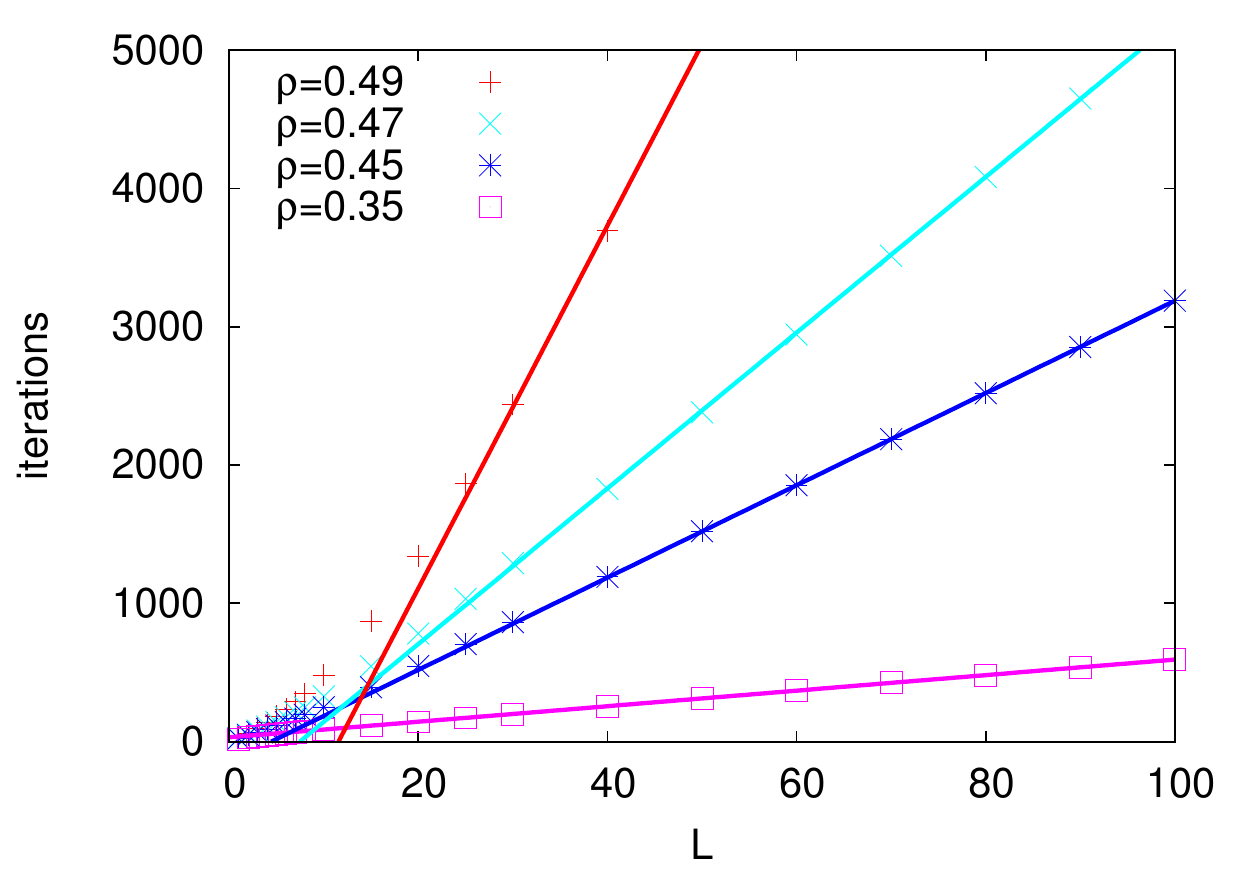}
    \hspace{0.5cm}
    (b)
    \includegraphics[scale=0.6]{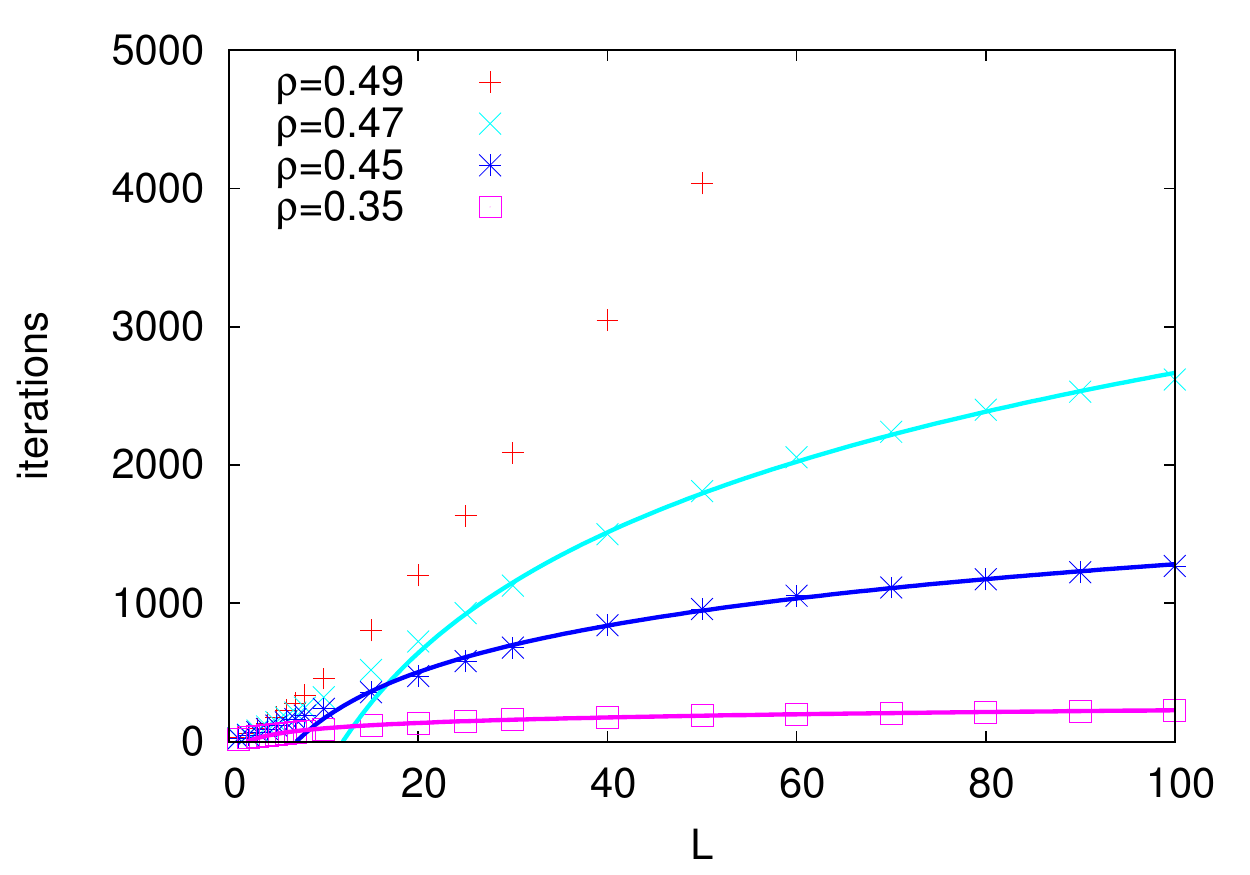}
        \hspace{0.5cm}
        \caption{Number of iterations needed for reconstruction with
          type (ii) seeding matrices (on the left) and type (iii)
          seeding matrices (on the right). With type (ii) matrices, a
          wave is propagating in the system with a constant speed,
          while for type (iii) matrices with long range interactions the speed
          is proportional to $L$, hence the total time scales as
          $\log{L}$. Left: we used $J=0.02, W=2$, $\alpha_{\rm
            seed}=1.0$, $\alpha_{\rm bulk}=0.5$. Right: we used
          $J=0.01$, $\alpha_{\rm seed}=1.0$, $\alpha_{\rm
            bulk}=0.5$. We used $\rho=\rho_0$ to make these data.  }
\label{fig:time_seeding}
  \end{center}
\end{figure}

\begin{figure}[!ht]
  \begin{center}
 \includegraphics[scale=0.4]{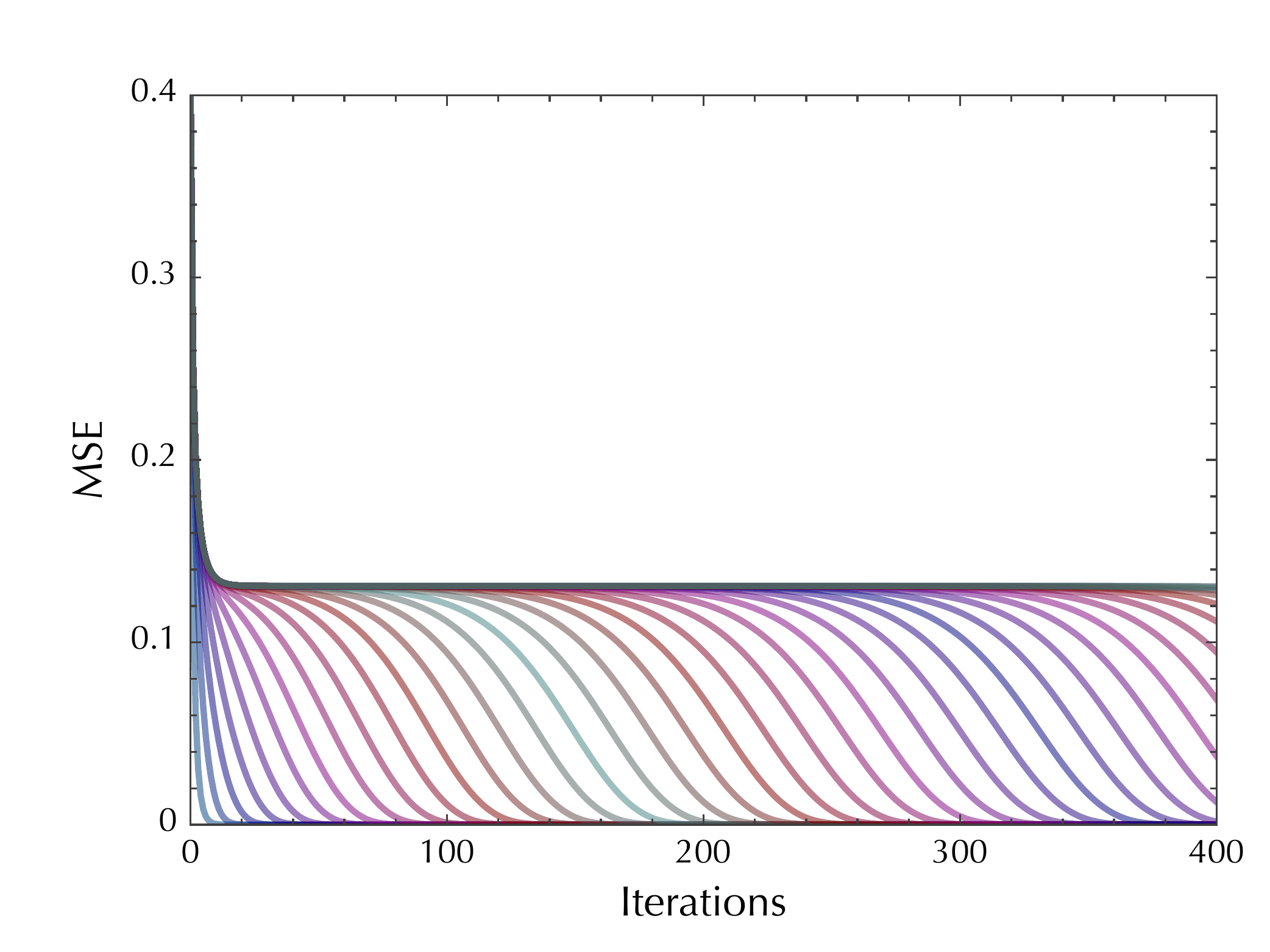}
   \includegraphics[scale=0.4]{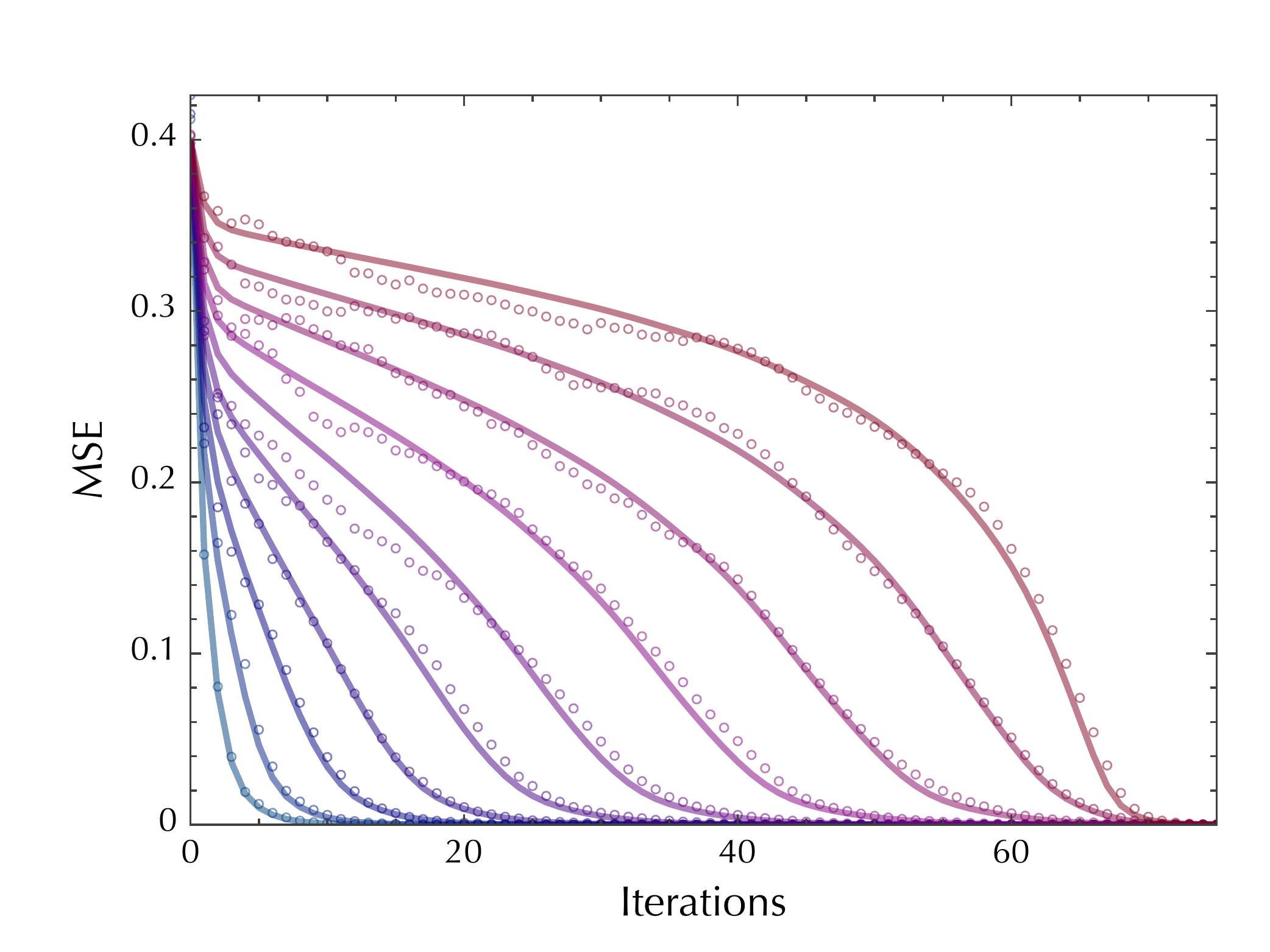}
   \caption{Left: Evolution of the mean-squared error in each block as
     a function of iteration time. Here we used type (ii) seeding
     matrices with $W=2$, $L=50$, $\alpha_{\rm seed}=1.0$,
     $\alpha_{\rm bulk}=0.5$, and $J=0.01$. With that type of matrix,
     the boundary of the reconstructed region is propagating as a
     localized wave.  Right: Same as Fig.~\ref{fig:wave_1}
     for a ``adversary-case'' signal having components $0,\pm 1$, with
     $N=10000$. We used $\alpha=0.6$ and $\rho_0=0.4$, with the
     seeding matrix of type (iv) with $L=10$, $\alpha_{\rm seed}=1.0$,
     $\alpha_{\rm bulk}=0.5$, $J=0.1$. Exact reconstruction is
     achieved even thought the signal model (Gauss-Bernoulli) does not
     correspond to the empirical signal distribution.}
\label{fig:seeding3}
\end{center}
\end{figure}

In Fig.~\ref{fig:time_seeding} we show how does the number of
iterations needed for exact reconstruction depends on the number of
blocks $L$ for different signal densities $\rho_0$. We see that in
case of the one-dimensional seeding matrix of type (ii) the number of
iterations depend linearly on the the number of blocks. The
boundary of the reconstructed region is propagating as a kind of
spatially localized wave at a constant speed, as illustrated in
Fig.~\ref{fig:seeding3}. On the other hand for the long-range
triangular matrices of type (iii) the number of iterations grows only
as logarithm of the number of blocks, $\log{L}$, (at least for large
$L$). The propagation of the reconstructed region does not really
correspond to a localized traveling wave, as visible from
Fig.~\ref{fig:wave_1} (b). In both cases the speed of the growth of
the seed (i.e. reconstructed region) is proportional to the
interaction strength between the first non-reconstructed block and the
seed. In the case of one-dimensionally coupled matrices this strength
does not depend on the position of the seed boundary. In the case of
triangular seeding matrix the strength is proportional to the size of
the already reconstructed region, hence $\delta L/\delta t \sim L$, which
gives the logarithmic dependence seen in Fig.~\ref{fig:time_seeding} .

In Fig.~\ref{fig:seeding3} right, and
Fig.~\ref{fig:time_evolution_seeded} right we show that the BP
reconstruction with seeding matrices works also in the case when the
signal model does not at all correspond to the actual signal
distribution. In the two figures the signal components are $0,\pm 1$,
whereas the signal model was still Gauss-Bernoulli. Since the
probabilistic approach is optimal for noiseless measurements even when
the signal distribution is not known (as proven in Sec.~\ref{Optimality}) the
seeding strategy is able to approach the information theoretic limit
$\alpha \to \rho_0$ also in this case.

We have not done expectation maximization learning in the data
presented in this section, but this strategy is also useful with the
seeding matrices and is included in our implementations. Its behavior
is analogous to the one in the case of homogeneous matrices, as
discussed in Sec.~\ref{Res:learning}.

\subsection{Seeding experiments for noisy measurements}\
\label{1D_noise}

Every CS method requires robustness with respect to the
measurement noise. In Sec.~\ref{Sec:Noisy} we analyzed the phase
diagram under measurement noise. In particular we showed existence of
two phase transitions $\alpha_c(\rho_0)$ and $\alpha_d(\rho_0)$ (see e.g. Fig.~\ref{fig_noise2}) such that for $\alpha
\notin {(\alpha_c,\alpha_d)}$ and for the signal model matching the
empirical signal distribution the belief propagation inference is as good
as the optimal Bayesian inference. In other words the final MSE
achieved by BP for $\alpha<\alpha_c$ or $\alpha>\alpha_d$ is the best
achievable for a given measurement matrix $\bF$. If a stronger noise
robustness is required then one would have to use a
different measurement protocol or much larger sampling rate $\alpha$.
The only region that is open to improvement is for measurement rates
$\alpha_c < \alpha < \alpha_d$. With seeding
we can indeed improve considerably the final MSE in this region.
The noise stability of the seeding strategy was touched already in
\cite{KrzakalaPRX2012}, see also \cite{DonohoJavanmard11} for a rigorous discussion.

\begin{figure}[ht]
  \begin{center}
    a)
    \includegraphics[scale=0.35]{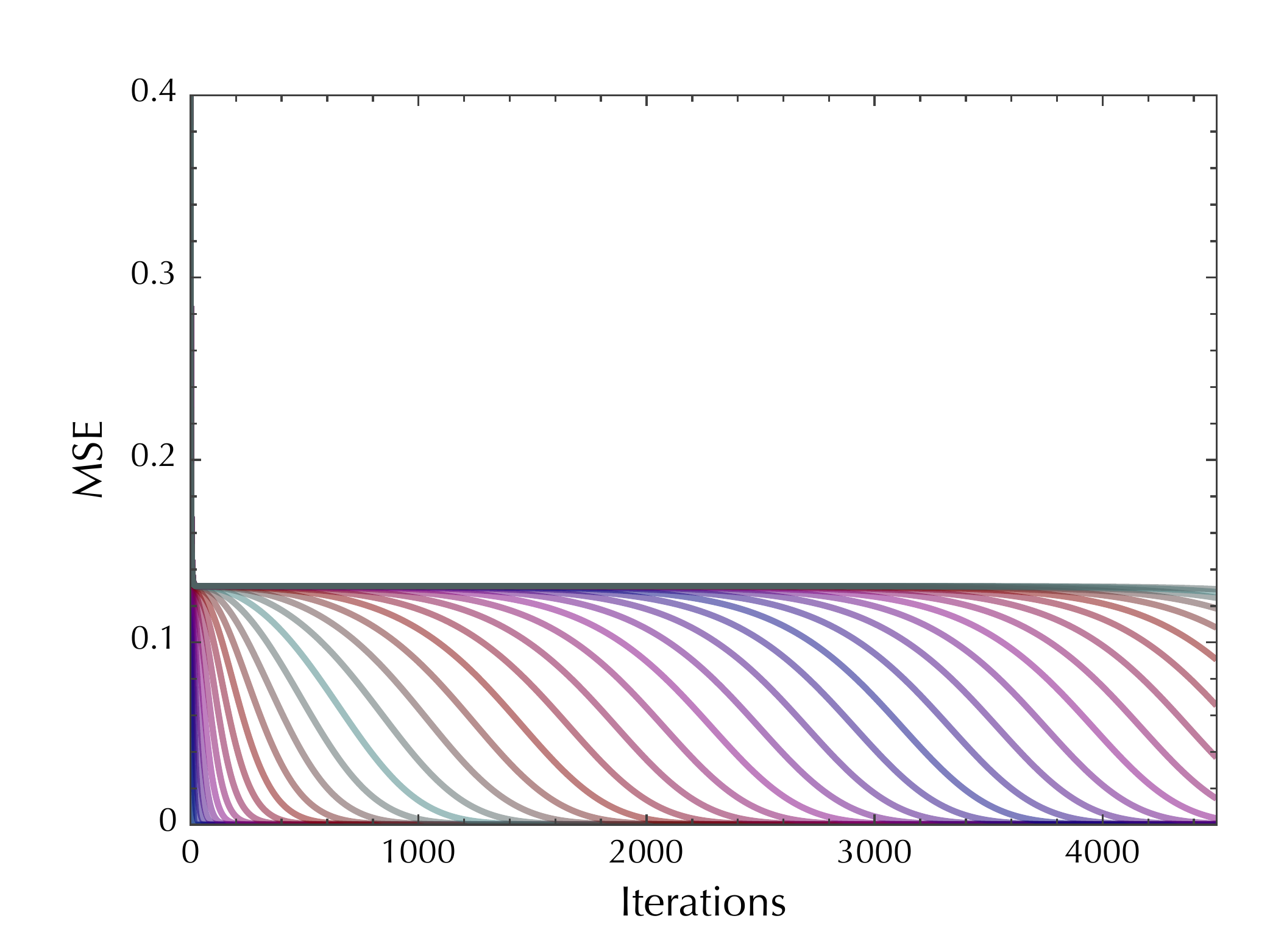}
    \hspace{0.5cm}
    b)
    \includegraphics[scale=0.35]{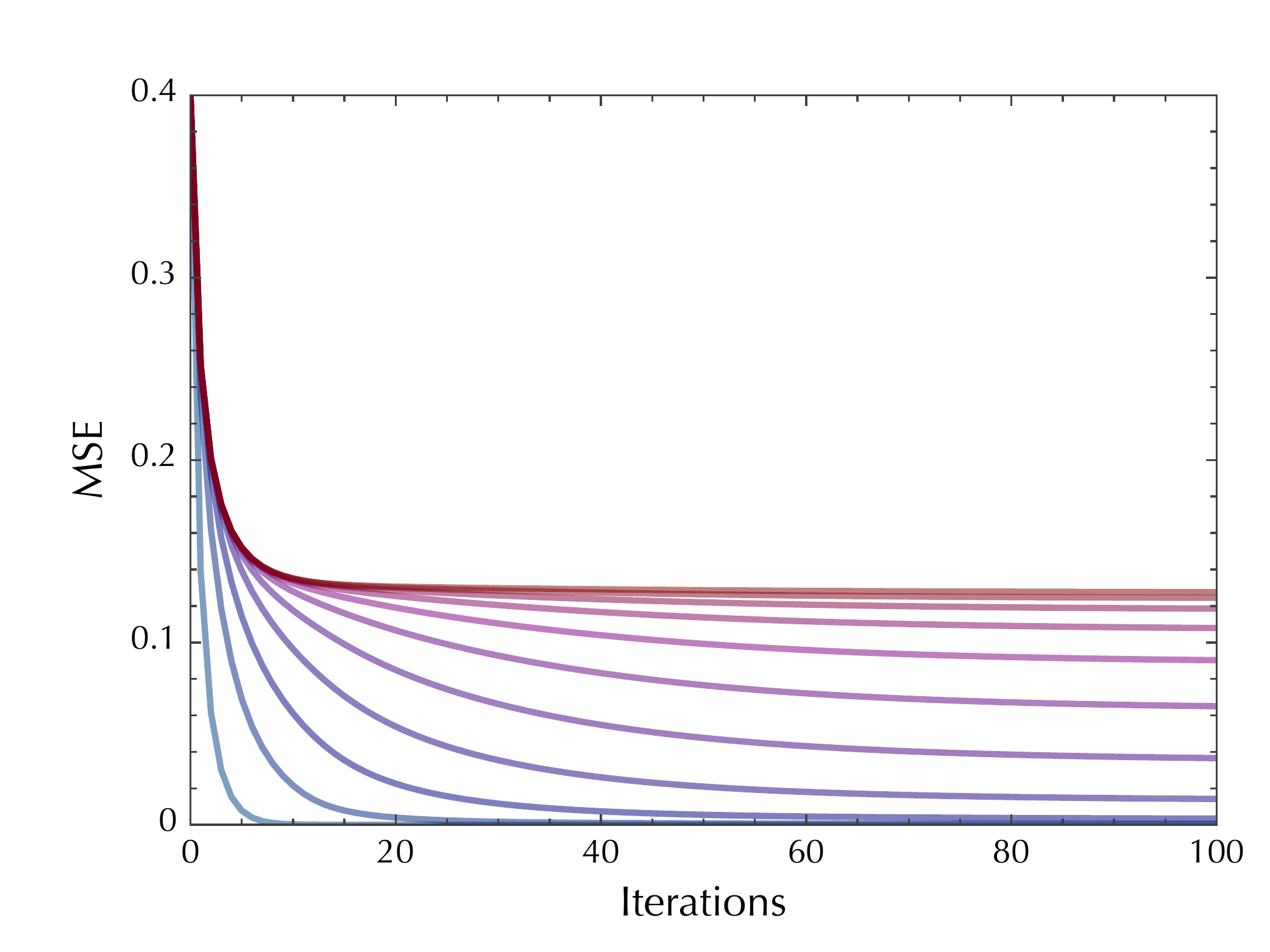}
        \hspace{0.5cm}
        \caption{Seeding matrices with noise: Evolution of the MSE in
          each block, as in Fig.~\ref{fig:seeding3}, but in the noisy
          case. Here we used type (iv) seeding matrices with $W=2$,
          $L=100$, $\alpha_{\rm seed}=1.0$, $\alpha_{\rm bulk}=0.5$,
          and $J=0.001$.  Left: Even with a large noise with standard
          deviation $\sqrt{\Delta}=10^{-3}$, the front wave is still
          propagating with a finite speed, leading to a reconstruction
          with a final MSE of the order of $\Delta$. This demonstrates
          the robustness of our approach with additive noise.  Right:
          When the noise (here $\sqrt{\Delta}=10^{-2}$) is too high
          (so that $\alpha_{\rm bulk}<\alpha_c$, see text) there is no such
          propagation. Here, only the very first blocks (the first one
          having $\alpha_{\rm seed}=1$ and its close neighbors) go to low MSE
          while the rest of the system stays far from the
          solution. Essentially all the changes are done in the $100$
          first iterations shown here. In this case, the seeding
          matrices does not bring improvement (and no other method
          could in this case).}
\label{fig:wave_noise}
 \end{center}
\end{figure}

The performance of the seeding strategy in the presence of noise can
be again studied using the replica/density evolution
equation. In Fig. \ref{fig:wave_noise} we illustrate the evolution of
the MSE for CS with noisy measurements for subsampling
rates $\alpha_c<\alpha<\alpha_d$ for which BP with the homogeneous matrices gives a
MSE much larger than the noise variance $\Delta$. Again, in order to have a working seeding
mechanism, the free entropy associated with the fixed point of BP
close to the solution must dominate the free entropy associated with
the meta-stable state. This is the case for measurement rates
$\alpha_c<\alpha_{\rm bulk}<\alpha_d$ and can thus be exploited.

Note, however, that in the presence of noise the free energy
difference between the global and local maxima is finite (whereas it
was diverging for the noiseless case), this means that the seeding
matrices need to be constructed with more care in order to saturate
the threshold. In particular the interaction width $W$ (see
Fig.~\ref{fig:1d_matrix}) has to grow when the threshold $\alpha_c$ is approached.

\section{Conclusion}

This paper presents a detailed analysis of the new strategy for
compressed sensing that we introduced in  \cite{KrzakalaPRX2012}.
With respect to this earlier work we have provided here a more detailed study of
the phase diagrams and of the associated phase transition for BP
reconstruction algorithm and for the (intractable) optimal
reconstruction. We have treated in detail the case of noisy measurements and we
have shown that our approach presents excellent stability with
respect to noise, in the sense that the BP algorithm (with seeding if needed)
is able to reconstruct the signal with mean-squared error as low as
the optimal inference algorithm based on exhaustive enumeration (which
is of course not computationally tractable).  We have discussed
reconstruction in the case of mismatching signal model and signal distribution
and we have shown that in the noiseless case this mismatch does not pose a
serious problem. We have introduced and studied new types of seeding measurement
matrices with which we were also able to achieve reconstruction at
almost optimal reconstruction rates.

\acknowledgments This work has been supported in part by the EC Grant
‘‘STAMINA’’, No. 265496, and by the Grant DySpaN of ‘‘Triangle de la
Physique.’’ We thank Richard Morris for his comments on our
manuscript. 

\appendix

\newpage
\section{Derivation of the replica analysis for block matrices} \label{appendix:free_entropy}
Here, we rederive the replica analysis for the
seeding matrices described in the main text
Sec.~\ref{Sec:Seeding}. This follows closely the derivation presented
in Sec.~\ref{replicas}, we only need add the block indices.
To evaluate the average of the replicated partition function (\ref{ave_Zn}) for the
block matrices $\bF$ we introduce the order parameters per block
\begin{eqnarray}
m_p^a & = & \frac{1}{N_p}\sum_{i \in B_p}x_i^as_i, \ a=1,2,\dots, n\\
Q_p^a & = & \frac{1}{N_p}\sum_{i \in B_p}(x_i^a)^2, \ a=1,2,\dots,n\\
q_p^{ab} & = & \frac{1}{N_p}\sum_{i \in B_p}x_i^ax_i^b, \  a<b
\end{eqnarray}
where $B_{p}$ represents the index of the variables in block $p=1,\dots,L_c$.

We introduce a Dirac delta function that fixed the order parameters and
we make use of the following integral representation for delta function:
\begin{eqnarray}
1 & = & \int \prod_{a,p}{\rm d}\hat{Q}_p^a {\rm d} Q_p^a\prod_{a\neq b,p}{\rm d}\hat{q}_p^{ab}{\rm d}q_p^{ab}\prod_{a,p}{\rm d}\hat{m}_p^a {\rm d}m_p^a e^{\frac{1}{2}\sum_{p=1}^{L_c}\sum_{a=1}^n\hat{Q}_p^a(N_p Q_p^a-\sum_{i \in B_p}(x_i^a)^2)}\nonumber\\
& & e^{-\frac{1}{2}\sum_{p=1}^{L_c}\sum_{a \neq
    b}\hat{q}_p^{ab}(N_pq_p^{ab}-\sum_{i \in B_p}x_i^ax_i^b)
}e^{-\sum_{p=1}^{L_c}\sum_{a=1}^n\hat{m}_p^a(N_pm_p^a-\sum_{i \in
    B_p}x_i^ax_i^0)}\, .
\label{delta_fuction}
\end{eqnarray}
Inserting Eq.~(\ref{delta_fuction}) into the expression of
$\mathbb{E}_{\bF,\bs,{\bf \xi}}(Z^n)$, Eq.~(\ref{ave_Zn}), we get
\bea
& &\expect_{\bF,\bs,\bf{\xi}}(Z^n)  =  \int \prod_{a,p}{\rm
  d}\hat{Q}_p^a{\rm d}Q_p^a{\rm d}\hat{m}_p^a{\rm d}m_p^a \e^{\sum_{p=1}^{L_c}N_p\[[\frac{1}{2}\hat{Q}_p^aQ_p^a-
\hat{m}_p^a{\rm d}m_p^a\]]}
\prod_{a\neq b,p}{\rm d}\hat{q}_p^{ab}{\rm d}q_p^{ab}\e^{-\frac{1}{2}\sum_{p=1}^{L_c}N_p\hat{q}_p^{ab}q_p^{ab}}  \int \prod_{i,a}{\rm d}x_i^a \nonumber \\
&& \prod_{i,a} \left[ (1-\rho) \delta(x_i^a) + \rho \phi(x_i^a) \right]\prod_i\[[(1-\rho_0)\delta(s_i)+\rho_0\phi_0(s_i)\]] \prod_{\mu} \frac{1}{\sqrt{2\pi \Delta_\mu}}\expect_{\bF,\bf{\xi}} \[[e^{-\frac{1}{2\Delta_\mu}\sum_{a=1}^n(\sum_{i=1}^NF_{\mu i}s_i +\xi_{\mu}- \sum_{i=1}^N F_{\mu i}x_i^a)^2}\]]\nonumber\\
& & \prod_{p=1}^{L_c}\e^{-\frac{1}{2}\sum_a\hat{Q}_p^q(\sum_{i\in
    B_p}x_i^a)^2+\frac{1}{2}\sum_{a \neq b}\hat{q}_p^{ab}\sum_{i \in
    B_p}x_i^ax_i^b+\sum_a\hat{m}_p^a\sum_{i \in B_p}x_i^a s_i} \nonumber
\eea
When averaging $Z^n$,  we again need to evaluate the quantity $X_\mu$,
defined in Eq.~(\ref{Xmu}). We define $u_{\mu}^a=\sum_{i=1}^NF_{\mu i}x_i^a$, with
$a=\{0,1,\ldots, n\}$ where 0 corresponds to the index of the signal,
$x_i^0=s_i$. The quantities then obey joint Gaussian distribution with $\expect_{\bF,\bf{\xi}} (u_{\mu}^a)=0$ and
\bea
& &\expect_{\bF,\bf{\xi}} (u_{\mu}^0u_{\mu}^0) =  \rho_0 \overline{s^2} \sum_{p=1}^{L_c} J_{I(\mu)p}n_p \, ,  \quad
\expect_{\bF,\bf{\xi}} (u_{\mu}^au_{\mu}^0)  = \sum_{p=1}^{L_c} J_{I(\mu)p}n_p m_p^a \, , \\
& & \expect_{\bF,\bf{\xi}} (u_{\mu}^au_{\mu}^a)  = \sum_{p=1}^{L_c} J_{I(\mu)p}n_p Q_p^a \, , \quad
\expect_{\bF,\bf{\xi}} (u_{\mu}^au_{\mu}^b)  =  \sum_{p=1}^{L_c} J_{I(\mu)p}n_p Q_p^{a} \, .
\eea
Under the replica symmetric ansatz the replicas are considered
equivalent, i.e.
\be
m_p^a=m_p, \ \  q_p^{ab}=q_p, \ \ Q_p^{a}=Q_p \, .\\
\ee
We introduce $\tilde{\rho}_q,\tilde{m}_q,\tilde{q}_q,\tilde{Q}_q$ as follows
\begin{equation}
\tilde{\rho}_q=\rho_0\overline {s^2}\sum_{p=1}^{L_c} J_{qp}n_p,\quad  \tilde{m}_q=\sum_{p=1}^{L_c} J_{qp}n_pm_p, \quad \tilde{q}_q=\sum_{p=1}^{L_c} J_{qp}n_p q_p,\quad \tilde{Q}_q=\sum_{p=1}^{L_c}J_{qp}n_pQ_p \, . \label{def_tilde-}
\end{equation}
And thus $v_{\mu}^a=u_{\mu}^0-u_{\mu}^a+\xi_{\mu}$, $a=1,2, \ldots, n$ are also joint Gaussian distributed with zero means and
\begin{eqnarray}
G_{aa}=\expect_{\bF,\bf{\xi}}(v_{\mu}^av_{\mu}^a) & = &
\tilde{Q}_{I(\mu)}+\tilde{\rho}_{I(\mu)}-2\tilde{m}_{I(\mu)}+\Delta_0,
\quad a=1,2, \ldots, n\, , \\
G_{ab}=\expect_{\bF,\bf{\xi}}(v_{\mu}^av_{\mu}^b) & = &
\tilde{q}_{I(\mu)}+\tilde{\rho}_{I(\mu)}-2\tilde{m}_{I(\mu)}+\Delta_0,
\quad a<b\, ,
\end{eqnarray}
where $G$ is the inverse covariance matrix.
For the block matrices we have
\be
\text{det}(\one+\frac{G}{\Delta}) = e^{n\left[\frac{\tilde{q}_{I(\mu)}-2\tilde{m}_{I(\mu)}+\tilde{\rho}_{I(\mu)}+\Delta_0}{\tilde{Q}_{I(\mu)}-\tilde{q}_{I(\mu)}+\Delta}+\text{log}(1+\frac{\tilde{Q}_{I(\mu)}-\tilde{q}_{I(\mu)}}{\Delta})\right]} \, .
\ee
From here following the same steps as for derivation of
Eq.~(\ref{free_rep}) we obtain
 \begin{eqnarray}
  & & \expect_{\bF,\bs,\bf{\xi}}Z^n    =    \int\prod_{p}\text{d}\hat{Q}_p\text{d}Q_pd\hat{q}_p\text{d}q_p\text{d}\hat{m}_p\text{d}m_p\nonumber\\
  & & \text{exp}\Bigg(nN\Bigg\{\frac{1}{2}\sum_{q=1}^{L_r} n_1\alpha_{q1} \left[ \frac{\tilde
    q_q-2\tilde m_q+\tilde \rho_q+\Delta_0}{\tilde
    Q_q-\tilde q_q+\Delta} + \log{(\Delta+\tilde Q_q-\tilde q_q)}\right]
+ \sum_{p=1}^{L_c} n_p\left( \frac{Q_p\hat Q_p}{2}- m_p\hat m_p +
  \frac{q_p \hat q_p}{2} \right)\nonumber\\ & & + \sum_{p=1}^{L_c}n_p\int {\rm d}s \left[(1-\rho_0)\delta(s) + \rho_0 \phi_0(s)
\right]\int{\cal D}z \log \int {\rm d}x \, e^{-\frac{\hat Q_p+\hat
        q_p}{2}x^2 + x(\hat m_p s + z \sqrt{\hat q_p})} \left[
      (1-\rho)\delta(x) +\rho \phi(x) \right] \Bigg\}\Bigg)
\label{seeded:Z^n}
\end{eqnarray}
where $\alpha_q=M_q/N$. From here we obtain the expression of free entropy in  Eq.~(\ref{free_seeded}).

\section{Phase diagram of the $\ell_1$ reconstruction for seeding matrices} \label{appendix:L1_seeded}
In this section, we apply the well-known $\ell_1$ norm reconstruction for the seeding matrix (i) in Fig.~\ref{fig:1d_matrix}, i.e.,
for the coupling matrix, $J_{p,p}=1,J_{p,p-1}=J_1,J_{p_1,p}=J_2$ and others are zeros, and see if the delicate designed matrix can also provide substantial improvement for the reconstruction limit.

In order to study the $\ell_1$ norm, we use the large $\beta$ limit of
the problem defined by the partition function
\be
Z=\int \prod_{i=1}^N  \left( dx_i \;
  e^{-\beta |x_i|}
\right)
\prod_{\mu=1}^M \delta\left(\sum_i F_{\mu
    i}(x_i-s_i)\right)
\ee

We can use all our previous replica computation with the substitution in the local measure
of $(1-\rho)\delta(x_i)+\rho \phi_0(x_i) $ by $ e^{-\beta |x_i|} $.  In the case of our seeding matrix $\bF$, this gives
$Z=\int \e^{nN\Phi}$ with
\bea
     && \Phi(\{Q_p\}_{p=1}^L,\{q_p\}_{p=1}^L,\{m_p\}_{p=1}^L,\{\hat Q_p\}_{p=1}^L,\{\hat q_p\}_{p=1}^L,\{\hat m_p\}_{p=1}^L) =\nonumber\\
     & &  -\frac{1}{2}\sum_{p=1}^L
\alpha_p \left[     \frac{\tilde q_p-2\tilde m_p+\tilde \rho_p}{\tilde
    Q_p-\tilde q_p} + \log{(\tilde Q_p-\tilde q_p)} \right] +
\sum_{p=1}^L\left( \frac{Q_p\hat Q_p}{2} - m_p\hat m_p + \frac{q_p
    \hat q_p}{2} \right)\nonumber \\ && + \sum_{p=1}^L\int {\cal D}z
\int {\rm d}s  \left[(1-\rho)\delta(s) + \phi_0(s) \right]
\log{\left\{  \int {\rm d}x \, e^{-\frac{\hat Q_p+\hat q_p}{2}x^2 +
      x(\hat m_p s + z \sqrt{\hat q_p})}e^{-\beta |x|}\right\}}
\eea
where we always use (\ref{def_tilde}):
\be
\tilde{\rho}_p=\rho_0 \langle s^2\rangle \sum_{q=1}^L J _{pq},\quad
\tilde{m}_p=\sum_{q=1}^L J_{pq}m_q, \quad
\tilde{q}_p= \sum_{q=1}^L J_{pq} q_q,\quad
\tilde{Q}_p=\sum_{q=1}^L J_{pq}Q_q \, .
\ee

In the large $\beta$ limit, we assume the scaling for the order parameters as follows:
\bea
\hat Q_p+\hat q_p=\beta \hat R_p\ \ ,\ \ \hat q_p=\beta^2\hat
r_p\ \ ,\ \ \hat m_p=\beta \hat \mu_p\nonumber\\
Q_p=O(1) \ , \ q_p=O(1)\ , \ m_p=O(1)\ , \ Q_p-q_p=O(1/\beta)
\eea
and we write specifically $r_p=\beta(Q_p-q_p)$.
Therefore, the free entropy $\Phi$ scales linearly in $\beta$
\bea\nonumber
\frac{\Phi}{\beta}&=&
-\frac{1}{2}\sum_{p=1}^L
\alpha_p \left[     \frac{\tilde q_p-2\tilde m_p+\tilde \rho_p}{\tilde
    r_p} \right]
+ \sum_{p=1}^L\left( \frac{q_p\hat R_p}{2} - m_p\hat \mu_p -
 \frac{r_p  \hat r_p}{2} \right)\\
 &-&\sum_{p=1}^L \int {\cal D}z \int {\rm d}s
 \left[(1-\rho_0)\delta(s) + \phi_0(s) \right]
\min_x
\(( \frac{\hat R_p}{2}x^2-(\hat \mu_ps+\sqrt{\hat r_p}z)x+|x| \)) \, .
\eea
It is easy to check that, for $L=1$, this gives back the free energy
written e.g. by Kabashima et al. \cite{KabashimaWadayama09}.

In order to study the transition, we assume the following scaling when
one is near to the regime of exact retrieval of the signal:
\be
\forall p\in\{1,\dots,L\}\; : \ \ \hat \mu_p\to \infty \; , \ \ \hat
r_p=1/\lambda_p^2
\ee

With the above scaling  we find that the saddle point equations of order parameters
$r_p, E_p=q_p-2m_p+\rho_0\langle s ^2 \rangle, \hat \mu_{p}$ and $\hat r_p$ are independent
of the distribution of  the nonzero elements in signal $\phi_0(x)$, as long as $\phi_0(x)=\phi_0(-x)$. For $L=1$, i.e., the canonical
matrix $\bF$ ,  the saddle point equations are given as :
\bea
 r&=&\frac{1}{\hat \mu}\left[\rho_0+2(1-\rho_0)\int_{\lambda}^\infty
   Dz\right] \, ,
\label{uneq1}\\
E&=&
\frac{1}{\lambda^2 \hat\mu^2 }
\left[
2(1-\rho_0)\int_{\lambda}^\infty
Dz\; (z-\lambda)^2
+\rho_0 (1+\lambda^2)
\right] \, ,
\label{uneq2}\\
\hat \mu&=&\frac{\alpha}{r} \,
\label{uneq3}\\
\hat r&=& \frac{\alpha}{r^2}[q-2m+\rho_0\langle
s^2\rangle] =\frac{\hat \mu^2}{\alpha}[q-2m+\rho_0\langle
s^2\rangle] \, .
\label{uneq4}
\eea
They can be simplified further as a closed system of two variables $\alpha, \lambda$:
\bea
\alpha & =  & \rho_0+2(1-\rho_0)\int_{\lambda}^\infty
   Dz \, , \nonumber\\
 \alpha & =  & 2(1-\rho_0)\int_{\lambda}^\infty
Dz\; (z-\lambda)^2
+\rho_0 (1+\lambda^2) \, .
\eea
They give the critical value of
$\alpha$ for a given value of $\rho_0$, these are the equations of
\cite{KabashimaWadayama09,DonohoMaleki09}.

For the seeding matrix,  $L\geq 2$,  due to the fact that the final result
does not depend on $\phi_0$ (for symmetric ones), we thus take $\phi_0$  as a centered Gaussian distribution of
variance one.
After some work we get:
\bea
r_p&=&\frac{2}{\hat\mu_p}\left[(1-\rho_0) H\left(\frac{1}{\sqrt{\hat
        r_p}}\right)+\rho_0 H\left(\frac{1}{\sqrt{\hat
        r_p+\hat \mu_p^2}}\right)\right] \, ,\\
\nonumber
E_p&=&\frac{1}{\hat\mu_p^2}\left[
2(1-\rho_0) \hat r_p \psi\left(\frac{1}{\sqrt{\hat
        r_p}},\frac{1}{\sqrt{\hat
        r_p}}\right)
+2 \rho_0 (\hat r_p+\hat \mu_p^2)
\psi\left(\frac{1}{\sqrt{\hat
        r_p+\hat \mu_p^2}},\frac{1}{\sqrt{\hat
        r_p+\hat \mu_p^2}}\right)\right.\\
&&\left.
-4\rho_0 \hat\mu_p^2 H\left(\frac{1}{\sqrt{\hat
        r_p+\hat mu_p^2}}\right)+\rho_0\hat\mu_p^2\right] \, ,\\
\hat \mu_p&=& \sum_q J_{qp} \frac{\alpha_q}{
  \sum_s J_{qs} r_s} \, , \\
\hat r_p&=& \sum_q J_{qp} \frac{\alpha_q}{ (\sum_s J_{qs} r_s)^2}
\left(\sum_s J_{qs} E_s\right) \, ,
\eea

where
\be
H(a)=\int_a^\infty Dz
\ee
and
\be
\psi(a,b)=\int_a^\infty Dz\;
(z-b)^2=(1+b^2) H(a)+\frac{(a-2b)}{\sqrt{2\pi}}e^{-a^2/2} \, .
\ee
The following table gives the reconstruction threshold $\rho_0$
obtained for the same values of the $\alpha$ with probabilistic reconstruction (left) and $\ell_1$
(right). The $\ell_1$ results are obtained by optimizing over $J_1,J_2$
in the window $[0.03,1]$.
We notice that the results of $\ell_1$  with optimal $J_1, J_2$ are slightly worse
than the results of $\ell_1$ with just one block $L=1$.
 \begin{table}[!ht]
\begin{tabular}{||c|c|c|c|c|c|c||}
  \hline
  $\rho_0^{BP}$ & $\alpha$ & $\alpha_{\rm seed}$ & $\alpha_{\rm bulk}$ & $J_1$ & $J_2$ & $L$ \\
  \hline
  $0.1$ & $0.130$ & $0.3$ & $0.121$ & $40$ & $1.2$ & $20$ \\
  $0.2$ & $0.227$ & $0.4$ & $0.218$ & $10$ & $0.8$ & $20$ \\
  $0.3$ & $0.328$ & $0.6$ & $0.314$ & $8$ & $0.4$ & $20$ \\
  $0.4$ & $0.426$ & $0.7$ & $0.412$ & $4$ & $0.4$ & $20$ \\
  $0.6$ & $0.624$ & $0.9$ & $0.609$ & $2$ & $0.2$ & $20$ \\
  $0.8$ & $0.816$ & $0.95$ & $0.809$ & $2$ & $0.2$ & $20$ \\
\hline
\end{tabular}
\quad\quad\quad\quad\quad\quad
\begin{tabular}{||c|c|c|c|c|c|c|c||}
  \hline
  $\rho_0^{\ell_1}$ & $\alpha$ & $\alpha_{\rm seed}$ & $\alpha_{\rm bulk}$ & $J_1$ & $J_2$ &
  $L$ & $\rho_0(L=1)$\\
  \hline
  $0.014$ & $0.130$ & $0.3$ & $0.121$ & $0.03$ & $0.31$ & $20$ & 0.016 \\
  $0.056$ & $0.227$ & $0.4$ & $0.218$ & $0.03$ & $0.31$ & $20$ &0.059\\
  $0.096$ & $0.328$ & $0.6$ & $0.314$ & $.097$ & $0.57$ & $20$ & 0.100\\
  $0.145$ & $0.426$ & $0.7$ & $0.412$ & $.03$ & $0.57$ & $20$ & 0.150\\
  $0.278$ & $0.624$ & $0.9$ & $0.609$ & $.175$ & $0.57$ & $20$ &0.283\\
  $0.476$ & $0.816$ & $0.95$ & $0.809$ & $.175$ & $0.31$ & $20$ & 0.481 \\
 \hline
\end{tabular}
\caption{Parameters used for the probabilistic BP reconstruction of the Gaussian
  signal (left) and with the seeded $\ell_1$. \label{sBEP-param}}
\end{table}

Altogether, this demonstrates that the gain in performance using seeding matrices is really specific to the Bayes inference approach and
hence  it is the combination of the probabilistic approach, the message passing reconstruction with parameter learning, and the seeding
design of the measurement matrix that is able to reach the best possible performance.

\section{Equations for a mixture of Gaussians}
\label{appendix:mixture}

We consider here the case when the signal model is a mixture of $G$
Gaussians
\be
\phi(x)=\sum_{a=1}^G w_a {\cal N}(\overline x_a,\sigma^2_a)\, ,
\ee
where $w_a$ are non-negative weights $\sum_{a=1}^G w_a=1$. The
functions $f_a$ and $f_c$ (\ref{f_a_gen}-\ref{f_c_gen}) needed by the BP algorithm are then

\bea
f_a(\Sigma^2,R) &=& \frac{  \rho \sum_{a=1}^G w_a
e^{-\frac{(R-\overline x_a)^2}{2(\Sigma^2+\sigma_a^2)}}
\frac{\Sigma}{(\Sigma^2+\sigma_a^2)^{\frac{3}{2}}} (\overline x_a \Sigma^2
+ R \sigma_a^2) }{     (1-\rho)
e^{-\frac{R^2}{2\Sigma^2}} + \rho \sum_{a=1}^G w_a \frac{\Sigma}{\sqrt{\Sigma^2+\sigma_a^2}}
e^{-\frac{(R-\overline x_a)^2}{2(\Sigma^2+\sigma_a^2)}} } \, ,\\
f_c(\Sigma^2,R) &=& \frac{  \rho \sum_{a=1}^G w_a
e^{-\frac{(R-\overline x_a)^2}{2(\Sigma^2+\sigma_a^2)}}
\frac{\Sigma}{(\Sigma^2+\sigma_a^2)^{\frac{5}{2}}} \left[ \sigma_a^2
  \Sigma^2 (\Sigma^2 + \sigma_a^2)+ (\overline x_a \Sigma^2
+ R \sigma_a^2)^2 \right]}{     (1-\rho)
e^{-\frac{R^2}{2\Sigma^2}} + \rho \sum_{a=1}^G w_a \frac{\Sigma}{\sqrt{\Sigma^2+\sigma_a^2}}
e^{-\frac{(R-\overline x_a)^2}{2(\Sigma^2+\sigma_a^2)}} } - f_a^2 \, .
\eea

For a signal that itself is a mixture of Gaussians
\be
\phi_0(x)=\sum_{a=1}^{G_0} w^0_a {\cal N}(\overline x^0_a,(\sigma_a^0)^2)\, ,
\ee
the density evolution equations (\ref{eq_m_gen}-\ref{eq_q_gen}) simplify into single-Gaussian-integral equations
\bea
     E &=& \rho_0 \overline{s^2} -2 \rho_0 \sum_{a=1}^{G_0} w^0_a \overline x_a^0 \int {\cal
       D}z  f_a(  \frac{1}{\hat m} , z\sqrt{ (\sigma_a^0)^2  + \frac{\hat
         q}{\hat m^2}  }+ \overline x_a^0)  -2 \rho_0 \sum_{a=1}^{G_0} w^0_a
     \hat m (\sigma_a^0)^2  \int {\cal
       D}z  f_c( \frac{1}{\hat m} , z\sqrt{ (\sigma_a^0)^2  + \frac{\hat
         q}{\hat m^2} }+ \overline x_a^0)   \, , \nonumber \\
   &+&  (1-\rho_0)   \int {\cal D}z   \, f^2_a( \frac{1}{\hat m} ,
     z\frac{\sqrt{\hat q} }{\hat m}) + \rho_0 \sum_{a=1}^{G_0} w^0_a  \int {\cal D}z
     f^2_a( \frac{1}{\hat m}, z\sqrt{ (\sigma_a^0)^2  + \frac{\hat
         q}{\hat m^2}  }+
     \overline x_a^0)  \, ,  \label{eq_E_gen_GM} \\
     V &=&  (1-\rho_0)   \int {\cal D}z   \, f_c( \frac{1}{\hat m} ,
     z\frac{\sqrt{\hat q}}{\hat m}) + \rho_0 \sum_{a=1}^{G_0} w^0_a  \int {\cal D}z f_c( \frac{1}{\hat m}, z\sqrt{(\sigma_a^0)^2  + \frac{\hat
         q}{\hat m^2} }+ \overline x_a^0)
   \, , \label{eq_V_gen_GM}
\eea
where we used integration per-parts to obtain the simplification in
the first equation. We took advantage of the fact that a double
Gaussian integral of a function that depends only on a sum of the
Gaussian variables can be written as a single Gaussian integral with
variance being the sum of variances and mean being the sum of means. We remind $1/\hat m =(\Delta+V)/\alpha$, and $\hat q/\hat m^2 =
(\Delta_0+E)/\alpha$.

Under the optimal Bayesian inference when $\phi_0(x)=\phi(x)$,
$\rho_0=\rho$, $\Delta=\Delta_0$
the system of two equations reduces into a single one, since $E=V$ and
$\hat q = \hat m$.

\newpage
\bibliographystyle{nature}
\bibliography{refs}

\end{document}